\documentclass[]{aa}  

\usepackage{graphicx}
\usepackage{derivative}
\usepackage[dvipsnames]{xcolor}
\usepackage{float}
\usepackage{rotating}
\usepackage{booktabs}
\usepackage{mathtools}
\usepackage[thinc]{esdiff}
\usepackage{placeins}
\usepackage{bm}
\usepackage{longtable}
\usepackage{adjustbox}
\usepackage{supertabular}
\usepackage{siunitx}
\usepackage[hang,flushmargin]{footmisc}

\usepackage{comment}
\usepackage{multirow}
\usepackage{tabularx}
\usepackage{url}

\usepackage{natbib}

\usepackage[unicode]{hyperref}
\hypersetup{colorlinks=true,
            linkcolor=blue,
            citecolor=blue,
            filecolor=magenta,
            urlcolor=magenta}
            
\usepackage{footmisc}

\usepackage{txfonts}

\begin{document}

   \title{Hidden worlds: a non-transiting candidate planet in the Neptunian desert around the solar-type pulsator KIC~9139163}
   \titlerunning{Hidden worlds}

    \author{S.N.~Breton\inst{1,}\thanks{S.N.~Breton and A.~Dyrek contributed equally to this work and can both claim this article as first authors.}
           \and
           A.~Dyrek\inst{2,*}
           \and
           C.~Allende~Prieto\inst{3,4}
           \and
           A.~Bonfanti\inst{5}
           \and
           H.J.~Deeg\inst{3,4}
           \and
           R.A.~Garc\'{i}a\inst{6}
           \and 
           E.~Pallé\inst{3,4}
           \and
           F.~Pérez~Hernández\inst{3,4}
           \and
           O.~Benomar\inst{7,8,9}
           \and
           L.~Borg\inst{6}
           \and
           I.~Carleo\inst{10}
           \and
           C.~Gourvès\inst{6}
           \and
           A.F.~Lanza\inst{1}
           \and
           K.~Masuda\inst{11}
           \and
           S.~Mathis\inst{6}
           \and
           S.~Mathur\inst{3,4}
           \and
           D.B.~Palakkatharappil\inst{6}
           \and
           A.R.G.~Santos\inst{6, 12}
           \and 
           F.~Selsis\inst{13}
          }
    \institute{INAF – Osservatorio Astrofisico di Catania, Via S. Sofia, 78, 95123 Catania, Italy
    \and
    Space Telescope Science Institute, Baltimore, Maryland, USA \\
    \email{sylvain.breton@inaf.it, adyrek@stsci.edu}
    \and
    Instituto de Astrof\'isica de Canarias (IAC), V\'ia L\'actea S/N, La Laguna 38200, Tenerife, Spain 
    \and
    Departamento de Astrof\'isica, Universidad de la Laguna, La Laguna 38200, Tenerife, Spain 
    \and
    Space Research Institute, Austrian Academy of Sciences, Schmiedlstrasse 6, A-8042 Graz, Austria
    \and
    Universit\'e Paris-Saclay, Universit\'e Paris Cit\'e, CEA, CNRS, AIM, 91191, Gif-sur-Yvette, France
    \and
    Department of Astronomical Science, School of Physical Sciences, SOKENDAI, 2-21-1 Osawa, Mitaka, Tokyo 181-8588, Japan
    \and
    Solar Science Observatory, National Astronomical Observatory of Japan, 2-21-1 Osawa, Mitaka, Tokyo 181-8588, Japan
    \and
    New York University Abu Dhabi, Center for Space Science, PO Box 129188, Abu Dhabi, UAE
    \and
    INAF – Osservatorio Astrofisico di Torino, Via Osservatorio 20, 10025 Pino Torinese, Italy
    \and
    Department of Earth and Space Science, Osaka University, Osaka 560-0043, Japan
    \and 
    Instituto de Astrof\'isica e Ci\^encias do Espa\c{c}o, Universidade do Porto, CAUP, Rua das Estrelas, PT4150-762 Porto, Portugal
    \and
    Laboratoire d'astrophysique de Bordeaux, University of Bordeaux, CNRS, Pessac, France
    }

   \date{}

 \abstract{
  Close-in substellar companions experience strong tidal and magnetic interactions with their host stars and are therefore subject to fast orbital evolution. Moreover, due to the strong irradiation of their surface, they exhibit orbital phase modulations of significant amplitudes, allowing to characterise their atmospheric {radiative properties}. In particular, this means that companions can be detected and characterised even in the absence of transits. The solar-type pulsator KIC~9139163 exhibits in its light curve a stable 0.6-day modulation for which the best explanation is the presence of a close-in non-transiting companion that we therefore attempt to characterise. We combine \textit{Kepler} and TESS photometric data with spectroscopic observations obtained with the High Accuracy Radial velocity Planet Searcher for the Northern hemisphere (HARPS-N). We validate KIC~9139163 surface stellar rotation period and we use the HARPS-N spectra to refine the stellar modelling. The analysis of the radial velocities obtained with HARPS-N provides a companion mass $M_p \sin i = 7.3 \pm 1.4 \, \mathrm{M}_{\oplus}$. {Using a joint grid fitting of the \textit{Kepler} and TESS phase curves with an ultra-hot exoplanet model, we infer a planetary radius of $2.43 \pm 0.14 \, \mathrm{R}_\oplus$, which, combined with the measured mass and retrieved inclination, implies a bulk density consistent with a hot water-rich world. This places the non-transiting companion candidate of KIC~9139163 within the Neptunian desert, a regime where planets are expected either to have lost their primordial hydrogen/helium envelopes or to harbour metal-enriched atmospheres. We further detect significant variations in amplitude between the \textit{Kepler} and TESS phase curves, obtained six years apart, as well as a secular increase in amplitude over the \textit{Kepler} baseline. 
  Our joint fit favours a model with two distinct longitudinal cloud offsets over a single-offset scenario. Both datasets indicate a moderate-to-high geometric albedo and low-to-moderate heat redistribution.
  The opposite phase offsets observed in the \textit{Kepler} and TESS datasets suggest a time-variable longitudinal brightness distribution, potentially driven by evolving cloud properties or changes in scattering behaviour. {Nevertheless, given the larger noise level and contamination affecting the TESS photometry, this interpretation remains tentative.} These results point towards a dynamic and evolving reflective component in the atmosphere of the candidate planet.}
  While making KIC~9139163 an interesting candidate for future ground follow-ups, it also suggests that searching for other non-transiting planets around fast stellar rotators in space-borne photometric surveys might provide new insights into the physics of the Neptunian planets located in the desert. 
 }


   \maketitle
   \nolinenumbers

\section{Introduction \label{section:introduction}}

Studies of non-transiting objects phase curve modulations have an important potential for optical to infrared spatial photometric and spectroscopic surveys, which main goal is to search for and characterise exoplanets, such as the \textit{Kepler}/K2 missions \citep{Borucki2010,Howell2014}, the Transiting Exoplanet Survey Satellite \citep[TESS,][]{Ricker2015}, the CHaracterising ExOPlanet Satellite \citep[CHEOPS,][]{Benz2021,Fortier2024}, the Hubble Space Telescope (HST), the James Webb Space Telescope \citep[JWST,][]{Gardner2023JWST}, the PLAnetary Transits and Oscillations of stars mission \citep[PLATO,][]{Rauer2025}, and the Atmospheric Remote-sensing Infrared Exoplanet Large-survey \citep[Ariel,][]{Tinetti2018}. 
Granting the wealth of additional information transits and eclipses are able to provide concerning orbital and planetary parameters, the analysis of data obtained from transiting systems has been widely privileged over the past decades. Nevertheless, a simple geometrical argument demonstrates that non-transiting exoplanets are expected to be much more numerous than their transiting counterparts. The detection of a few hundreds of such planets in data of the \textit{Kepler} mission was predicted by \citet{2003ApJ...595..429J} and several attempts to detect them have been performed over the past years. 
\citet{Faigler2011} designed the BEaming, Ellipsoidal and Reflection/heating periodic modulations (BEER) algorithm to look for signature of non-transiting companions in stellar photometric data. 
\citet{Shporer2011} showed that it was possible to use the BEER algorithm to detect non-transiting exoplanets and \citet{Faigler2012} exploited its ability to detect seven non-transiting low-mass companions in \textit{Kepler} light curves.
\citet{gaulme_possible_2010} and \citet{Millholland2017} identified more than 60 non-transiting Hot-Jupiter candidates. \citet{LilloBox2021} performed a radial velocity (RV) follow-up of 10 of these targets, among which they claimed the confirmation of three planetary companions. {Recently, \citet{Cullen2024} identified 27 non-transiting candidates in TESS photometric light curves, suited for RV follow-up in the up-coming years.}
Providing phase curve modulations without transiting events (which usually introduce complex harmonic patterns in the Fourier spectrum of the light curve), these systems constitute unique astrophysical laboratories to study planetary atmospherical {radiative properties} \citep[e.g.][]{Selsis2011, Maurin2012, Parmentier2018} as well as star-planet interactions \citep[e.g.][]{Mathis2018,Strugarek2018}.
Their characterisation is therefore a challenge that needs to be undertaken by modern exoplanetary science. This can only be achieved by characterising reference systems that will pave the way for future analysis. 
A key aspect of this characterisation is to combine the analysis of photometric modulations with RV measurements obtained from ground-based high-resolution spectroscopic instruments in order to at least get constraints on the $M_p \sin i$ parameter, where  $M_p$ is the mass of the companion and $i$ the orbit inclination relative to the observer.  
{Recently, \citet{gourves_non-transiting_2025} presented a catalogue of 88 targets as being possible hosts for non-transiting exoplanets. In their catalogue, the brightest and most suited target to explore the signature of a possible companion is KIC~9139163, a hot main-sequence solar-type pulsator with a 8.329 magnitude in the \textit{Kepler} photometric band. 
The putative signature for the presence of the orbital companion is a modulation at period $P \sim 0.6$~days.
In this work, we therefore attempt characterising the non-transiting candidate planet of KIC~9139163, taking advantage of the precise stellar characterisation that the combination of asteroseismology and high-resolution spectroscopy can offer in this case. We also propose to discuss the current limitations and caveat of such an analysis that require to be explored in the future.}   

Observed by the \textit{Kepler} mission in short and long cadence modes, KIC~9139163 is a F-type main-sequence solar-like pulsator that has been abundantly scrutinised in the literature. It is part, with KIC~9139151, of the wide binary system HD~176071.
\citet{Bruntt2012} used spectroscopic observations to derive fundamental parameters and chemical abundances. 
A thorough asteroseismic analysis of the acoustic modes (p modes) properties was performed by \citet{Appourchaux2012a,Appourchaux2012b}, \citet{Corsaro2014}, \citet{Benomar2015},  \citet{Lund2017}, and \citet{Hall2021} while \citet{Metcalfe2014}, \citet{SilvaAguirre2017}, and \citet{Creevey2017} provided stellar structure parameters and ages from modelling with several stellar evolution pipelines. In particular, \citet{Benomar2015} reported a possible evidence of differential rotation between the convective envelope and the top of the radiative zone. 
\citet{Salabert2018} and \citet{Santos2019b} analysed the p-mode frequency variations to look for magnetic activity signature.  
\citet{Sowicka2017} detected a unique transit-like event during \textit{Kepler} second quarter of observation, Q2.
\citet{Clarke2018} and \citet{Yang2019} classified it as a flaring star with eight flaring events during the \textit{Kepler} mission while \citet{Pezzotti2026} was able to measure its X-ray luminosity from the ROentgen Survey with an Imaging Telescope Array observation \citep[eROSITA,][]{Predehl2021}. 
Using a combination of frequency analysis methods enhanced by the random forest classification \citep[see][]{Breton2021}, \citet{Santos2021} reported a surface rotation period of $6.18 \pm 0.62$ days. In addition to this rotational signature, the light curve modulation exhibits a threshold crossing event (TCE) with a period of $\sim$0.6 day, but, in the absence of discernible transits related to this periodicity, KIC~9139163 is not considered as a Kepler Object of Interest (KOI). 
The existence of this signal is the reason why KIC~9139163 was removed from the sample considered by \citet{Breton2023} when they looked for the signature of gravity modes (g modes) in the low-frequency range of the periodogram of F-type stars. 

In this work, we provide a first attempt at characterising the 0.6-day modulation detected in KIC~9139163 light curve due to the possible presence of a planetary companion. The layout of the paper is as follows. In Sect.~\ref{sec:data}, we describe the photometric and spectroscopic observations we used for our study. {In Sect.~\ref{sec:preliminary_considerations} we present arguments to justify that the presence of a non-transiting planet is the most plausible scenario to explain the 0.6-day modulation}. In Sect.~\ref{sec:star} we provide updated stellar parameters for KIC~9139163 combining asteroseismology and our new spectroscopic observations into stellar evolution models. In Sect.~\ref{sec:rv_analysis}, we use the RV time series to put a constraint on the mass and the time of orbital conjunctions of the candidate planet, while in Sect.~\ref{sec:photometric_planet_model}, we exploit the photometric phase curve to provide a first attempt at characterising the candidate planet. In Sect.~\ref{sec:discussion}, we discuss some of the perspectives opened by the study of such non-transiting systems for past, current and future space mission as well as for ground follow-up spectroscopic surveys, and we draw the main conclusions of this exploration work in Sect.~\ref{section:conclusion}.    

\section{Considered data \label{sec:data}}

\subsection{Kepler and TESS Photometry}

\begin{figure}[ht!]
    \centering
    \includegraphics[width=.48\textwidth]{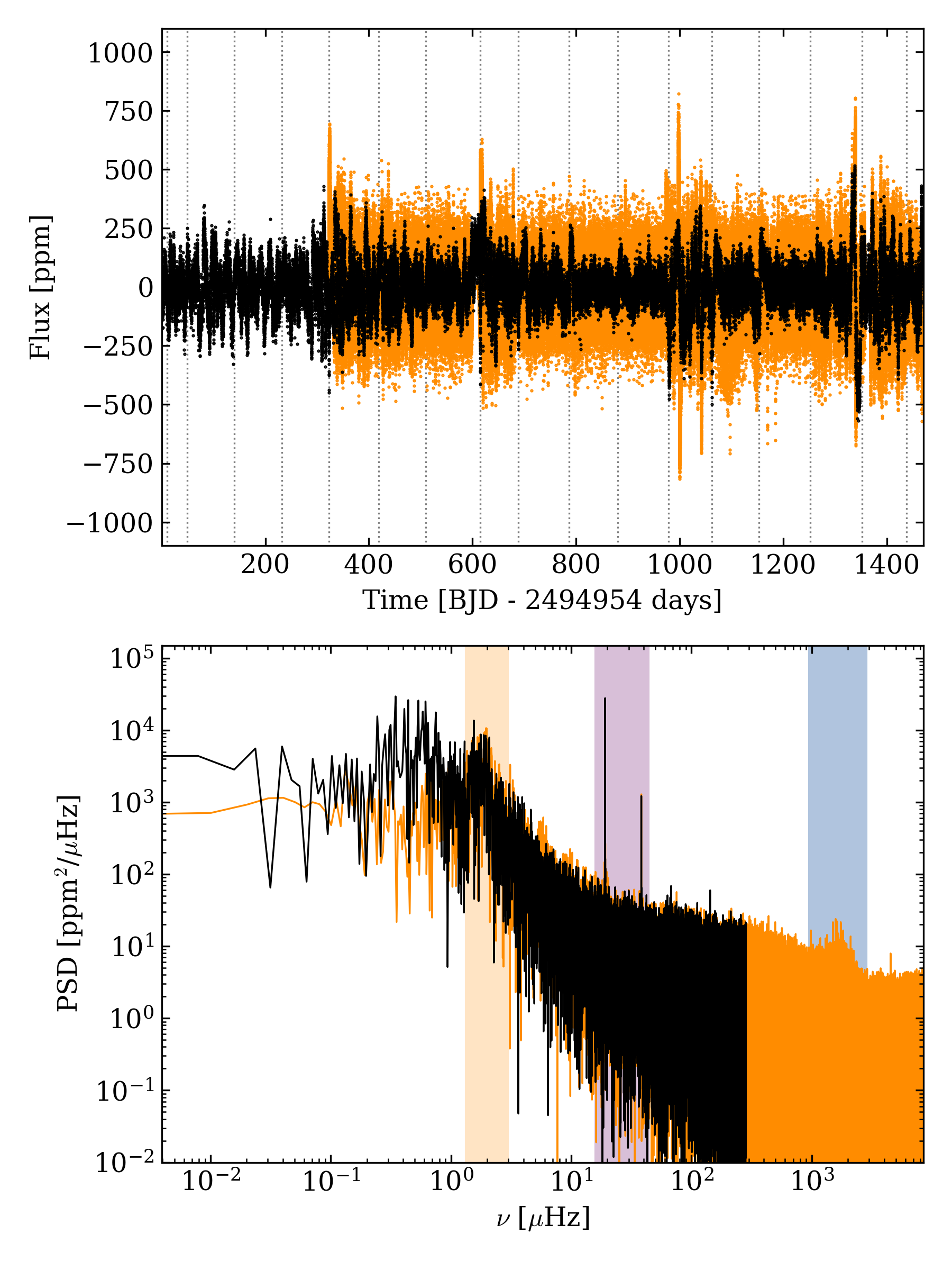}
    \caption{{Photometric variability of KIC~9139163 during the \textit{Kepler} mission.} \textit{Top:} Short (orange) and long (black) cadence KIC~9139163 \textit{Kepler} light curves. The vertical dotted lines indicate the position of the \textit{Kepler} quarters. \textit{Bottom:} Corresponding PSDs. The frequency region of the surface rotational modulation is highlighted in light orange, the 0.6-day signal and its first harmonic in purple and the p-mode hump in blue.
    }
    \label{fig:kic9139163_lc_psd}
\end{figure}

KIC~9139163 was observed in long cadence (sampling $\mathrm{dt}$ $\approx 30$~min) mode along the full extent of the \textit{Kepler} nominal mission (1470 days, from May 2009 to May 2013), and almost continuously in short cadence ($\mathrm{dt} \approx 58.8$~s) mode from Quarter 5 to Quarter 17 (1147 days, from March 2010 to May 2013). The data was calibrated with KADACS and the products (light curves) are called KEPSEISMIC. We analysed the KEPSEISMIC data\footnote{The data is available at: \url{https://archive.stsci.edu/prepds/kepseismic/}.} that were reduced and calibrated following the methodology described in \citet{Garcia2011,Garcia2014,Pires2015}, yielding the light curves represented in Fig.~\ref{fig:kic9139163_lc_psd}.
The corresponding power spectral density (PSD) is also shown. The p-mode power hump can be seen in the short-cadence PSD while two strong harmonics associated with the candidate non-transiting companion signal are visible at 19.14 and 38.28~$\mu$Hz, corresponding to periods of 0.60474 and 0.30230 days. {In order to verify that this signal is not an artefact of the data reduction, we confirm that it is also visible in the Pre-search Data Conditioning Simple Aperture Photometry light curve \citep[PDCSAP,]{Stumpe2012,Smith2012}.}
In what follows, we adopt $P = 0.60474$~day as the reference value for the orbital period of the putative planet. By fitting in the PSD a Lorentzian profile centred at 19.14~$\mu$Hz, we obtain for $P$ a small uncertainty of $\num{5e-6}$~days. We finally note that, at a frequency of around 1.9~$\mu$Hz, a slight power excess likely related to the surface rotational modulation of the star is also visible, corresponding to $P_{\rm rot} \approx 6$~days. This latter feature will be analysed and commented in a wider extent in Sect.~\ref{sec:surface_rotation}.

KIC~9139163 has also been observed during the Transiting Exoplanet Survey Satellite \citep[TESS,][]{Ricker2015} mission\footnote{In TESS sectors 14, 26, 40, 41, 53, 54, 55, 75, 81, and 82. Data is available via the MAST archive: \url{https://mast.stsci.edu/portal/Mashup/Clients/Mast/Portal.html}.} as TIC~164670309. However, due to the large TESS pixel size, the KIC~9139163 light curve is contaminated by the contribution from its wide binary companion KIC~9139151 \citep[which is also a seismic late F-type star, see e.g.][]{Lund2017}. {Nevertheless, the peak at $P = 0.60474$~day remains clearly visible, as illustrated in Fig.~\ref{fig:tess_lc_psd}.}



\subsection{Spectroscopic observation with HARPS-N \label{RV_main}}

Between March $26^{\rm th}$ 2022 (UT) and August $23^{\rm rd}$ 2022 (UT), 59 spectra of KIC~9139163 were collected with the High Accuracy Radial velocity Planet Searcher for the Northern hemisphere \citep[HARPS-N: $\lambda$\,$\in$\,(378--691)\,nm, R$\approx$115\,000,][]{Cosentino2012} mounted at the 3.58-m Telescopio Nazionale Galileo (TNG) of Roque de los Muchachos Observatory in La Palma, Spain, under the observing program CAT22A\_48. The exposure time was set to 300--480 seconds, based on weather conditions and scheduling constraints leading to an signal-to-noise ratio (SNR) per pixel of 55--105 at 550\,nm. 

The spectra were extracted using the off-line version of the HARPS-N Data Reduction Software \citep[DRS,][]{Cosentino2014}, version  {\tt3.7}. {Absolute radial velocities (RVs) and spectral activity indicators (Full Width at Half-Maximum, contrast and Bisector Span of the cross-correlation function, as well as the Mount-Wilson $S$ index of chromospheric activity)} were measured using an on-line version of the DRS, the YABI tool\footnote{Available at: \url{http://ia2-harps.oats.inaf.it:8000}.}, by cross-correlating the extracted spectra with a G2 mask \citep{Baranne1996}. In particular, we obtain $S = 0.143 \pm 0.002$, which matches the previous observation from \citet{Karoff2019}. 


\section{Preliminary considerations on the nature of the 0.6-day signal \label{sec:preliminary_considerations}}

{
\subsection{The signal as seen by \textit{Kepler} and TESS over more than 15 years of observation}

The combination of \textit{Kepler} and TESS observations allows us to explore the characteristics of the signal over more than 15 years of observation, from 2009 to 2024. After applying a high-pass filter with cutoff at $1.15$~days (10~$\mu$Hz), the 0.60474-day binned phase-folded modulation of the \textit{Kepler} and TESS light curves is shown in Fig.~\ref{fig:kepler_tess_phase}. While the phase of the signal in \textit{Kepler} remains stable during the four years of operation of the spacecraft, the amplitude of the modulation, estimated through the fit of a sinusoidal function to each quarter, increases over time. 

Between 2014 and 2019, there is a lack of observation to monitor the evolution of the signal. The first TESS observation dates back to July 2019. Contrary to what was seen in the \textit{Kepler} datasets, the TESS light curve does not exhibit significant variation of the phase curve amplitude. Similarly to \textit{Kepler}, the phase appears to remain stable. For both \textit{Kepler} and TESS, in order to evaluate how the level of stellar activity vary at the same time, we compute the standard deviation of the low-frequency component of the light curve, $\sigma_{\rm low}$, below $1.15$~days. In this case, therefore, we apply a low-pass filter with the same $1.15$~days cutoff, and we take the same binning factor as for the corresponding phase curve. This allows us to show that the amplitude variation of the 0.6-day signal is not correlated with low-frequency variability that might be induced by stellar activity. We also underline that the larger $\sigma_{\rm low}$ values in the TESS light curves do not necessarily reflects an increase in intrinsic stellar magnetic activity, but may instead be caused by a reduced instrumental stability at low-frequency from TESS, with respect to \textit{Kepler}. Nevertheless, the most striking feature between the \textit{Kepler} and TESS phase curves is that the phase and/or the amplitude of the phase-curve signal underwent a complete reversal between 2014 and 2019. Even if \textit{Kepler} and TESS observing passbands differ, their overlap is sufficient to rule out the possibility that the phase reversal in the signal comes only from a wavelength-dependent variability. 

We finally note that the TESS phase curve is one order of magnitude noisier than \textit{Kepler}, as reported in the literature \citep{lu_astraea_2020}. Indeed, TESS has higher shot noise because it uses cameras with 10 cm apertures, whereas \textit{Kepler} has a 95 cm mirror. TESS pixels are larger compared to \textit{Kepler} (respectively 21 arcsec $\rm px^{-1}$ and 4 arcsec $\rm px^{-1}$), which increases the background light and contamination from nearby stars. Plus, TESS suffers thermal changes and data downlink ramp effect every 13.7 days \citep{ricker_transiting_2015, taaki_search_2025}. To ensure that this reversal is genuine, the timings of \textit{Kepler} and TESS datasets were examined with great care. Both mission's timings are provided in Barycentric Julian Date (BJD) using the Barycentric Dynamical Time (TDB) system. However, TESS data are expressed in Barycentric TESS Julian Date (BTJD), that corresponds to BJD - 2457000, where 2457000 is the BTJD reference date, specified in the header of each file under the \texttt{BJDREFI} keyword. We therefore convert the timings of the TESS data from BTJD\_TDB to BJD\_TDB by adding the BTJD reference date to each time stamp. 

\begin{figure}[ht!]
    \centering
    \includegraphics[width=.48\textwidth]{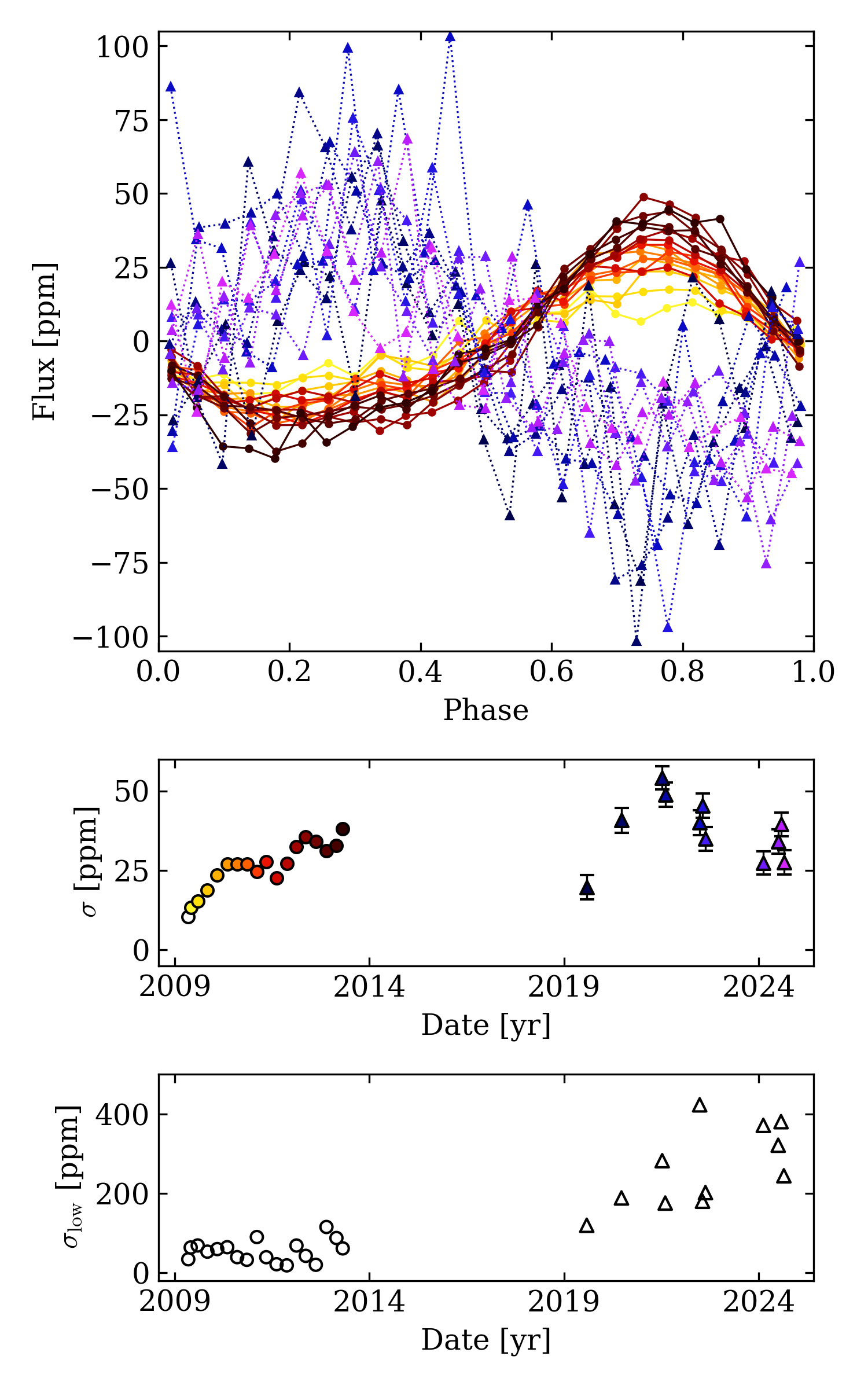}
    \caption{{Multi-year time variability of the \textit{Kepler} and TESS signals.} \textit{Top:} 35-min binned phase curve of the 0.6-day signal for each \textit{Kepler} quarter (dots connected by thick lines) and each TESS sector (triangle connected by dotted lines). Each individual phase curve is colour-coded to improve readability. A high-pass filter with cutoff at $1.15$~days was applied before computing the phase curve. \textit{Middle:} Fitted amplitude for each \textit{Kepler} (dots) and TESS (triangles) phase curve. The colour-coding scheme is the same as in the top panel. The errorbar for \textit{Kepler} are smaller than for the marker symbol. \textit{Bottom:} Standard deviation $\sigma_{\rm low}$ for the corresponding low-frequency component of the signal (period longer than $1.15$~days) in each \textit{Kepler} quarter (dots) and TESS sector (triangles).  
    }
    \label{fig:kepler_tess_phase}
\end{figure}

\subsection{No explanation from intrinsic stellar variability of solar-type stars}

For main-sequence solar-type stars, no photometric periodic modulation is expected in this range of frequencies. In the specific case of KIC~9139163, the surface rotation period is of about 6 days, therefore at significantly lower frequency. The light curve modulation associated with stellar surface rotation is quasi-periodic rather than strictly periodic due to active-region evolution and latitudinal differential rotation. As a result, the corresponding signal in the Fourier spectrum exhibits a spread of power around the main frequency \citep[see e.g.][]{Santos2019,Santos2021,Breton2024,Degott2025}. While it might be envisioned that this variability is a surface manifestation from $g$ modes trapped in the radiative interior of the star, their surface amplitudes are not expected to reach such high values in F-type stars with solar-type oscillations \citep{Breton2022simuASH,Breton2026simuASH}. Moreover, \citet{Breton2023} analysed the light curves of 34 \textit{Kepler} solar-type pulsators with global properties analogous to KIC~9139163, and did not find evidence for similar signatures in the 10-30~$\mu$Hz range.

\subsection{Ruling out the contamination possibility \label{sec:contamination_possibility}}

As pointed out by \citet{Pezzotti2026}, there is a faint \textit{Gaia} target 5 arcseconds away from KIC~9139163. Unfortunately, there is no information available on this source in the \textit{Gaia} database beyond the apparent magnitude (15.71 in the \textit{Gaia} G band while KIC~9139163 magnitude in this band is 8.27, about 1000 times brighter). One might therefore wonder if the 0.6-day modulation might originate from this secondary source. Given the period stability of the signal, the most probable explanation would be that the faint source is a pulsating star. Three classes of classical pulsators have oscillation eigenmodes with period between 0.5 and 1 day: classical  Cepheids, RR Lyrae, and $\gamma$ Dor pulsators \citep[see e.g.][]{Aerts2010}. 

However, almost all of the galactic classical Cepheids are located in the thin disc and only one is within the \textit{Kepler} field \citep[e.g.][]{Derekas2012}. RR Lyrae stars are generally, but not exclusively, associated with globular clusters, and less than 50 of them have been studied in the \textit{Kepler} field \citep[e.g.][]{Plachy2021}. In addition, RR Lyrae stars with pulsations with a 0.6-day period, referred as RR~Lyrae~ab \citep{Bailey1902}, usually have a phase curve signature that is not sinusoidal \citep[see][]{Smith2004}. More importantly, classical Cepheids and RR~Lyrae have large-amplitude variability, of at least half a magnitude. This means that even diluted in KIC~9139163 variability, their pulsation would create a modulation in the light curve with an amplitude of at least 1250 ppm, significantly much larger than what is observed.

Concerning $\gamma$ Dor pulsators, they are more common in the \textit{Kepler} field than classical Cepheids and RR Lyrae \citep[e.g.][]{Tkachenko2013,VanReeth2016,Li2020}, and diluted in the KIC~9139163 signal, their apparent brightness variation could match the amplitude we observe. Nevertheless, given that their pulsations are, from a dynamical point-of-view, unstable modes (that is, from a theoretical point-of-view, reacting to the perturbation of the equilibrium state, their amplitude grows exponentially and reach a saturated excited state), their amplitude should not vary significantly and their phase is expected to remain coherent over time. This is not what is observed for KIC~9139163 (we would expect the same phase coherence for classical Cepheids or RR Lyrae pulsations). 

Finally, RV follow-up can also play a critical role in ruling out contamination as the signal origin. In particular, the optical fibre of HARPS-N has a 1 arcsecond aperture, which is five time smaller than the separation between KIC~9139163 and the faint object.

\subsection{The exoplanet scenario: atmospheric variability with a decadal timescale}

The scenario we formulate to explain the features observed in the \textit{Kepler} and TESS phase curves is the following: the signal corresponds to the reflection-emission phase curve of a close non-transiting planet, with atmospheric features evolving on a dynamical timescale of about a decade. In particular, the contribution of a non-static cloud structure would result in a longitudinal shift of the planetary bright spot, explaining the change of phase evidenced from the \textit{Kepler} and TESS observations. Similar evidence for time-dependent atmospheric dynamics have been found in planets such as WASP-121b \citep{Changeat2024}. In the following sections, we analyse the photometric and spectroscopic data in our possession in the light of such a scenario, and we discuss the ensuing implications. 


}

\section{Deriving updated stellar parameters \label{sec:star}}

\subsection{Asteroseismology \label{sec:asteroseismology}}

Using the \texttt{apollinaire} module\footnote{The documentation for \texttt{apollinaire} is available at: \url{https://apollinaire.readthedocs.io/en/latest}.} for which a complete description can be found in \citet{Breton2022apollinaire}, we perform a new Bayesian asteroseismic analysis of KIC~9139163 short cadence data. The details of the analysis are provided in Appendix~\ref{appendix:asteroseismic_analysis}. In particular, from our analysis we obtain an inclination of the stellar rotation axis, $\mathcal{I}$, of $26.5_{-2.4}^{+2.8}$ degrees, a value that is close to the $28.0_{-4.0}^{+3.5}$ degrees and $33.5_{-3.0}^{+3.0}$ measured by \citet{Benomar2015} and \citet{Hall2021}, respectively.


\subsection{Atmospheric parameters from spectroscopy \label{sec:spectroscopy}}

We created a custom grid of synthetic spectra using 
\texttt{Synple}\footnote{The source code for \texttt{Synple} is available at: \url{https://github.com/callendeprieto/synple}.} \citep{2021arXiv210402829H}, based on 
Kurucz model atmospheres \citep{1979ApJS...40....1K, 2012AJ....144..120M}, accounting for the effective temperature, $T_{\rm eff}$, the surface gravity $\log g$, the metallicity $\rm [Fe/H]$, the alpha-element abundance $\rm [\alpha/H]$, the microturbulence $\xi$, and the macroturbulence $\zeta_{\rm RT}$. Our grid spans the ranges 
$5500 \le T_{\rm eff} \le 6500$ K (steps of 250 K), 
$3 \le \log g \le 5$ (steps of 0.5),  
$-0.5 \le$ [Fe/H]$\le +0.5$ (steps of 0.25), 
$-0.5 \le$[$\alpha$/Fe]$ \le +0.5$ (steps of 0.25), 
$0.5 \le \xi \le 2.5$ km s$^{-1}$ (steps of 0.5 km s$^{-1}$), 
$3 \le \zeta_{\rm RT} \le 5$ km s$^{-1}$ (steps of 0.25 km s$^{-1}$), 
and covering $480-520$ nm. 

The analysis was performed with the FERRE\footnote{The source code for FERRE is available at: \url{https://github.com/callendeprieto/ferre}.} code \citep{2006ApJ...636..804A}, in a way 
strictly differential relative to the Sun. FERRE was run using the Nelder-Mead algorithm
\citep{NeldMead65} and cubic B\'ezier interpolation, while the continuum was normalised on the fly using a running mean filter with a width of 0.45 nm. 
We interpolated a model for the solar parameters ($T_{\rm eff}=5777$ K, $\log g=4.437$, [Fe/H] = [$\alpha$/Fe] $=0$, $\xi = 1.1$ km s$^{-1}$, $\zeta_{\rm RT} = 3.7$) from the grid and divided the entire grid of models by the ratio of the solar spectrum reflected off the moon and observed with the same instrument \citep{2013A&A...560A..61M} to the interpolated model. With this strategy the solar spectrum is matched perfectly and the answer corresponds exactly to the desired parameters. The spectra of KIC 9139163 were then analysed with the same grid. After dropping a few of outliers with lower signal-to-noise ratio, we get excellent consistency and the mean values given in Table 1, in addition to $\xi =1.5 \pm 0.1$~km s$^{-1}$,  and $\zeta_{\rm RT} =5.4 \pm 0.2$~km s$^{-1}$.

\subsection{Stellar modelling \label{sec:stellar_modelling}}

The global surface parameters derived from spectroscopy are used as input for the stellar modelling, as well as the p-mode frequencies fitted during the seismic analysis. 
The stellar modelling was performed using the IACgrid \citep[see][]{Gonzalez-Cuesta2023,Breton2023} considering as input observables the fitted mode frequencies (see Sect.~\ref{sec:asteroseismology} and Appendix~\ref{appendix:asteroseismic_analysis}), the spectroscopic $T_\mathrm{eff}$, $\log g$, $\rm [Fe/H]$ derived in Sect.~\ref{sec:spectroscopy}, and the \textit{Gaia} luminosity from \citet{Berger2020}. 
The IACgrid is constituted of models ran with the Module for Experiments in Stellar Astrophysics \citep[MESA,][]{Paxton2011,Paxton2013,Paxton2015,Paxton2018,Paxton2019}, using standard MESA input physics, except for OPAL opacities \citep{Iglesias1996}.
The mode frequencies computed for the models of the grid \citep[with the Aarhus adiabatic oscillation package, ADIPLS,][]{Christensen-Dalsgaard2008b} are corrected for surface effects according to what is described in \citet{Perez-Hernandez2019}.

The stellar parameters of KIC~9139163 that we adopt in this work are summarised in Table~\ref{tab:global_parameter_punto}.
The stellar mass $M_\star$, radius $R_\star$, $\log g$, and age we provide in Table~\ref{tab:global_parameter_punto} are the ones of the best-fitting model obtained on the IACgrid. 
{As illustrated in Fig.~\ref{fig:stellar_modelling_comparison}, our values are consistent with the set of parameters published by \citet{SilvaAguirre2017}.}

\begin{table}[ht!]
    \centering
    \caption{Stellar parameters of KIC~9139163 and their origin. 
    }
    \adjustbox{max width=0.49\textwidth}{
    \centering
    \begin{tabular}{lll}
        \hline
        \hline
        Parameter & Value & Origin \\
        \hline
        && \\
        & Age and structure & \\
        \hline
        && \\
         Age (Gyr) & $1.554 \pm 0.072$ & stellar modelling \\
         $M_\star$ ($\rm M_\odot$) & $1.390 \pm 0.013$ & stellar modelling  \\
         $R_\star$ ($\rm R_\odot$) & $1.558 \pm 0.006$ & stellar modelling  \\
         $\log g_{\rm model}$ (dex) & $4.196 \pm 0.001$ & stellar modelling \\
         $\log g_{\rm spectro}$ (dex) & $4.169 \pm 0.050$ & spectroscopy \\
         && \\
         & Stellar atmospheric parameters & \\
         \hline
         && \\
         $T_\mathrm{eff}$ (K) & $6358 \pm 10$ & spectroscopy \\
         $\rm [Fe/H]$ (dex) & $0.10 \pm 0.05$ & spectroscopy \\
         $\rm [\alpha / Fe]$ (dex)  & $-0.046 \pm 0.050$  & spectroscopy \\
         $\xi$ (km.s$^{-1}$) & $1.5 \pm 0.1$  & spectroscopy \\
         $\zeta_{\rm RT}$ (km.s$^{-1}$) & $5.4 \pm 0.2$  & spectroscopy \\
         && \\
         &Stellar rotation and activity& \\
         \hline
         && \\
         $P_\mathrm{rot, phot}$ (day) & $6.3 \pm 2.4$ & photometry \\
         $S_\mathrm{ph}$ (ppm) & $82.5 \pm 35.6$ & photometry \\
         $S$ index  & $0.143 \pm 0.002$ & spectroscopy \\
         && \\
         & Other & \\
         \hline
         && \\
         $\mathcal{I}$ ($^o$) & $26.5_{-2.4}^{+2.8}$ & mode fitting \\
         \hline
    \end{tabular}
    }
    \tablefoot{The uncertainties provided here have to be considered as internal uncertainties that does not account for all systematics (see Appendix~\ref{appendix:modelling_comparison}). Both $\log g$ obtained from modelling and spectroscopy are reported for completeness. The parameter $\log g_{\rm model}$ incorporates the seismic constraints inputted in the grid fitting and should be favoured. Given the stellar rotation period measured from photometry, the macroturbulence broadening is mainly accounting for rotational broadening, with no significant consequences on the inferred abundances.}
    \label{tab:global_parameter_punto}
\end{table}

\subsection{Surface rotation \label{sec:surface_rotation}}

\begin{figure}[ht!]
    \centering
    \includegraphics[width=.48\textwidth]{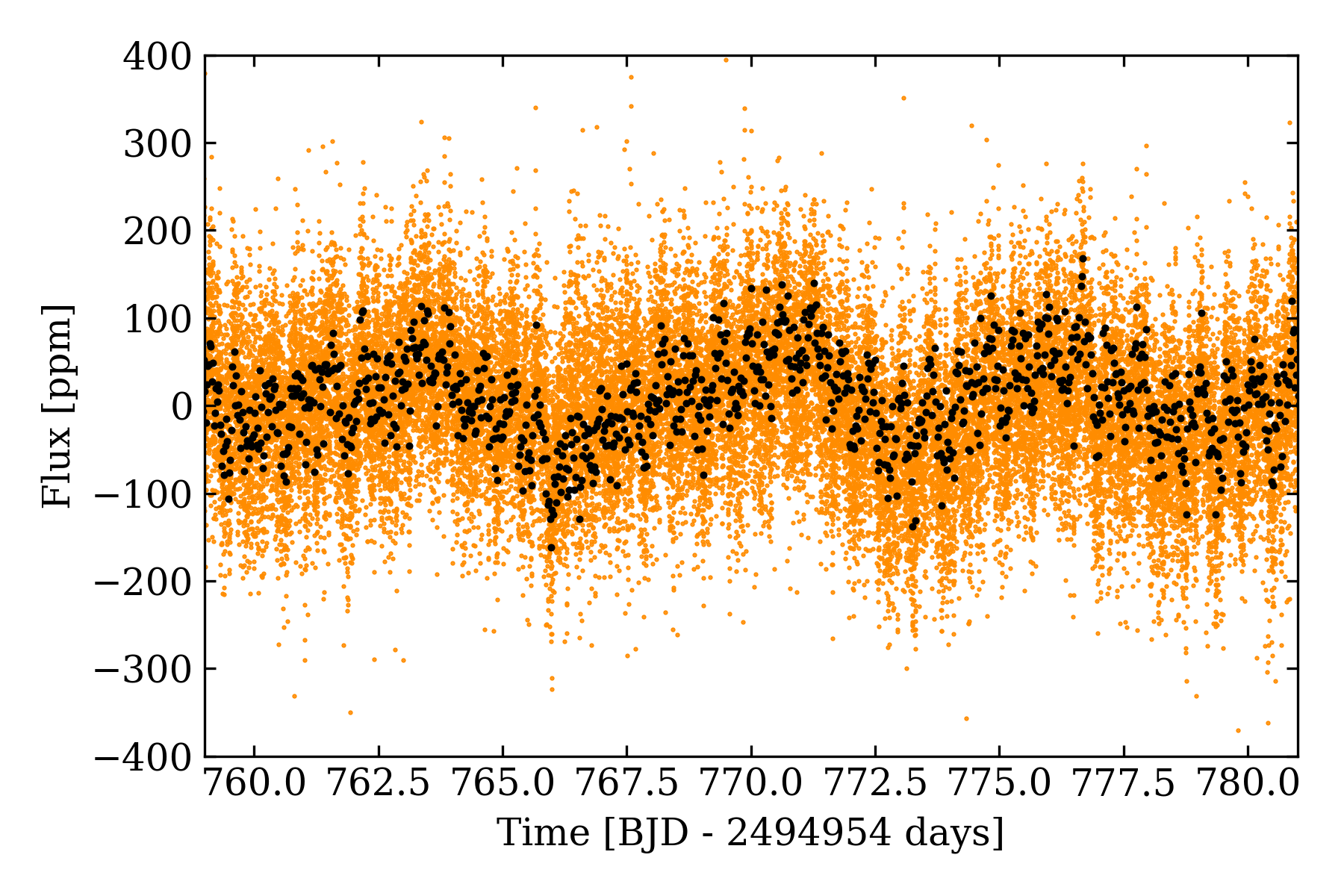}
    \caption{20-day segment of the short (orange) and long (black) cadence KIC~9139163 \textit{Kepler} light curve.}
    \label{fig:kic9139163_lc_zoom}
\end{figure}

\begin{figure}[ht!]
    \centering
    \includegraphics[width=0.49\textwidth]{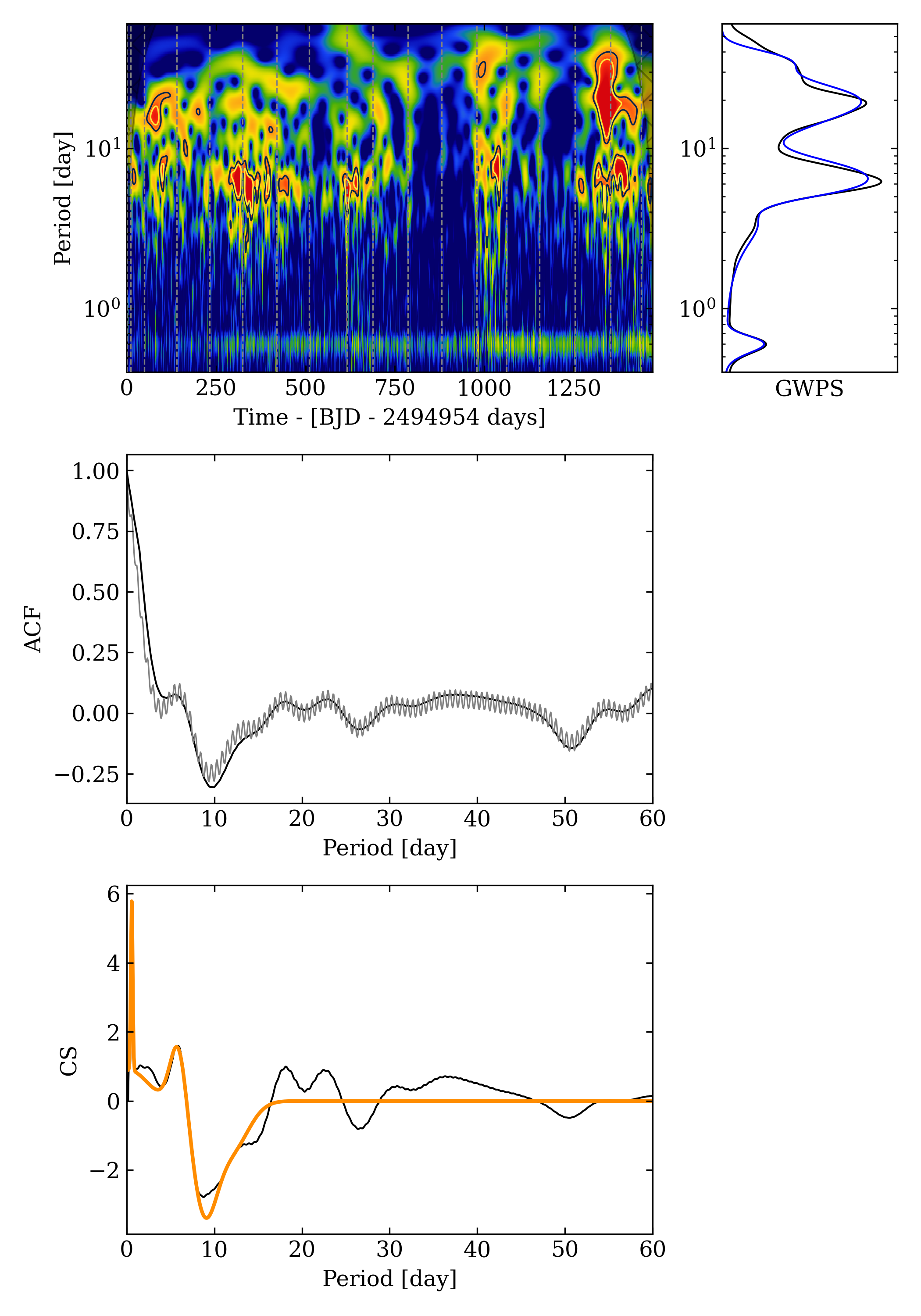}
    \caption{{Surface rotation diagnostics for KIC~9139163.} \textit{Top:} WPS (left panel) and GWPS (right panel, in black) computed from the long-cadence \textit{Kepler} light curve of KIC~9139163. The \textit{Kepler} quarters are highlighted with dashed vertical lines. The fitted Gaussian profiles are overplotted on the GWPS in blue. \textit{Middle:} ACF of the same light curve, filtering out the 0.6 day modulation (black) and including it (grey). \textit{Bottom:} CS (black) obtained from the GWPS and the ACF. The fitted Gaussian profiles are overplotted in orange.
    }
    \label{fig:kic9139163_rotation}
\end{figure}

A surface rotation period close to six days was provided for KIC~9139163 in \citet{Benomar2015} and \citet{Santos2021}.
This six-day modulation is clearly visible in the light curve, as we illustrate it by showing a $\sim$20-day section of the light curve, in Fig.~\ref{fig:kic9139163_lc_zoom}. In particular, in this figure, both 0.6-day and 6-day modulations are visible in the short cadence data.
The power excess related to the rotational modulation of KIC~9139163 is visible in the PSD (see highlighted area in the bottom panel Fig.~\ref{fig:kic9139163_lc_psd}), but difficult to characterise due to the large spread in power and the low SNR relatively to the stellar activity background. 
Given its importance for the characterisation in the system, we provide in this section additional evidence that the measurement of this {surface rotation period} in the light curve is robust. The analysis methodology follows the procedure described in \citet{Santos2019} through the \texttt{star-privateer}\footnote{The documentation is available at: \url{https://star-privateer.readthedocs.io/en/latest/}.} implementation \citep{Breton2024}. It is applied on KIC~9139163 long-cadence light curve. It is summarised here briefly and illustrated in Fig.~\ref{fig:kic9139163_rotation}.

First, the wavelet power spectrum \citep[WPS,][]{Torrence1998,Liu2007,Mathur2010} is computed, using a Morlet wavelet as the mother wavelet. Projecting the WPS along the period axis yields the global wavelet power spectrum (GWPS). Significant periodicities of the GWPS are extracted by fitting an ensemble of Gaussian profiles. 
The WPS, GWPS, and the corresponding fitted profiles are represented in the top panel of Fig.~\ref{fig:kic9139163_rotation}. 
The WPS and GWPS exhibit a strong modulation around six days, together with a wide modulation at longer period, between 10 and 20 days, that is the filtering cutoff of the light curve. 
The contribution of the 0.6-day modulation is also visible. It is important to note that this modulation appears consistently in each \textit{Kepler} quarters, ruling out the possibility that it is the signature of a contaminant star appearing only in some quarters when the pixel mask of KIC~9139163 is modified.
The light curve auto-correlation function \citep[ACF,][]{McQuillan2013} is then computed for both cases: filtering out the 0.6 modulation and keeping it in the signal. 
The profiles of the two ACFs obtained this way are compared in the middle panel of Fig.~\ref{fig:kic9139163_rotation}. When it has not been filtered out, the 0.6-day modulation signature is clearly visible in the ACF as it is one of the largest peaks in the PSD (see Fig.~\ref{fig:kic9139163_lc_psd}).
Finally, by multiplying the GWPS by the ACF with the 0.6-day signal filtered out, we obtain the composite spectrum \citep[CS,][]{Ceillier2017}. Similarly to what we did for the GWPS, we fit the significant peaks of the CS with Gaussian profiles. The CS obtained for KIC~9139163 is represented in the bottom panel of Fig.~\ref{fig:kic9139163_rotation}. 

For each of the considered methods, a signal related to the 6-day modulation is clearly visible. 
In this work, we adopt as reference for the surface rotation period the value we measure with the GWPS, $P_\mathrm{rot,phot} = 6.3 \pm 2.4$~days. We compute the photometric activity index $S_\mathrm{ph}$ as the mean value of the standard deviations of light curves segments of length $5\times P_\mathrm{rot,phot}$ \citep{Mathur2014b,Salabert2016,Salabert2017}. The photon noise is estimated with the prescription from \citet{Jenkins2010} and subtracted from each value of $S_\mathrm{ph}$. We finally get $S_{\rm ph} = 82.5 \pm 35.6$~ppm, from which we can conclude that KIC~9139163 is moderately active in visible light \citep[for reference, the solar variability during activity minima is a few tens of ppm, and around 300 ppm at activity maximum,][]{Salabert2016}.
We finally note that the stellar rotation period and the putative orbital signal are approximately in a 10:1 commensurability which could be the results of star-planet magnetic interactions \citep{Lanza2022a,Lanza2022b}.

\section{Radial velocity analysis \label{sec:rv_analysis}}

As {discussed in Sect.~\ref{sec:preliminary_considerations}}, the most straightforward hypothesis for such a stable signal with quasi-sinusoidal shape is the presence of a super-close non-transiting exoplanet \citep{gourves_non-transiting_2025} imprinting the observed brightness variation with its phase curve. This section and the following are therefore devoted to the determination of the characteristics of this candidate companion in terms of mass and radius based on the spectroscopic HARPS-N and photometric \textit{Kepler} data presented in Sect.~\ref{sec:data}. 

\subsection{The challenge of analysing non-transiting planetary signals}

The analysis of non-transiting exoplanetary phase curves is challenging as no transits nor eclipses are available within the data to constrain the orbital parameters. In fact, the primary transit depth gives access to the stellar and planetary radii ratio $(R_{p} / R_{\star})^2$. The four contact points $\tau_{I - IV}$ (only two in the case of grazing transit events) and the time of inferior conjunction $t_{\rm inf}$ are key for determining the impact parameter $b$ and therefore the semi-major axis $a$ as well as the inclination angle of the orbital plane with respect to the observer's line of sight, $i$ \citep{winn_transits_2014}. In this work, we use $t_{\rm inf}$ to refer to the time of inferior conjunction, when the planet {is in front of the star with respect to} the line of sight of the observer. {We also define the time of superior conjunction, $t_{\rm sup} = t_{\rm inf} + P/2$, when the planet passes behind the star in the line of sight of the observer.} With no transit event displayed in the photometric data, one can only access the phase curve modulations as a combination of planetary reflection and emission, Doppler boosting and ellipsoidal distortion \citep[e.g.][]{Shporer2017}.  While reflection and emission are mostly driven by atmospheric processes and stellar irradiation, Doppler boosting and ellipsoidal distortion are induced by the gravitational interaction between the two bodies, hence directly linked to their masses. In some cases, the signal seems even dominated by ellipsoidal distortion \citep[see for example Table~B.2 from][]{gourves_non-transiting_2025} which means that the actual periods corresponds to twice the one of the harmonics with largest power. However, in this case, we would expect the contribution from the planetary atmosphere to be still visible at twice the period of the main harmonics. This is not the case here (see Fig.~\ref{fig:kic9139163_lc_psd}), and we can therefore confidently assume that the only possibility for the orbital period is $P=0.60474$~day. RV data are critical to consider before attempting a comprehensive analysis of the photometric data as they are able to provide a constraint both on the companion mass and on the time of orbital conjunctions. In this section, we therefore start by analysing the RVs obtained from HARPS-N.  

\begin{figure}[ht!]
    \centering
    \includegraphics[width=.48\textwidth]{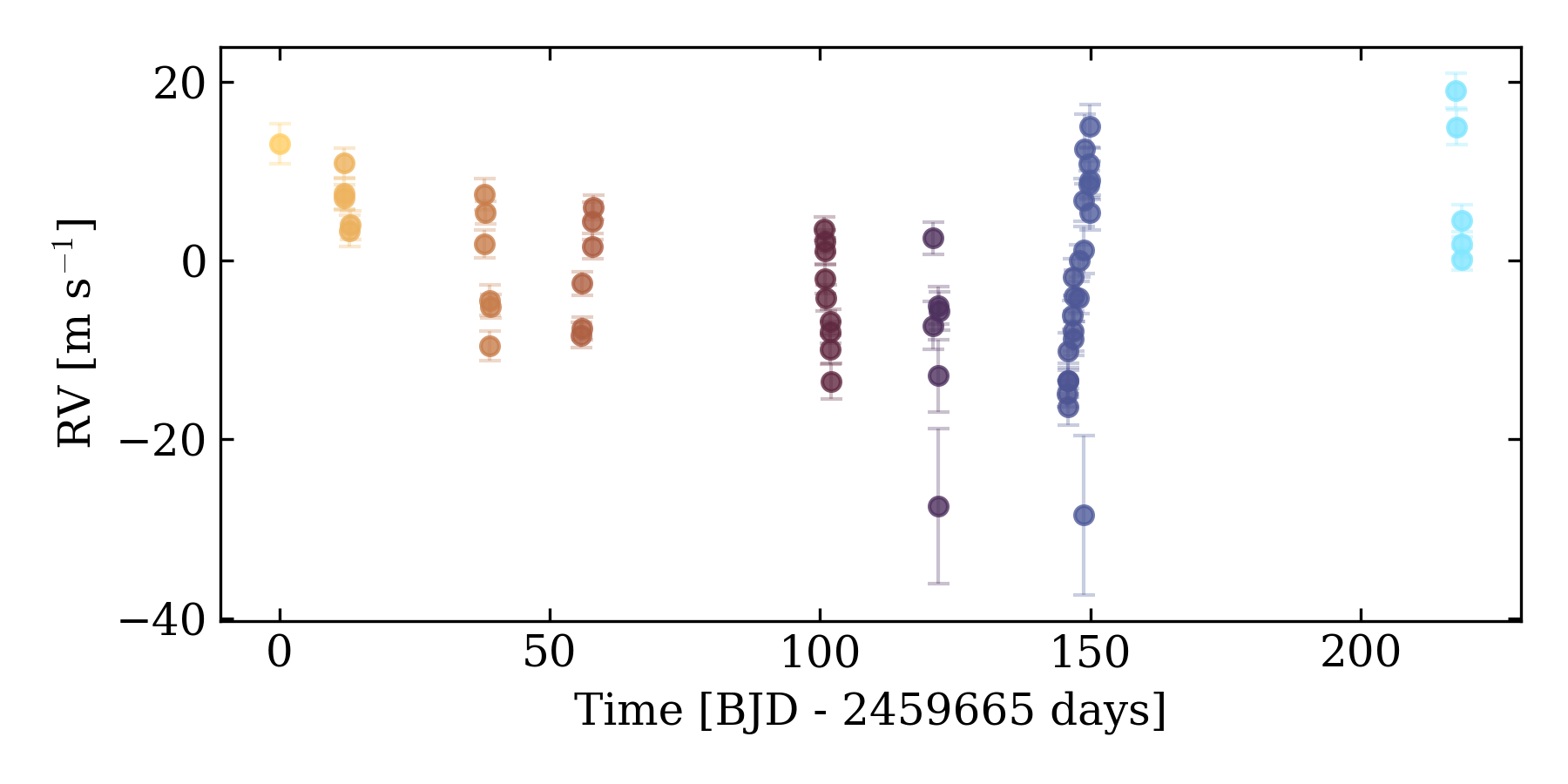}
    \caption{RV time series obtained from the SN and bp data reduction of the HARPS-N spectra.}
    \label{fig:kic9139163_rv_data}
\end{figure}

\begin{figure}[ht!]
    \centering
    \includegraphics[width=.48\textwidth]{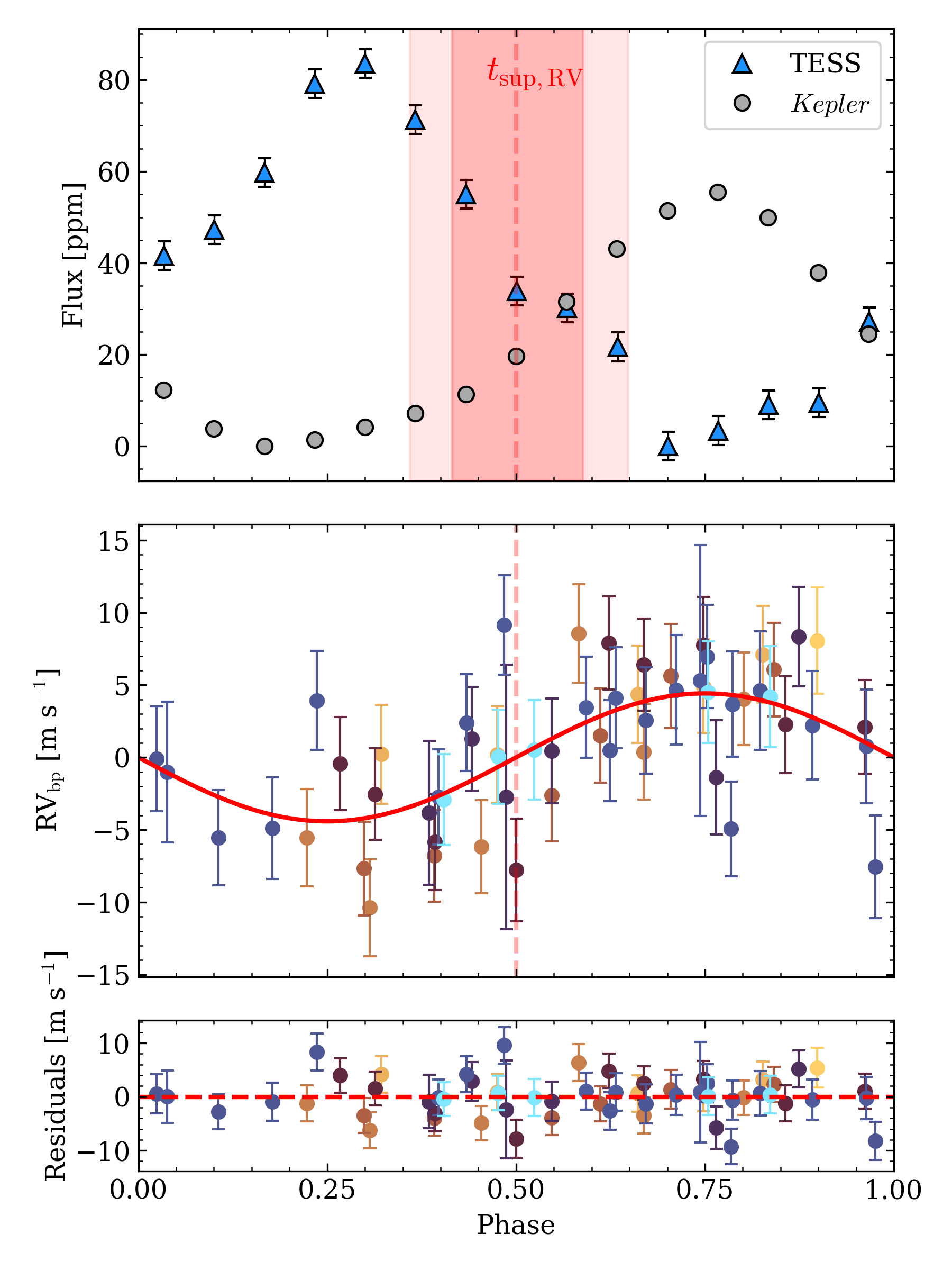}
    \caption{
    {Results of the RV analysis.}
    \textit{Top:} {Phase folded \textit{Kepler} (grey dots) and TESS (blue triangles) light curves after 58-min rebinning}. The median value for the phase of orbital {superior} conjunction, {$t_{\rm sup}$}, obtained with the RV analysis is shown with the vertical red dashed lines. The red areas highlight the corresponding 3 and 5 $\sigma$ intervals. The relative \textit{Kepler} flux was offset in order to have it positive everywhere. {The errorbars for the \textit{Kepler} data are smaller than the data points.}
    \textit{Middle:} Phase-folded data points at 0.60474 day after analysing the RV time series. The best fit obtained with the Bayesian exploration of the posterior probability is shown in red.
    \textit{Bottom:} Corresponding residuals.
    }
    \label{fig:kic9139163_rv_data_analysis}
\end{figure}

\subsection{RV extraction and analysis \label{sec:rv_extraction_analysis}}

Our RV dataset was obtained by applying the skew normal (SN) and breakpoint (bp) methods. The corresponding RV time series is shown in Fig.~\ref{fig:kic9139163_rv_data}. Following \citet{simola2019}, by fitting skew normal functions \citep{azzalini1985} onto the cross correlation functions (CCFs) extracted from the HARPS-N spectra, we obtained a SN-based RV time series $\mathrm{RV_{SN}}(t)$ along with the CCF-related activity indicators, that is the full width at half maximum $\mathrm{FWHM_{SN}}$, the contrast $A_{\mathrm{SN}}$, and the skewness $\gamma$ \citep[for further details see e.g.][and references therein]{bonfanti2023,bonfanti2025}. Active regions on the stellar photosphere evolve over time, which may possibly result in different correlation patterns between the RVs and the activity indicators \citep[e.g.][]{dewarf2010,dumusque2011,borgniet2015}. Therefore, we applied the bp method as described in \citet{simola2022} to look for the possible locations along the RV time series where the correlations of the RV data against $\mathrm{FWHM_{SN}}$, $A_{\mathrm{SN}}$, and $\gamma$ change in a statistically significant way. {To this end, we applied a hybrid approach consisting of a BIC-based \citep[Bayesian information criterion;][]{schwarz1978} model selection to detect the optimal breakpoints, followed by a Markov Chain Monte Carlo (MCMC) sampling to perform the analysis of the subsequent RV time series. When statistically justified, the stellar activity removal is significantly more effective if the RV time series is split into piecewise stationary segments identified by the bp method and the RV detrending is performed on each segment rather than on the whole time series \citep[e.g.][]{simola2022,bonfanti2023,luque2023}. Splitting our RV time series at observation 46 (i.e. 1 breakpoint) is supported by the BIC.}

The RV$_{\rm SN}$ time series was then analysed within an MCMC framework using the \texttt{MCMCI} code developed by \citet{bonfanti2020} that detrends the RV time series via low-order polynomials of ($t$, $\mathrm{FWHM_{SN}}$, $A_{\mathrm{SN}}$, $\gamma$) while fitting Keplerian signals, further accounting for the breakpoints. We first performed several \texttt{MCMCI} mini-runs (10\,000 steps) where we varied only one polynomial order at a time to assess the optimal set of polynomial orders following the BIC minimisation criterion. It turned out that the best detrending baseline consists of a linear trend against time and $\mathrm{FWHM_{SN}}$, a second-order polynomial against $\gamma$, and a third-order polynomial against $A_{\mathrm{SN}}$. After that, we performed four independent runs made of 100\,000 steps each (burn-in 20\%) to check their mutual convergence via the Gelman-Rubin statistic \citep[$\hat{R}$;][]{gelman1992}. {We fixed the orbital eccentricity at $e = 0$ and we imposed a Gaussian prior on the period following our PSD analysis in Sect.~\ref{sec:data}, while the time of inferior conjunction $t_{\rm inf}$ and the RV semi-amplitude $K$ were subject to uniform unbounded priors (except for the physical limits). The MCMC runs converged successfully ($\hat{R}\lesssim1.0002$ for all the jump parameters) and, from the posterior distributions, we inferred the following median values along with their uncertainties at the 1$\sigma$ level: $t_{\rm inf, RV}=2\,459\,664.753_{-0.017}^{+0.018}$ BJD and $K=4.42 \pm_{-0.83}^{+0.84}$\,m\,s$^{-1}$. Given the stellar mass, this implies $M_p\sin{i}=7.28_{-1.37}^{+1.38}, \mathrm{M}_{\oplus}$ with its 3$\sigma$ upper limit $(M_p\sin{i})_{3\sigma}=11.4\,\mathrm{M}_{\oplus}$. The BIC supports the inclusion of a Keplerian signal in the RV model. A comparable MCMC run without a Keplerian component yields $\Delta\mathrm{BIC}=+25$. We note that, by finding the 0.6-day signature in the HARPS-N data, we definitely confirm that the signal originates from KIC~9139163 and not a contaminant, as mentioned in Sect.~\ref{sec:contamination_possibility}.} The phase-folded and detrended $\mathrm{RV_{bp}}$ time series along with the best-fit Keplerian model is displayed in the third row  of Fig.~\ref{fig:kic9139163_rv_data_analysis}. The values for the $\rm RV_{\rm SN}$ and $\rm RV_{\rm bp}$ time series are given in Table~\ref{tab:RVsnbp}. {The very low values of the Gelman-Rubin statistic and the quick drops of the autocorrelation times visible in the trace plots of Fig.~\ref{fig:rv_autocorrelation} confirm the consistent convergence of the four independent runs and the efficient exploration of the parameter space.}

We further explored the possibility of an eccentric orbit by imposing uniform priors on $(e\sin{\omega};\,e\cos{\omega})$, being $\omega$ the argument of periastron \citep{anderson2011}; we obtained $e=0.14_{-0.10}^{+0.17}$ and $\omega=349_{-130}^{+100}\,^{\circ}$. The eccentricity is consistent with zero at $\sim$\,$1.4\sigma$, and the other jump parameters are fully compatible with the circular case, but with slightly greater uncertainties. In addition, $\Delta\text{BIC} = \text{BIC}_{e\neq0} - \text{BIC}_{e=0}=+8$ does not favour the eccentric scenario, therefore we decided to adopt the parameters obtained assuming circular orbit.

\begin{table}[h!]
    \centering
    \caption{Results of the RV analysis.}
    \centering
    \begin{tabular}{cc}
    \hline\hline
      $K_{\rm mean}$ (m s$^{-1}$) & $4.42_{-0.83}^{+0.84}$ \\
      $M_p \sin i$ ($\rm M_\oplus$) & $7.28_{-1.37}^{+1.38}$  \\
      $K_{\rm 3 \sigma}$ (m s$^{-1}$) & 6.93  \\
      $(M_p \sin i)_{\rm 3\sigma}$ ($\rm M_\oplus$) & 11.4  \\
      $t_{\rm inf, RV}$ (day) & $2\,459\,664.753_{-0.017}^{+0.018}$  \\
     \hline
    \end{tabular}
    \label{tab:rv_analysis_results}
\end{table}

\subsection{System stability}

We briefly discuss here the system dynamical stability.
\citet{Hut1980} demonstrated that a binary system was able to reach a tidal equilibrium only if the total angular momentum $L \approx L_o + L_\star$ of the system was above a critical threshold, $L_\mathrm{crit}$ verifying 
\begin{equation}
\begin{split}
   L_\mathrm{crit} &= 4 \left[\frac{1}{27} G \frac{M_\star^3 M_p^3}{M_\star + M_p} (I_\star + I_p) \right]^{1/4} \\
   &\approx 4 \left[\frac{1}{27} G M_\star^2 M_p^3 I_\star \right]^{1/4} \; ,
\end{split}
\end{equation}
where $G$ is the gravitational constant, $L_o$ is the orbital angular momentum and $L_\star$ is the stellar rotational angular momentum. $I_\star$ and $I_p$ are the stellar and planetary momentum of inertia, respectively.
Given the upper mass limits that we obtained above, no stable orbital configuration is possible when accounting for tides, which suggests that the system might be experiencing orbital decay. 

We also computed the Roche limit to assess tidal disruption risks for the companion candidate. Using both the point-mass approximation \citep{paczynski_evolutionary_1971} and the incompressible-body model \citep{jeans_1919}, we find $R_L \sim 0.0004$ AU, which is well within the stellar radius ($R_L < R_\star$). For a period of 0.60474 day, the semi-major axis is $a = 0.0155 \pm 0.00015$ AU, calculated following the prescription of \cite{seager_unique_2003}. The companion would therefore survive at this orbital distance regardless of its mass. The system is dynamically stable with respect to tidal disruption.

\section{A model for an ultra-hot exoplanet \label{sec:photometric_planet_model}}

Our goal is to model the optical phase-curve of KIC~9139163 in order to interpret the phase offset detected in the \textit{Kepler} light curve (see Fig.~\ref{fig:kic9139163_rv_data_analysis}), in combination with the RV constraints. The constraint we obtained in Sect.~\ref{sec:rv_analysis} on the time of inferior conjunction allows us to fold the \textit{Kepler} data appropriately, revealing a peak in the light curve occurring after the timing of the secondary eclipse, indicative of a phase-offset. Using a model of a planetary phase-curve, we explore the parameter space of cloud coverage and heat redistribution that can produce the observed offset, accounting for eastward and westward displacement of the planetary emission and reflection. 

\subsection{Description of the model}\label{sect:planet_model_description}

The model assumes the simplest possible configuration: a planet on a circular orbit, tidally locked in a 1:1 spin-orbit resonance, consistent with the expected tidal evolution of short-period planets \citep{mathis_tidal_2018}. The observed phase curve is modelled as the sum of an atmospheric component, comprising both reflected stellar light and thermal emission \citep[e.g.][]{Selsis2011}. The planet is treated as having a simplified, one-layer, radiative atmosphere with no vertical structure, but a temperature that varies horizontally across the surface (with longitude and latitude). 
The orbital phase $\phi$ is defined as
\begin{equation}
\label{eq:phase}
\phi = \left(t - t_{\rm inf} \right)/P \; ,
\end{equation}
where $t$ is the observation time and {$t_{\rm inf}$ is the time of inferior conjunction inferred from the fit to the RV data, {$t_{\rm inf, RV}$}}. The phase $\phi$ spans the interval $[0, 1]$ and is chosen such that $\phi = 0.5$ at secondary eclipse. 

We discretise the planetary surface into a $180 \times 90$ grid of longitude $\Phi$ and latitude $\Theta$, spanning respectively $[-\pi, \pi]$ and $[-\pi/2, \pi/2]$. 
For a synchronously rotating planet, the substellar point is fixed at longitude $\Phi = 0$, the dawn terminator at $\Phi = -\pi / 2$, and the dusk terminator at $\Phi = \pi / 2$. All thermal and reflective properties are initially 
defined in this intrinsic planetary frame.

\subsection{Reflection component}\label{sect:reflected_component}

The reflection component of the planetary flux is described as a Lambertian sphere with an additional longitudinal offset parameter, $\eta_c$
\begin{equation}
I_{\mathrm{refl}}\left(\Phi, \eta_c\right) = A_g \;\max\bigl[0, \cos(\Phi + \eta_c)\bigr] \left( \frac{R_\star}{a} \right)^2 \cdot I_{\mathrm{\star}}\left(T_{\mathrm{eff}}\right) \; ,
\label{eq:phase_curve_model_reflection}
\end{equation}
where $A_g$ is the geometric albedo, $a$ is the semi-major axis, $\Phi$ is the longitude and $R_{\star}$ is the stellar radius. For simplicity, we assume that this longitudinal pattern is uniform in latitude. $I_{\mathrm{\star}}$ is the stellar specific intensity, computed from Planck’s law at the stellar effective temperature $T_{\rm eff}$, and integrated over the \textit{Kepler} bandpass $\Delta \lambda_{Kepler}$ 
\begin{equation}
I_{\mathrm{\star}}(T_{\mathrm{eff}}) = \int_{\Delta\lambda_{telescope}} \tau_{\it{telescope}}(\lambda)\frac{2h c^2}{\lambda^5} \cdot \left(\mathrm{exp}\left(\frac{h c}{\lambda k_B T_{\mathrm{eff}}}\right) - 1\right)^{-1} \, d\lambda \; .
\label{eq:planck_law}
\end{equation}
where $\tau_{\it{telescope}}(\lambda)$ is the response function of either the \textit{Kepler}\footnote{See: \url{https://keplergo.github.io/KeplerScienceWebsite/data/kepler_response_hires1.txt}.} or the TESS missions\footnote{See: \url{https://heasarc.gsfc.nasa.gov/docs/tess/data}. }.

In this work, unlike self-consistent models \citep[e.g.][]{webber_effect_2015, Parmentier2018} where the phase offset is produced by longitudinal variations in albedo, we assume a constant albedo. The phase shift is instead only captured by the $\eta_c$ parameter. While $\eta_c$ is not fully self-consistent, it allows the model to reproduce the observed westward shift of the reflection peak, which could not be matched by the self-consistent model. {A self-consistent model would use the zenith angle $\theta$, defined as the angle between the local vertical direction at each point on the surface and the incoming stellar radiation. In Eq.~\ref{eq:phase_curve_model_reflection}, $\max\bigl[0, \cos(\Phi + \eta_c)\bigr]$ would be replaced by $\cos \theta$. The limitation in reproducing the observed light curve arises from using only the Lambertian reflection assumption, which has the cosine dependence on the stellar zenith angle and therefore strongly weights the reflected flux toward the substellar region.} As a result, even large longitudinal variations in albedo could not produce a substantial displacement of the reflected-light centroid, preventing the model from reproducing the observed phase offset. {The $\eta_c$ parameter therefore allows us to capture non-Lambertian and anisotropic scattering effects associated with inhomogeneous cloud coverage.} The concept of a cloud-induced offset is similar to that described by \citet{angerhausen_comprehensive_2015} (see Eq.~22), although the formalism differs slightly.


\subsection{Thermal emission component}\label{sect:thermal_emission}
We compute the phase-dependent atmospheric temperature defined as in \cite{jansen_detection_2020}
\begin{equation}
T_{\mathrm{atm}} (\Phi, \Theta, \varepsilon) =  \left(\left( \mathcal{P}(\Phi, \varepsilon)\; T_0(\Theta)\right)^4 + T_{\rm int}^4\right)^{1/4} \; ,
\end{equation}
{where $T_0$ is the sub-stellar temperature and $\mathcal{P}$ is the thermal phase function, a temperature that is dimensionless, defined in \citet{cowan_model_2011}. $\varepsilon$ is a dimensionless constant that quantifies the atmospheric energy redistribution efficiency. $T_{\mathrm{int}}$ is the internal planetary temperature, a measure of the internal heat flux escaping from its deep interior, independent of heating from its host star. This parameter is one of the least constrained one in the literature with an order of magnitude of 100 K for Jupiter and $\sim50$ K for Neptune \citep{guillot_interior_2004, scheibe_thermal_2021}. For exoplanets, it can theoretically take higher values to explain the ``radius inflation'' problem, where hot Jupiters or Neptune-like planets are larger than predicted by standard models \citep{thorngren_intrinsic_2019}, which we don't consider here. In this model, we only consider super-rotating winds, with $\varepsilon > 0$ and an internal temperature of 100 K. The sub-stellar temperature $T_0$ is given by}
\begin{equation}
T_0(\Theta) = T_{\rm eff} \left( \frac{R_{\star}}{a} \right)^{1/2} \left(1 - A_g\right)^{1/4} \; \cos^{\frac{1}{4}}{\Theta};,
\end{equation}

In the case of circular orbits, $\mathcal{P}$ is given as the solution to the ordinary differential equation
\begin{equation}
\frac{\mathrm{d}\mathcal{P}}{\mathrm{d}\Phi} = \frac{1}{2\varepsilon} \; \left( \cos\Phi + |\cos\Phi| - \mathcal{P}^4 \right)  \; ,
\label{eq:ODE_P}
\end{equation}
Our model builds on the formalism introduced by \citet{cowan_model_2011, hu_semi-analytical_2015, jansen_detection_2020}. 
Because Eq.~(\ref{eq:ODE_P}) cannot be integrated in closed form, we compute $\mathcal{P}$ numerically using the ODE solver provided in \texttt{scipy}. The equation is integrated over the bounds $\Phi \in [-\pi/2,\, 3\pi/2]$, corresponding to a complete cycle from dawn back to dawn. The longitudinal temperature asymmetry produces an eastward hotspot offset (towards the evening terminator), that depends on the efficiency of heat redistribution. The integration is initialised using the approximate dawn-side value derived by \citet{cowan_model_2011},
\begin{equation}
\mathcal{P}_{\rm dawn}(\varepsilon) \equiv \mathcal{P} (-\pi/2, \varepsilon) \simeq 
\left(
\, \pi + \left( \frac{3\pi}{\varepsilon} \right)^{4/3}
\right)^{-1/4} \; ,
\end{equation}


{
The emitted specific intensity is computed similarly to Eq.~(\ref{eq:planck_law})
\begin{multline}
I_{\mathrm{em}}(\Phi, \Theta, \varepsilon) = \int_{\Delta\lambda_{telescope}} \tau_{\it{telescope}}(\lambda)\frac{2h c^2}{\lambda^5} \cdot \\\left(\mathrm{exp}\left(\frac{h c}{\lambda k_B T_{\mathrm{atm}}(\Phi, \Theta, \varepsilon)}\right) - 1\right)^{-1} \, d\lambda \; .
\end{multline}

\subsection{Composite phase-curve}
To compute the phase-curve, we transform from the planetary reference frame to the observer’s viewing geometry \citep[e.g.][]{Selsis2011, Maurin2012}. {At each orbital phase, the observer’s direction is determined by the system inclination $i$ and orbital phase $\phi$, where the term $\cos\alpha_\phi$ represents the cosine of the angle $\alpha$ at phase $\phi$ between the observer's line of sight and the normal to the surface element of each cell of the discretised planet. The sum is taken over all grid cells that are visible to the observer,
that is, when $\cos\alpha_\phi > 0$.} The total flux from the candidate planet in the \textit{Kepler} observational bandpass as a function of the orbital phase $\phi$ is
\begin{multline}
F_{\mathrm{total}}\left(\phi, \Phi, \Theta, \varepsilon\right) = \sum_{\mathrm{visible}} \left(I_{\mathrm{refl}}(\Phi, \eta_c) + I_{\mathrm{em}}(\Phi, \Theta, \varepsilon)\right) \cdot \\ \mathcal{A}(\Phi, \Theta) \cdot \cos \alpha_{\phi} \; ,
\end{multline}
where $\mathcal{A}(\Phi, \Theta)$ is the area element of each grid cell, in square radian (sr).
Finally, we compute the planet-to-star power contrast as
\begin{equation}
P_{\mathrm{atm}}(\phi, \Phi, \Theta, \varepsilon) = 10^{6} \; \frac{F_{\mathrm{total}}(\phi, \Phi, \Theta, \varepsilon)}{F_{\star, \rm{disk}}} \cdot \left(\frac{R_p}{R_\star}\right)^2\; \mathrm{ppm},
\label{eq:planet_star_contrast}
\end{equation}
where $F_{\star, \mathrm{disk}}$ is the stellar flux of a uniformly bright stellar disk in the same bandpass expressed by 
\begin{equation}
F_{\star, \rm{disk}} = \pi \int_{\Delta\lambda_{telescope}} I_{\star}\left(\lambda, T_{\mathrm{eff}}\right)\;  d\lambda \; .
\end{equation}

\subsection{Mass-dependent photometric effects}

Photometric effects due to gravitational interactions can be included alongside the atmospheric contributions. These consist of tidal ellipsoidal distortion and Doppler boosting \citep[e.g.][]{Shporer2017}. Both effects depend on the candidate planets’s mass and can be quantified through the $M_p \sin i$ parameter derived from the RV analysis. The total power contrast can thus be written as
\begin{multline}
P_{\mathrm{norm}}(\phi, \Phi, \Theta, \varepsilon, A_{\mathrm{boost}}, A_{\mathrm{ellip}}) = P_{\rm{atm}}\left(\phi, \Phi, \Theta, \varepsilon\right) \\ + P_{\star}\left(\phi, A_{\mathrm{boost}}, A_{\mathrm{ellip}}\right) \; ,
\end{multline}
where $P_{\mathrm{atm}}$ represents the atmospheric contribution described previously (see Eq.~\ref{eq:planet_star_contrast}), and $P_{\star}$ corresponds to the stellar flux modulation due to ellipsoidal distortion and Doppler boosting. The amplitudes of these effects, $A_{\mathrm{ellip}}$ and $A_{\mathrm{boost}}$, are defined by Eqs.~(\ref{eq:A_ellip}) and (\ref{eq:A_beam}), respectively.
Following \citet{Shporer2017} and \citet{LilloBox2021}, the stellar modulation term is expressed as
\begin{equation}
P_{\star}\left(\phi, A_{\mathrm{boost}}, A_{\mathrm{ellip}}\right) = -A_{\rm {ellip}} \cos (4 \pi \phi)+A_{\rm {boost}} \sin (2 \pi \phi) \; ,
\end{equation}
with the amplitude of the ellipsoidal distortion defined as
\begin{equation}
\begin{aligned}
A_{\rm {ellip}} & =13 \alpha_{\rm {ellip }} \sin i\left(\frac{R_{\star}}{R_{\odot}}\right)^3 \\
& \times\left(\frac{M_{\star}}{M_{\odot}}\right)^{-2}\left(\frac{P}{\rm { day }}\right)^{-2}\left(\frac{M_{\mathrm{p}} \sin i}{M_{\mathrm{J}}}\right) \mathrm{ppm},
\label{eq:A_ellip}
\end{aligned}
\end{equation}
and the amplitude of the Doppler boosting given by
\begin{equation}
A_{\rm {boost }}=2.7 \alpha_{\rm {boost }}\left(\frac{P}{\rm { day }}\right)^{-1 / 3}\left(\frac{M_{\star}}{M_{\odot}}\right)^{-2 / 3}\left(\frac{M_{\mathrm{p}} \sin i}{M_{\mathrm{J}}}\right) \mathrm{ppm}.
\label{eq:A_beam}
\end{equation}
The coefficients $\alpha_{\mathrm{ellip}}$ and $\alpha_{\mathrm{boost}}$ depend on the stellar effective temperature and are parameterised following \citet{Millholland2017}
\begin{equation}
\begin{split}
\alpha_{\mathrm{ellip}} &= -\left(2.2 \times 10^{-4} \mathrm{~K}^{-1}\right) \; T_{\mathrm{eff}} + 2.6 \; , \\
\alpha_{\mathrm{boost}} &= - \left(6 \times 10^{-4} \mathrm{~K}^{-1}\right) \; T_{\mathrm{eff}} + 7.2 \; .
\end{split}
\end{equation}

\subsection{Grid-fitting of the photometric data}\label{sect:grid-fitting}

To constrain the physical properties of the candidate non-transiting planet, we performed a grid-based exploration of the parameter space using the photometric phase-curve model described above. The grid is parametrised by the planetary radius $R_p$, the geometric albedo $A_g$, the heat redistribution efficiency $\varepsilon$, the orbital inclination $i$, and the cloud-induced longitudinal offset $\eta_c$. We adopted a fixed orbital period of $P = 0.60474$~day and the  upper limit on the planetary mass from the RV analysis, $(M_p \sin i)_{3\sigma} = 11.4 \;M_{\oplus}$. The grid consists of $10 \times 10 \times 10 \times 45 \times 10$ points, uniformly sampling the following ranges: $R_p \in [0.1, 0.7]\,R_{\rm J}$, $A_g \in [0.05, 0.99]$, $\varepsilon \in [0.05, 0.99]$, $i \in [17^\circ, 62^\circ]$, and $\eta_c \in [-90^\circ, 90^\circ]$. The upper inclination limit corresponds to the grazing-transit boundary for a circular orbit, $i=62^\circ$. The lower limit of $i=17^\circ$ was determined empirically: for inclinations below this value, the observer views nearly the same planetary hemisphere throughout the orbit, resulting in a phase curve that is flat and the amplitude scales only with the planetary radius, preventing the reproduction of the observed photometric modulation. Using the adopted upper limit on $M_p \sin i$, we computed the expected amplitudes of Doppler boosting and ellipsoidal distortion across the inclination range. In all cases, these effects contribute at the level of a few ppm or less and therefore have a negligible impact on the total modelled phase-curve amplitude, varying up to $A_{\mathrm{ellip}} \simeq 0.2$~ppm and $A_{\mathrm{boost}} \simeq 2$~ppm for an inclination of 62$^{\circ}$. 

{We then compute two separate grids of models, one with the \textit{Kepler} response function, and one with the TESS response function. Both grids have the same parametrisation, bounds and sampling. Next, we perform a joint grid fitting using the \texttt{UltraNest} nested sampler \citep{buchner_ultranest_2021}, fixing the orbital period at $P = 0.60474$~day. We fit for the planetary radius, the inclination and a phase offset $\delta_{\rm phase}$, which accounts for the uncertainty in the time of inferior conjunction derived from the RV fit, treating these as shared parameters between the two datasets. Subsequently, for each dataset independently, we fit for the geometric albedo, the heat redistribution efficiency, the cloud offset, and the amplitude offset $\Delta A$, to account for a possible vertical shift of the light curves after normalisation \citep{Garcia2011}. We apply wide uniform priors to all parameters, bounded by the minimum and maximum values defined by the grid. All priors are listed in Table~\ref{fig:bestfit}. We adopt conservative priors for the phase offset $\delta_{\rm phase}$ by taking the 5$\sigma$ error bars on the time of inferior conjunction (see Fig.~\ref{fig:kic9139163_rv_data_analysis}). We allow the additive photometric offset $\Delta A$ to vary within $\pm5\sigma_{\rm mean}$.}

\subsection{Interpretation of fitted and derived parameters}

{The retrieved parameters are summarised in Table~\ref{tab:grid_fit_results}. Fig.~\ref{fig:bestfit} shows the solution at $t_{\rm inf, RV}$, as a normalised light curve versus orbital phase, with the observed \textit{Kepler} and TESS data displayed in grey and the best-fit model in blue. The lower panel of each plot presents the residuals, defined as the data–model difference, which are centred around zero and show no significant large-scale trends. The posterior distributions are shown in Fig~\ref{fig:corner_joint_fit_photometry}.} 

{The posterior distributions reveal several degeneracies and secondary solutions that help interpret the fitted parameters. The inclination is skewed towards high values, with a secondary tail extending between $\sim50^{\circ}$ and $55^{\circ}$, and a maximum probability density close to the upper prior boundary, where the system would become transiting. This preference for high inclinations implies a lower planetary mass and a configuration closer to the line of sight, rather than a more massive companion on a highly inclined orbit. Notably, the inferred inclination is higher than the stellar inclination ($\sim 26.5^\circ$, see Table~\ref{tab:global_parameter_punto}), suggesting a potential spin–orbit misalignment, which may have implications for the system’s dynamical history and tidal evolution \citep{Albrecht2012, mazeh_photometric_2015}. A correlated tail is observed in the planetary radius, spanning $\sim0.225$ to $0.240 \;R_{\rm J}$, indicating a degeneracy between inclination and radius. This trade-off is also visible in other parameters: a secondary local solution emerges at lower inclinations (though these inclinations are still $>50^\circ$), associated with larger radii, higher heat redistribution in the \textit{Kepler} dataset (approaching $\varepsilon \sim 0.9$), and slightly more negative photometric offsets. This local minimum appears as a small, distinct feature in the two-dimensional posterior distributions. This weak constraint on the redistribution factor is expected, as it primarily governs the thermal emission component of the phase curve. Given the relatively high albedos inferred for both datasets, the reflected light dominates the observed signal, with the thermal emission contributing roughly two orders of magnitude less to the total contrast. As a result, strong constraints on the redistribution are not required by the data. To properly capture these alternative solutions, we adopted conservative uncertainties based on the 5$^{\rm th}$ and 95$^{\rm th}$ percentiles of the marginal posterior distributions, corresponding to wide (90\%) credible intervals, instead of the more commonly used 16$^{\rm th}$ and 64$^{\rm th}$ percentiles. For the TESS data, the larger uncertainties reduce the impact of inclination on the retrieved parameters, and the posterior distributions remain broadly consistent across the explored parameter space. While additional local minima may be present, they remain close to the median solution, except for the redistribution factor, which is the least constrained parameter. In particular, for TESS, the redistribution factor spans nearly the full prior range ($0$–$1$), whereas the \textit{Kepler} data favour lower values, but still exhibit a secondary preference for high redistribution at lower inclinations.}

{Finally, the phase offset $\delta_{\rm phase}$ is well constrained and consistent with the value inferred from the RV solution within $2\sigma$. A degeneracy with the cloud offset parameters might be expected since both can shift the phase curve relative to the time of inferior conjunction. The data instead favour a shift in the location of the phase-curve maximum rather than a global phase shift. This behaviour supports a scenario in which longitudinally asymmetric, highly reflective structures dominate the observed variability in both datasets.
The derived parameters and their uncertainties were obtained using a Monte Carlo propagation of the posterior distributions by drawing samples from the asymmetric constraints on the fitted parameters and propagating them through the physical relations. Similarly to the retrieved parameters, the reported values correspond to the 5$^{\rm th}$, 50$^{\rm th}$, and 95$^{\rm th}$ percentiles of the resulting distributions. As such, planetary masses are obtained both from the RV-measured $M_p \sin i$, by adopting the retrieved inclination. The planetary density is computed from the derived masses and retrieved radius. The planetary temperatures and semi-major axis are listed as reference system parameters.} 

{The joint fit yields a planetary radius of $R_p = 0.2168^{+0.0072}_{-0.0081}\;R_{\rm J}$, corresponding to $2.43 \pm 0.14\;R_{\oplus}$, and an inclination of $i = 59.7{^{+1.5}_{-6.3}}^{\circ}$. Combined with the RV constraints, this implies a planetary mass of $M_p = 8.4^{+2.8}_{-2.7}\;M_{\oplus}$. The retrieved cloud offsets differ significantly between the two datasets, indicating a longitudinally asymmetric brightness distribution that evolves over time. To assess the significance of this behaviour, we compared the nominal model, which includes two independent cloud offsets, to a simplified model with a single offset. The two-offset model is strongly favoured, with a log-Bayesian evidence difference of $\Delta \ln Z = 431$, providing decisive support\footnote{A model is strongly favoured over the other if the ln-difference of the Bayesian evidence is $>$5 \citep{kassRaftery95}.} for the presence of distinct cloud structures. Similarly, we compared our nominal model to a simplified model with a single redistribution factor to test the hypothesis of a time-invariant heat redistribution, i.e. a constant longitudinal temperature structure. 
The two-redistribution-factor model is not favoured against the single-redistribution-factor scenario, with a log-Bayesian evidence difference of $\Delta \ln Z = 0.049$\footnote{For any $\Delta\ln Z < 2$, \cite{kassRaftery95} states that there's no evidence against the null hypothesis.}. This result can be interpreted in two ways: first, the redistribution factor is the least well-constrained parameter, as reflected by the extended tails in its posterior distributions; second, the shift in the phase-curve peak is more likely driven by reflected light rather than thermal emission, since the redistribution factor primarily governs the longitudinal temperature structure of the planet and hence its emission component.}

\begin{figure}[htbp!]
    \centering
    \includegraphics[width=0.5\textwidth]{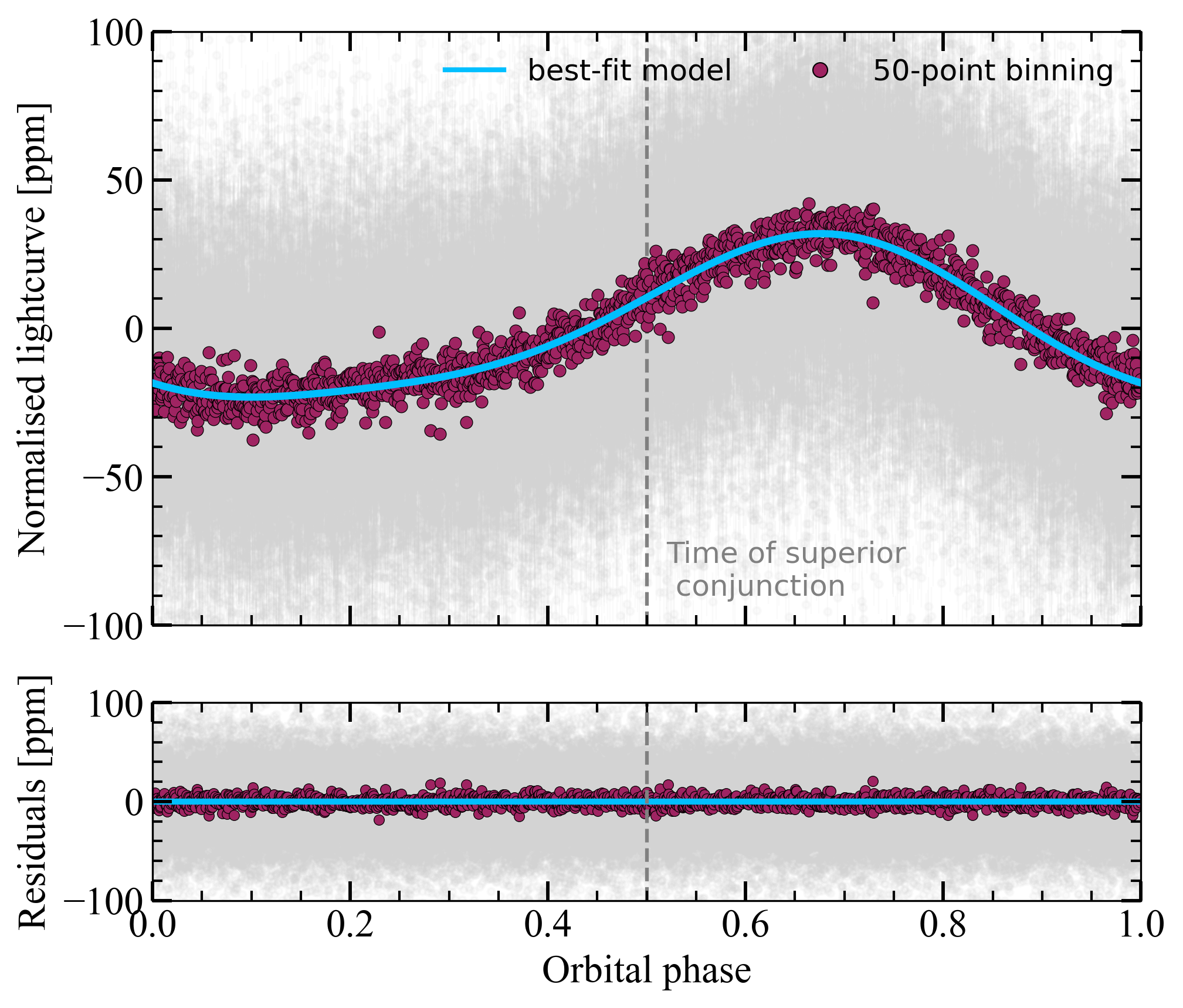}
    \includegraphics[width=0.5\textwidth]{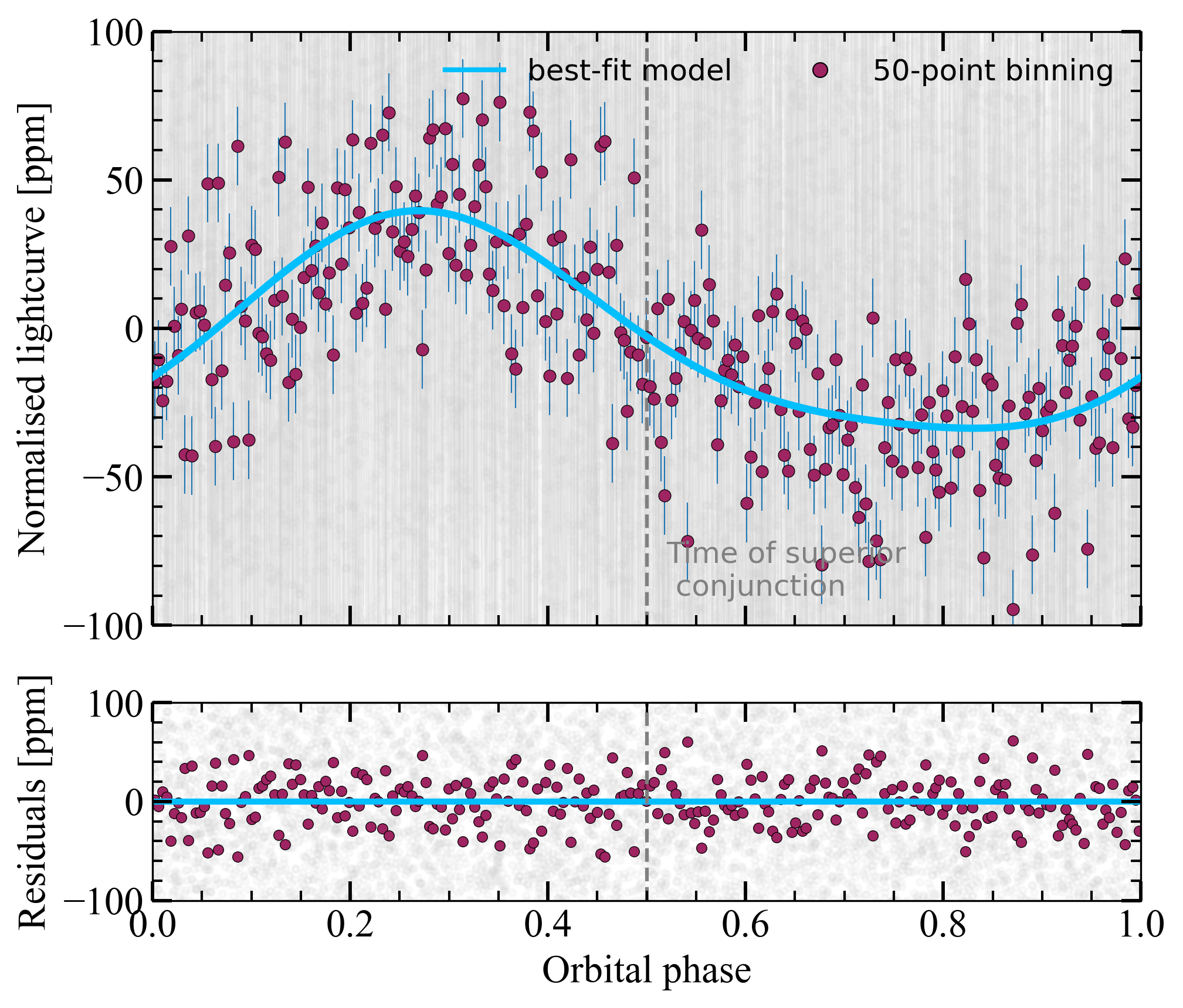}
    \caption{{Joint-fit of the \textit{Kepler} and TESS data.} \textit{Top:} Phase-folded \textit{Kepler} light curve of KIC~9139163, together with the joint best-fit model. The red points correspond to a 50-point binning (that is $~\sim30\;$s). The residuals show the difference between the data points and the model and are free of large systematic trends. \textit{Bottom:} Phase-folded TESS light curve of the same target, together with the joint best-fit model, with a similar 50-point binning scheme (that is $~\sim 3\;$min).}
    \label{fig:bestfit}
\end{figure}


\begin{table*}[ht]
\centering
\caption{Retrieved and derived parameters for the non-transiting candidate planet from phase-curve fitting.}
\label{tab:grid_fit_results}
\begin{tabular}{lcc}
\hline\hline
Retrieved parameters & Priors & Values \\
\hline
\multicolumn{3}{l}{\textit{Joint}} \\
\hline
Planetary radius, $R_{p}$ [$R_{\mathrm{J}}$] & $\mathcal{U}[0.1, 0.7]$ & $0.2168^{+0.0072}_{-0.0081}$  \\
Inclination\tablefootmark{a}, $i$ [$^{\circ}$] & $\mathcal{U}[17, 62]$ & $59.7^{+1.5}_{-6.3}$ \\
Phase offset, $\delta_{\rm phase}$ [$^{\circ}$] & $\mathcal{U}[-54, 54]$  & $17.7^{+2.2}_{-5.7}$ \\
\hline
\multicolumn{3}{l}{\textit{Kepler}} \\
\hline
Geometric albedo, $A_{g, \rm Kepler}$          & $\mathcal{U}[0.05, 0.99]$ & $0.5689^{+0.048}_{-0.016}$ \\
Heat redistribution, $\varepsilon_{\rm Kepler}$  & $\mathcal{U}[0.05, 0.99]$ & $0.16^{+0.63}_{-0.043}$ \\
Cloud offset, $\eta_{c, \rm Kepler}$ [$^\circ$]  & $\mathcal{U}[-90, 90]$ & $86.4^{+2.3}_{-8.3}$ \\
Photometric offset, $\Delta A_{\rm Kepler}$ [ppm]         & $\mathcal{U}[-500, 500]$  & $-29.35^{+0.99}_{-2.4}$ \\
\hline
\multicolumn{3}{l}{\textit{TESS}} \\
\hline
Geometric albedo, $A_{g,\rm TESS}$                & $\mathcal{U}[0.05, 0.99]$ & $0.749^{+0.082}_{-0.052}$ \\
Heat redistribution, $\varepsilon_{\rm TESS}$  & $\mathcal{U}[0.05, 0.99]$ & $0.52^{+0.42}_{-0.42}$ \\
Cloud offset, $\eta_{c, \rm TESS}$ [$^\circ$]  & $\mathcal{U}[-90, 90]$ & $-62^{+4.2}_{-5.8}$ \\
Photometric offset, $\Delta A_{\rm TESS}$ [ppm]         & $\mathcal{U}[-10000, 10000]$ & $-36.9^{+2.3}_{-2.9}$ \\

\hline
\multicolumn{3}{l}{Derived parameters} \\
\hline
Planetary radius, $R_{p}$ [$R_{\oplus}$] & \multicolumn{2}{c}{$2.43\pm0.14$} \\
Planetary mass, $M_{p}$ [$M_{\oplus}$]& \multicolumn{2}{c}{$8.4^{+2.8}_{-2.7}$} \\
Planetary mass, $M_{p}$ (3$\sigma$-RV) [$M_{\oplus}$] & \multicolumn{2}{c}{{$<13.2^{+1.1}_{-0.2}$}} \\
Planetary density, $\rho_p / \rho_{\oplus}$ & \multicolumn{2}{c}{$0.59^{+0.24}_{-0.20}$} \\
Planetary density, $\rho_p / \rho_{\oplus}$ (3$\sigma$-RV) & \multicolumn{2}{c}{$0.56^{+0.23}_{-0.19}$} \\

\hline
\multicolumn{3}{l}{System parameters} \\
\hline
Equilibrium temperature, $T_{\mathrm{eq}}$ [K] & \multicolumn{2}{c}{$2780$--$2960$} \\
Dayside temperature, $T_{\mathrm{day}}$ [K] & \multicolumn{2}{c}{$2840$--$3750$} \\
Nightside temperature, $T_{\mathrm{night}}$ [K] & \multicolumn{2}{c}{$1400$--$2750$} \\
Semi-major axis, $a$ [AU] &  \multicolumn{2}{c}{$0.0158$} \\
\hline
\end{tabular}\\
\tablefoot{The retrieved parameters are those directly fitted to the light curve, while the derived parameters are computed quantities inferred from the best-fit solutions, with all error bars given at 2$\sigma$ levels. These values represent the plausible range of physical photometric properties consistent with the \textit{Kepler} and TESS data, and the non-transiting geometry. For the temperatures, we computed the maximum ranges obtained by combining a Bond albedo in the range 0.1--0.3 and a heat redistribution efficiency between 0 and 1, see Sect.~\ref{sect:albedo_temperature}.
\tablefoottext{a}{The upper inclination limit corresponds to the grazing-transit boundary for a circular orbit, $i=62^\circ$. The lower limit of $i=17^\circ$ corresponds to flat phase curves with the amplitude scaling only with the planetary radius, preventing the reproduction of the observed photometric modulation.}
}

\end{table*}

\subsection{A candidate planet in the Neptunian desert}

The \textit{Kepler} phase-curve analysis indicates that the favoured configuration is a phase-curve with a pronounced offset of the flux maximum relative to the time of inferior conjunction, occurring after the secondary eclipse. Such an offset can be interpreted as the presence of inhomogeneous reflective structures on the planetary dayside, most plausibly high-albedo clouds located west of the substellar point. Similar westward phase offsets have been reported for a few ultra-hot exoplanets and are commonly attributed to cloud formation or survival on the cooler evening terminator, where condensates may persist despite extreme irradiation \citep[e.g.][]{munoz_probing_2015, angerhausen_comprehensive_2015, webber_effect_2015,
placek_analyzing_2017, Parmentier2018}. 

{A more intriguing result arises when considering the TESS phase curve, which instead displays an offset of the flux maximum occurring before the secondary eclipse, i.e. an eastward shift relative to the substellar point.
{Although the TESS data are noisier and do not allow us to robustly detect long-term amplitude variations}, the persistence of an eastward offset suggests that the longitudinal brightness distribution differs from that inferred from the \textit{Kepler} data. This apparent reversal in the direction of the phase offset cannot be straightforwardly explained by a static cloud distribution alone and instead points towards a more complex, potentially time-variable atmospheric structure.}

The fitted geometric albedo is relatively high, favouring reflective atmospheres, while the heat redistribution efficiency is weakly constrained but mostly compatible with inefficient day/night energy transport. {Our grid-based modelling, which includes both reflected light and thermal emission, indicates that the phase curve is dominated by reflection in the optical, with thermal emission providing only a secondary contribution even in the TESS bandpass. As a consequence, the observed phase offsets primarily trace the distribution and scattering properties of the reflective component rather than the longitudinal displacement of a thermal hotspot.} To display the reflective offsets as well as the temperature structure of the companion candidate, the emission and reflection longitude-latitude maps, computed with the mean retrieved parameters and in both \textit{Kepler} and TESS bandpasses, are shown respectively in Fig~\ref{fig:emission_reflection_maps_Kepler} and Fig~\ref{fig:emission_reflection_maps_TESS}.

{Importantly, we find that a Lambertian phase function is insufficient to reproduce the observed phase-curve morphology, in particular the large longitudinal displacement of the flux maximum. This implies that the scattering properties of the atmosphere are significantly non-isotropic. In realistic cloud-forming atmospheres, scattering is expected to be anisotropic, particularly in the Mie scattering regime when particle sizes are comparable to or larger than the observed wavelengths, leading to pronounced forward scattering \citep{seager_exoplanet_2010}. The eastward offset observed in the TESS phase curve can be explained by the anisotropic Mie scattering from cloud particles with characteristic radii of order $\sim1-10\,\mu$m \citep{lee_dynamic_2016,
pinhas_signatures_2017}. In this regime, forward scattering enhances the reflected flux phases before the eclipse, shifting the apparent brightness maximum eastward even in the absence of a corresponding longitudinal redistribution of clouds \citep{munoz_probing_2015, Parmentier2018}.
In such a regime, the location of the phase-curve maximum does not necessarily coincide with the longitude of maximum cloud coverage, but instead depends on the angular scattering phase function.
Within this framework, the combination of a westward offset in \textit{Kepler} and an eastward offset in TESS can be interpreted as evidence for an evolution in cloud microphysics. During the \textit{Kepler} observing window, the westward offset is consistent with reflective clouds located on the cooler western limb, in agreement with expectations from general circulation models predicting eastward equatorial jets and condensation on the evening terminator \citep{parmentier_transitions_2016}. The observed increase in phase-curve amplitude over the four years of \textit{Kepler} photometry further suggests a progressive enhancement in the optical depth. Such an increase may arise from a higher cloud particle number density, growth of particle sizes (increasing the scattering cross-section), or an increase in the vertical extent of the clouds  \citep[e.g.][]{gao_aerosols_2021}. In particular, cloud radiative feedback can amplify initial longitudinal asymmetries: regions with enhanced cloud coverage reflect more stellar radiation, leading to local cooling and condensation \citep{parmentier_cloudy_2021}. This mechanism can, in principle, produce time-dependent cloud configurations, with transitions between different longitudinal states.
We now discuss the possibility that atmospheric transport plays a role in shaping the longitudinal distribution of clouds. Indeed, an additional contribution may arise from the relative ordering of radiative, advective, and microphysical timescales. The atmospheric temperature structure of irradiated planets is primarily governed by the ratio of radiative to advective timescales, with short radiative timescales leading to strong day–night contrasts and weak heat redistribution \citep[e.g.][]{perez-becker_catastrophic_2013, showman_atmospheric_2020}. In ultra-hot atmospheres, radiative timescales can become sufficiently short that the thermal structure is largely controlled by local irradiation rather than by atmospheric transport.
Cloud formation, however, is additionally governed by microphysical timescales \citep{gao_aerosols_2021}. When these timescales are comparable to the advective timescale, cloud formation may occur after the gas has been transported away from the local temperature minimum, before fully condensing. This decoupling between temperature and cloud distribution can introduce longitudinal offsets in the reflective component of the atmosphere and contribute to phase-curve asymmetries. This effect further complicates the interpretation of phase curves and may contribute to the observed differences between the \textit{Kepler} and TESS datasets.}

Nonetheless, in the absence of a substantial atmosphere, the observed phase offset could instead arise from spatial variations in surface or subsurface reflectivity, potentially driven by tidal heating or interior energy dissipation. In such a scenario, localised hot regions displaced from the substellar point could produce an asymmetric optical phase curve even without efficient heat transport. This interpretation would be consistent with a highly irradiated exoplanet with a thin atmosphere, possibly with a molten or partially molten surface \citep{selsis_effect_2013}.

{The inferred planetary radius $R_p \simeq 2.47\pm0.14\,R_\oplus$ places the candidate companion into the Super-Earth to Sub-Neptune regime \citep{bean_nature_2021}. Combined with the RV constraints, the corresponding mass range is consistent with a Neptune-like planet. The derived density is in the range $\rho_p /\rho_{\oplus} \simeq 0.55$--$0.6$, placing the candidate planet in the so-called Water World regime, compatible with a composition of $\sim$50\% H$_2$O \citep[see][]{luque_density_2022}. With a period of $P = 0.60474$~day, the candidate companion is highly irradiated and would fall in the newly defined Hot Water World triangle defined as a region in mass–radius space where close-in planets cannot retain H/He envelopes due to atmospheric escape. This implies that planets located within these regions must possess heavier, volatile-rich atmospheres \citep[see Fig.~1 of][]{egger_searching_2025}.} 

The possible identification of the candidate planet as an ultra-hot Sub-Neptune places it within the Neptunian desert, a region characterised by a scarcity of Neptune-sized planets on short-period orbits ($P \lesssim 3$--$5$ days, $M_p \sim 0.03$--$0.3~M_{\rm Jup}$, $R_p \sim 2$--6 $R_\oplus$). Figure~\ref{fig:mass_radius_period} shows the location of the KIC~9139163 candidate planet in both radius–period and mass–period diagrams, using the NASA Exoplanet Archive\footnote{See \url{https://exoplanetarchive.ipac.caltech.edu/}.} as a comparison sample. The candidate planet’s radius, derived from the grid-based fit, is marked by the blue star and lies well within the desert boundaries defined by \citet{mazeh_dearth_2016}, while the dotted lines indicate the updated lower desert limits proposed by \citet{Deeg2023}. The object also lies in the more restrictive definition of the Neptunian Desert by \citet{castro-gonzalez_mapping_2024}, which contains only about a twentieth of as many planets as the desert from \citet{mazeh_dearth_2016}. The mass–period diagram further supports this placement: the blue corresponds to the RV-derived mass, which is also consistent with the desert’s parameter space.

The location of the KIC~9139163 candidate planet in the midst of the Neptunian desert underscores its potential significance as a benchmark object for testing theories of atmospheric mass loss and planetary survival in high-irradiation environments. This desert is indeed thought to arise from intense photoevaporation and atmospheric mass loss, which strip away the volatile envelopes of sub-Jovian planets exposed to high stellar irradiation \citep{mazeh_dearth_2016, Deeg2023}. However, some exoplanets are found to survive in this desert \citep[e.g.][]{ berger_revised_2018, livingston_sixty_2018, 
persson_toi-2196_2022, osborn_toi-332_2023, naponiello_super-massive_2023}. The bottom panel of Fig.~\ref{fig:mass_radius_period} shows the mass-radius diagram of all close-in planets ($P_{\rm{orb}} < 5$ days) surviving in the desert. Planets circled with purple are accepted programs with the James Webb Space Telescope (JWST) for atmospheric characterisation. Notable examples of planets in the desert include TOI-849 b \citep{armstrong_remnant_2020}, a dense, short-period planet interpreted as a remnant core of a former gas giant, and NGTS-4 b \citep{west_ngts-4b_2019}, which is thought to retain a thin atmosphere despite residing deep within the desert. LTT 9779 b is the most thoroughly studied example, exhibiting a short orbital period (0.79 days), a high geometric albedo \citep[$A_g \sim 0.8$,][]{hoyer_extremely_2023, coulombe_highly_2025}, and a substantial atmosphere inferred through both transmission and emission spectroscopy \citep{dragomir_spitzer_2020, radica_muted_2024, ramirez_reyes_closer_2025}. 

{We additionally explored whether the observed phase-curve offsets could be reproduced through modifications of the atmospheric circulation rather than changes in the cloud distribution itself. In particular, we tested models with a negative heat redistribution parameter ($\varepsilon <0$), corresponding to a reversal of the usual eastward hotspot offset expected from super-rotating equatorial jets. Although such retrograde or westward-dominated circulation patterns are not generally predicted for canonical Hot-Jupiter atmospheres, theoretical and numerical studies have shown that atmospheric circulation can differ from simple eastward super-rotation under specific dynamical or magnetic regimes, like sub-synchronous rotation or magnetically induced westward winds \citep[e.g.][]{showman_atmospheric_2013, showman_atmospheric_2020}. This hypothesis has already been evoked to explain the shift of the phase-curve amplitude maximum of the Hot-Jupiter CoRoT-2b \citep{barstow_dynamical_2018}.  While most of these studies have focused on Hot-Jupiters, similar uncertainties remain for highly irradiated H/He-dominated Sub-Neptunes, whose lower surface gravities, enhanced atmospheric metallicities, and potentially different radiative timescales may lead to circulation regimes that differ from those of more massive gas giants. 
Within this framework, we investigated whether the eastward offset observed in the TESS phase curve could be reproduced using a single cloud configuration combined with a reversed thermal redistribution pattern. We find that a negative redistribution parameter together with a positive cloud offset can indeed reproduce a phase-curve morphology qualitatively similar to that observed with TESS, particularly for low geometric albedos where the thermal contribution becomes more significant. However, this configuration simultaneously produces a similar offset in the \textit{Kepler} bandpass, failing to reproduce the distinct behaviour observed between the two datasets. Conversely, increasing the albedo while maintaining a positive cloud offset and negative redistribution allows us to recover the \textit{Kepler} morphology, but again results in similar phase offsets in both \textit{Kepler} and TESS. 
These tests therefore indicate that modifications of the thermal redistribution alone cannot simultaneously explain both the wavelength dependence and the apparent temporal variability of the observed phase curves. While retrograde circulation remains a possible contribution to the observed morphology, the scenario with two distinct cloud offsets between the \textit{Kepler} and TESS epochs remains the only configuration capable of reproducing the full set of observations.}

\subsection{Albedo assumptions, planetary temperatures and cloud properties}\label{sect:albedo_temperature}

The interpretation of the retrieved parameters from the phase curve fits, also requires assumptions regarding the scattering properties of the atmosphere. In particular, and as stated in Sect.~\ref{sect:reflected_component}, the Lambertian sphere approximation is not strictly applicable in our case. A Lambertian phase function implies isotropic scattering and a fixed relation between the geometric albedo $A_g$ and the Bond albedo $A_B$ through
\begin{equation}
A_B = q\,A_g \; ,
\end{equation}
where $q$ is the phase integral. For a perfectly Lambertian sphere, $q=3/2$ \citep[e.g.][]{seager_exoplanet_2010}, such that $A_B=1.5\,A_g$. For the moderate to high geometric albedo inferred here ($A_g \sim 0.57-0.75$), this relation would formally imply $A_B>1$, which is unphysical. This demonstrates that the assumption of Lambertian scattering breaks down and that the geometric albedo measured in the \textit{Kepler} band cannot be directly translated into a bolometric Bond albedo.

More generally, non-Lambertian scattering, longitudinally inhomogeneous clouds, or wavelength-dependent reflectivity lead to smaller effective phase integrals \citep[$q<1.5$,][]{luger_analytic_2022}. Because geometric albedos derived from optical phase curves probe only a limited wavelength range, they do not uniquely constrain the bolometric Bond albedo. In highly irradiated atmospheres, strong absorption at infrared wavelengths combined with reflective optical clouds can, in principle, yield large optical geometric albedos without requiring large Bond albedos \citep{parmentier_transitions_2016}. For highly irradiated planets, both observational and theoretical studies typically infer Bond albedos in the range $A_B\sim0.05$--$0.3$ \citep[e.g][]{cowan_statistics_2011}. We therefore treat $A_B$ as an independent parameter when computing the planetary energy balance.

We define the globally averaged equilibrium temperature following the prescription of \cite{persson_toi-2196_2022} as
\begin{equation}
T_{\rm eq} = T_{\rm eff} \left(\frac{R_\star}{2a}\right)^{1/2}
\left(1-A_B\right)^{1/4} \; .
\label{eq:Teq}
\end{equation}
where $A_{\mathrm B}$ is the Bond albedo. Eq.~\ref{eq:Teq} assumes that the absorbed stellar energy is redistributed instantaneously and emitted uniformly over the entire planetary surface, leading to the factor of 2 in the denominator. While this global equilibrium temperature provides a reference for the planetary thermal state, local temperatures can deviate significantly from this value in the presence of high geometric albedo and non-uniform heat redistribution. To account for non-uniform heat transport between the dayside and nightside hemispheres, we additionally adopt the formalism of \cite{cowan_statistics_2011}, in which the atmospheric redistribution efficiency is parameterized by $\varepsilon$
\begin{equation}
T_{\rm day} = T_{\rm irr}\,(1-A_B)^{1/4}
\left(\frac{2}{3}-\frac{5}{12}\varepsilon\right)^{1/4} \; ,
\end{equation}
\begin{equation}
T_{\rm night} = T_{\rm irr}\,(1-A_B)^{1/4} \left(\frac{\varepsilon}{4}\right)^{1/4} \; ,
\end{equation}
with $T_{\rm irr} = T_{\rm eff} \left(\frac{R_\star}{a}\right)^{1/2}$. {These expressions provide approximate, globally averaged temperatures that are useful for physical interpretation, but are not used in the phase-curve model described in Sect.~\ref{sect:thermal_emission}, where the full longitudinal temperature structure is computed. We note that, unlike the analytic scaling relations above, the temperature structure used in our phase-curve model does not explicitly depend on the Bond albedo, as it is parametrised through $T_0$ and the redistribution efficiency $\varepsilon$. The expressions for $T_{\rm day}$ and $T_{\rm night}$ should therefore be regarded as illustrative estimates rather than direct counterparts of the model temperatures.} For representative Bond albedos $A_B = 0.1$--$0.3$ \citep[e.g][]{cowan_statistics_2011}, the dayside temperature is of order $T_{\rm day} \simeq 2840$--$3750$ K, but the night temperature spans a lower range $T_{\rm night} \simeq 1400$--$2750$ K (depending on no or full heat redistribution).

Cloud formation is governed by local atmospheric temperatures rather than by a single global or dayside-averaged value. Three-dimensional general circulation models of highly irradiated planets consistently predict strong longitudinal temperature gradients when heat redistribution is inefficient, with the western and eastern limbs being substantially cooler than the substellar region \citep{showman_atmospheric_2013, parmentier_cloudy_2021}. For irradiation levels comparable to those considered here, these models typically yield limb temperatures in the range $\sim2000$--$2500$~K, despite much higher temperatures at the substellar point.

These limb temperatures therefore overlap the condensation regimes of several refractory species, including silicates (MgSiO$_3$, Mg$_2$SiO$_4$), which are predicted to form clouds on the cooler limbs and nightsides of highly irradiated planets \citep{powell_transit_2019, parmentier_cloudy_2021, mansfield_revealing_2023, morris_observations_2024}. Such clouds can produce strong optical scattering and longitudinal asymmetries, naturally explaining phase-curve offsets and large geometric albedos without requiring high Bond albedos.

Although most observational studies of optical phase curves have focused on Hot-Jupiters \citep{helling_cloud_2021}, cloud formation and longitudinal inhomogeneities are governed primarily by local temperature–pressure conditions and atmospheric composition rather than the planetary mass. Theoretical studies predict that irradiated Sub-Neptunes, which are expected to possess metal-rich atmospheres and lower surface gravities, can sustain optically thick clouds \citep{gao_aerosols_2021}. While phase-curve observations of Sub-Neptunes remain scarce, clouds are not excluded in this regime. The resulting lower surface gravity and higher metal abundances favour cloud formation relative to giant planets. While atmospheric escape may efficiently remove light species under such extreme irradiation, theoretical studies indicate that metal-rich atmospheres can survive on Gyr timescales depending on the ratio of their initial core and envelope mass \citep{jin_planetary_2014, owen_atmospheric_2019, 
vissapragada_upper_2022}.

We note that, as in transmission spectroscopy of warm and hot Neptunes (e.g. GJ 1214b or GJ 436b), there exists a degeneracy between high-altitude clouds and high mean molecular weight atmospheres \citep{kempton_reflective_2023}. Consequently, the presence of clouds cannot be uniquely inferred. Nevertheless, the combination of high geometric albedo, longitudinal phase-curve offsets, inefficient heat redistribution, and physically plausible limb temperatures provides a self-consistent interpretation for a cloudy, highly irradiated Sub-Neptunian planet. In this respect, our candidate planet may be analogous to LTT~9779~b, which exhibits a possible substantial highly-reflective atmosphere despite residing deep within the Neptunian desert.

The diversity of atmospheric properties observed among planets in the Neptunian desert, ranging from highly reflective, cloud-rich atmospheres to stripped cores with minimal envelopes, suggests that multiple evolutionary pathways may operate in this extreme irradiation regime. The candidate planet of KIC~9139163 may represent an intermediate or transitional case, where reflective clouds persist despite intense stellar heating, or alternatively a nearly atmosphere-free object whose phase curve is shaped by surface or interior processes. These results therefore point to the importance of continued effort in interpreting modulations from short-period, ultra-hot exoplanets exhibiting contributions from both thermal emission and reflected light.

\begin{figure}[htbp!]
    \centering
    \includegraphics[width=0.5\textwidth]{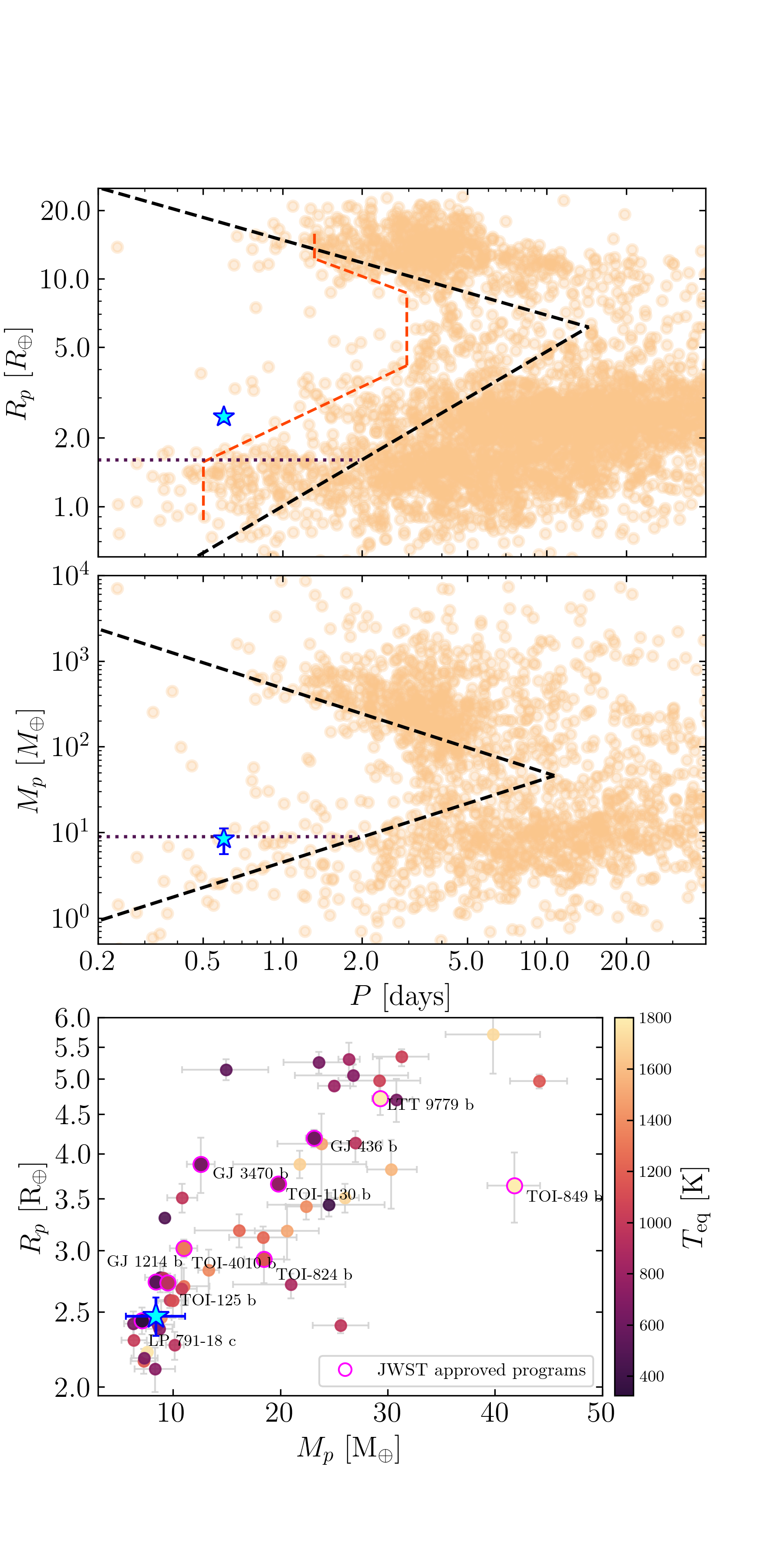}
    \caption{{Mass–radius properties of the KIC~9139163 planet candidate compared to exoplanets from the NASA Exoplanet Archive.} \textit{Top:} Radii versus period diagram of the known exoplanets, from the NASA Exoplanet Archive. The KIC~9139163 candidate's radius is indicated by the blue star. The dashed black lines show the boundaries of the Neptune Desert from \citet{mazeh_dearth_2016}. The horizontal dotted black lines show the updated lower limits for the desert from \citet{Deeg2023}, the red lines are the boundaries of the desert following the prescription of \cite{castro-gonzalez_mapping_2024}.
    \textit{Middle:} Same for the mass versus period diagram.
    \textit{Bottom:} KIC~9139163's candidate located on the mass-radius diagram along with all known exoplanets from the NASA exoplanet archive. The purple circled planets are JWST-approved programs for atmospheric characterisation. 
    }
    \label{fig:mass_radius_period}
\end{figure}

\section{Discussion \label{sec:discussion}}

{Additional RV follow-ups are needed to provide a definitive confirmation of the detection, but the strong photometric modulation in both \textit{Kepler} and TESS light curves shows that this system remains a promising candidate for follow-up observations.
Because of its short orbital period ($\sim$0.6 day), a full orbit could be covered in a single 14-hour observation with JWST, enabling time-resolved spectroscopy. Multiband JWST photometry could also help separate reflection from emission, while ground-based high-resolution spectroscopy may probe atomic and molecular lines through Doppler shifts. These approaches could provide confirmation of the planet that is still missing. Altogether, this system is a strong candidate for a dedicated non-transiting program with a lot of different facilities, leveraging its short orbital period and high atmospheric brightness to yield rich phase-resolved information in a single visit.}


\subsection{Additional insights from high-resolution spectroscopy}

High-resolution spectroscopy offers a compelling pathway to detect exoplanetary atmospheres by exploiting the planet’s Doppler-shifted spectrum as it orbits—allowing separation from the stellar and telluric signals via cross-correlation with atmospheric models \citep{borsa_gaps_2015, brogi_retrieving_2019, cabot_robustness_2019, borsa_gaps_2022}. This method could be applied to KIC~9139163 to confirm the planetary nature of the candidate through an attempt of detecting an atmosphere. To apply this method effectively, we would require continuous observations covering several hours. For our candidate planet with a $\sim$0.6-day orbit, observing windows up to 6 hours could provide enough time to cover a significant fraction of the orbital phase. Given the possible ultra-hot Sub-Neptunian nature of the candidate, species such as neutral iron (Fe\,\textsc{i}) may be detectable as well as H$_2$O and CO. High-temperature atmospheres in similar objects have exhibited clear Fe lines in emission or absorption. For example, high-resolution thermal emission spectroscopy of the ultra-hot Jupiter KELT-20b (MASCARA-2b) revealed robust Fe\,\textsc{i} features and a thermal inversion in its dayside atmosphere \citep{yan_detection_2022}. Similarly, observations of WASP-76b captured asymmetric Fe absorption arising from iron condensation and atmospheric winds on its cooler night side \citep{ehrenreich_nightside_2020}. Applying the same methodology to KIC~9139163 could enable detection of atomic and molecular lines, constraining atmospheric composition, temperature structure, and even dynamic behaviour. As the target is located in the northern hemisphere, observations with the near-infrared spectropolarimeter SPIRou \citep{Donati2018} on the Canada-France-Hawaii Telescope (CFHT) located in Hawaii (resolution of $\sim$70000), or the CARMENES spectrograph \citep[][]{Quirrenbach2014} in Spain (near-infrared resolution $\sim$80000), could provide some useful insights into the nature of this candidate companion. Also, the CO band can be covered by the iGRINS-2 instrument while the Fe\,\textsc{i} or H$_2$O lines can be observed with the MAROON-X instrument, both on the Gemini observatory. 

\subsection{Additional insights from infrared spectroscopy}

The emission of our ultra-hot candidate planet contributes $\sim$20\% of the total flux both in the \textit{Kepler} and TESS bandpasses, with the maximum of emission around $\sim$1.7 $\mu$m in the near infrared, considering both the star and the candidate planet as blackbodies. To disentangle reflection from emission, multi-wavelength observations bridging visible and near-IR can help us confirm the presence of a companion. Observations with JWST/NIRSpec, particularly the high-resolution G395H mode ($R\sim2700$) could help us to observe the peak of emission. Indeed, G395H covers 2.9–5.2 $\mu$m, where strong H$_2$O or Fe\,\textsc{i} lines should appear. With a 0.6‑day orbital period, a single 14‑hour continuous NIRSpec G395H time-series would cover the entire orbit, ensuring adequate phase coverage and the first non-transiting observation with JWST. NIRSpec G395H’s stability and spectral resolution \citep[highest resolution with JWST in the near-infrared][]{birkmann_near-infrared_2022} support cross-correlation methods to extract faint planetary signals from systematics, as validated in the case of the Hot-Jupiter WASP‑39b, where a CO detection at 6.6–7.5$\sigma$ was obtained \citep{esparza-borges_detection_2023}. These observations could allow us to detect H$_2$O and to confirm the planetary nature of the object.

\section{Conclusion \label{section:conclusion}}

In this work, we have studied in detail the 0.6-day periodic modulation exhibited by the \textit{Kepler} and TESS light curves of the hot solar-type pulsator KIC~9139163, attributed to the presence of a close-in non-transiting exoplanet. We presented an updated stellar model of KIC~9139163, combining \textit{Kepler} and TESS asteroseismic observables and new HARPS-N spectroscopic measurements. Our stellar modelling is consistent with results previously published in the literature. We were also able to use asteroseismology to confirm the stellar inclination measurements from \citet{Benomar2015} and \citet{Hall2021}. Subsequently, we performed an extended analysis of the photometric variability arising from the active regions' modulations in the \textit{Kepler} and TESS light curves, confirming the $\sim$6-day stellar surface rotation period provided in the catalogue from \citet{Santos2021}. 

We then presented the characterisation of a short-period, non-transiting candidate planet orbiting KIC~9139163 using the \textit{Kepler} and TESS photometry and HARPS-N radial velocity data. From the HARPS-N data, we derived a value of $M_p \sin i = 7.3 \pm 1.4 \,~M_\oplus$ for the mass of the companion. 
{The RV data allowed us to constrain the time of inferior conjunction, while the photometric phase modulation independently constrained the inclination to $i = 59.7^{+1.5}_{-.3}$ with a $62^{\circ}$ upper limit, imposed by the absence of transits. We inferred a planetary radius of $R_p \simeq 2.47\pm0.14\,R_\oplus$, placing the companion candidate in the Super-Earth to Sub-Neptune regime \citep{bean_nature_2021}. Combined with the RV constraints, the allowed masses are consistent with a Neptune-like planet. The extreme stellar irradiation, placed the object deep within the Neptunian desert \citep{mazeh_dearth_2016}. The \textit{Kepler} phase curve exhibits a clear westward offset of the flux maximum relative to the secondary eclipse, indicating a strongly asymmetric brightness distribution on the planetary dayside while the TESS phase curve exhibits an eastward offset of the flux maximum relative to the secondary eclipse. The fitted geometric albedos are relatively high ($A_g \sim 0.57$–$0.75$), favouring reflective atmospheres, while the heat redistribution efficiency is weakly constrained. Our joint modelling shows that the optical phase curves are dominated by reflected light, with thermal emission contributing only marginally. As a result, the observed phase offsets primarily trace the longitudinal distribution and scattering properties of reflective structures rather than thermal hot-spot shifts. The opposite offsets observed in \textit{Kepler} and TESS indicate that the brightness distribution is not static and likely evolves over time. This behaviour can be explained by several scenarios: (i) time-variable cloud coverage or microphysics, where changes in cloud formation, dissipation, or particle properties modify the longitudinal distribution of reflective material across the planet, leading to shifts in the observed brightness pattern; (ii) changes in particle size or composition that alter how clouds scatter light, in particular through anisotropic (Mie) scattering, where larger particles preferentially scatter light in the forward direction. In this regime, the observed brightness maximum can be shifted towards phases preceding the eclipse, even if the actual cloud coverage is not displaced eastward, because the directionality of scattering redistributes the reflected light in phase; or (iii) spatial variations in surface or subsurface properties in the case of a thin or absent atmosphere. {However, given the larger noise level and contamination affecting the TESS photometry, we emphasise the fact that these interpretations remain tentative.} Together, these results point towards a dynamic and potentially evolving reflective component as the main driver of the observed phase-curve variability. Plus, the position of the candidate exoplanet within the Neptunian desert illustrates the variety of atmospheric and surface compositions that could be compatible with highly-irradiated planets. While many close-in sub-Neptune-sized planets are expected to lose their primordial atmospheres, a small fraction appear capable of retaining reflective atmospheres or exhibiting emission patterns driven by surface or interior processes, similarly to LTT~9779b \citep{coulombe_highly_2025}. While the detection of possible variability offers valuable observational constraints on theoretical predictions, these scenarios remain speculative, and further observations are needed to clarify the nature of this candidate companion.}


This work opens a new science of characterising non-transiting companions, building upon the catalogue of \textit{Kepler} candidates published by \citet{gourves_non-transiting_2025}. Non-transiting exoplanets are key to help populate the region of ultra short-period orbiting fast rotators. Whether or not these candidates lay within the Neptunian desert is the next question that remains to be answered. While some non-transiting candidate planets have been proposed for the TESS sample \citep{Cullen2024}, we emphasise that the PLATO survey should contain a significant amount of systems analogous to KIC~9139163, for which a significant fraction will have stellar inclination measurements through asteroseismology \citep{Goupil2024}. Such targets are also promising for multi-wavelength photometric follow-up or IR spectroscopy, as such observations should help disentangling stellar and planetary contribution. In particular, they could allow us to detect unambiguously signatures from the companion atmosphere. Non-transiting exoplanets are far more numerous than transiting ones. Their study therefore opens the opportunity to bring new insights on the statistics of planetary populations in terms of mass, radius but also atmospheric composition.

\begin{acknowledgements}
The authors thank the anonymous referee for a series of useful comments that significantly helped improving the quality of the manuscript.
S.N.~Breton acknowledges support from PLATO ASI-INAF agreement no. 2022-28-HH.0 "PLATO Fase D" and from the MASTODINT INAF mini-grant. C.A.~Prieto acknowledges financial support from the Spanish Ministry of Science, Innovation and Universities (MICIU) projects PID2020-117493GB-I00, PID2023-149982NB-I00 and PID2023-146453NB-I00. R.A.~García, D.B.~Palakkatharappil, L.~Borg, S~.Mathis, and A.R.G.~Santos acknowledge support from PLATO and GOLF CNES grants. S.~Mathis acknowledges support from the Programme National de Planétologie (CNRS/INSU). S.~Mathis and C.~Gourvès gratefully acknowledge support from the European Research Council (ERC) under the Horizon Europe programme (S.~Mathis: 4D-STAR; Synergy Grant agreement N$^\circ$101071505; C.~Gourvès: ExoMagnets; Consolidator grant agreement N$^\circ$ 101125367). While partially funded by the European Union, views and opinions expressed are, however, those of the authors only and do not necessarily reflect those of the European Union or the European Research Council. Neither the European Union nor the granting authority can be held responsible for them. S.~Mathur and H.J.~Deeg. acknowledge support from the Spanish Ministry of Science, Innovation and Universities with grants  PID2019-107061GB-C66 and PID2023-149439NB-C41. S.~Mathur also acknowledges support from the same source through grant PID2019-107187GB-I00, the Ram\'on y Cajal fellowship no.~RYC-2015-17697, and through AEI under the Severo Ochoa Centres of Excellence Programme 2020--2023 (CEX2019-000920-S). A.R.G.~Santos acknowledges the support from Funda\c{c}\~ao para a Ci\^encia e a Tecnologia (FCT) through national funds by the grant UID/04434/2025 and work contract No. 2020.02480.CEECIND/CP1631/CT0001 (DOI: 10.54499/2020.02480.CEECIND/CP1631/CT0001). The authors want to thank C.~Damiani, B.~Klein, A.M.~Lagrange, C.~Moutou, G.~Nowak, and S.~Sulis for fruitful discussions. The authors also thank M.~Malin and M.~Holmberg for their insightful advice.  This paper includes data collected by the \textit{Kepler} mission, and obtained from the MAST data archive at the Space Telescope Science Institute (STScI). Funding for the \textit{Kepler} mission is provided by the NASA Science Mission Directorate. STScI is operated by the Association of Universities for Research in Astronomy, Inc., under NASA contract NAS 5–26555.
\\
\textit{Softwares:} MESA \citep{Paxton2011, Paxton2013, Paxton2015, Paxton2018, Paxton2019}, 
\texttt{Python} \citep{Python3}, \texttt{numpy} \citep{numpy,harris2020array}, \texttt{matplotlib} \citep{matplotlib}, \texttt{scipy} \citep{2020SciPy-NMeth}, \texttt{corner} \citep{corner}, \texttt{astropy} \citep{astropy:2013, astropy:2018, astropy:2022}, \texttt{apollinaire} \citep{Breton2022apollinaire}, \texttt{star-privateer} \citep{Breton2024},
\texttt{pymc} \citep{exoplanet:pymc3}, \texttt{exoplanet} \citep{exoplanet:joss}, \texttt{pytensor} \citep{exoplanet:theano}, 
\texttt{exo\_k} \citep{leconte_spectral_2021}.
\end{acknowledgements}

\bibliographystyle{bibtex/aa} 
\bibliography{biblio.bib} 

\appendix

\section{TESS light curve for KIC 9139163 \label{appendix:tess_binary}}

{KIC~9139163 was observed by TESS under the identifier TIC~164670309. Using the \texttt{pytadacs} module \citep[][]{Garcia2024pytadacs}, we have extracted the TESS data of KIC~9139163 for sectors 14, 26 (1800-s cadence), 40, 41, 53, 54, 55 (600-s cadence), 75, 81, and 82 (200-s cadence). Fig.~\ref{fig:tess_lc_psd} shows the corresponding time series and PSD. We note that the noise level of the TESS observation is significantly higher than for \textit{Kepler} (see Fig.~\ref{fig:kic9139163_lc_psd}). Nevertheless, the 0.60474-period is clearly visible, corresponding to the sharp peak at 19.14~$\mu$Hz, although the second harmonic at 38.28~$\mu$Hz is not apparent given the S/N of the data.}

\begin{figure}[ht!]
    \centering
    \includegraphics[width=0.99\linewidth]{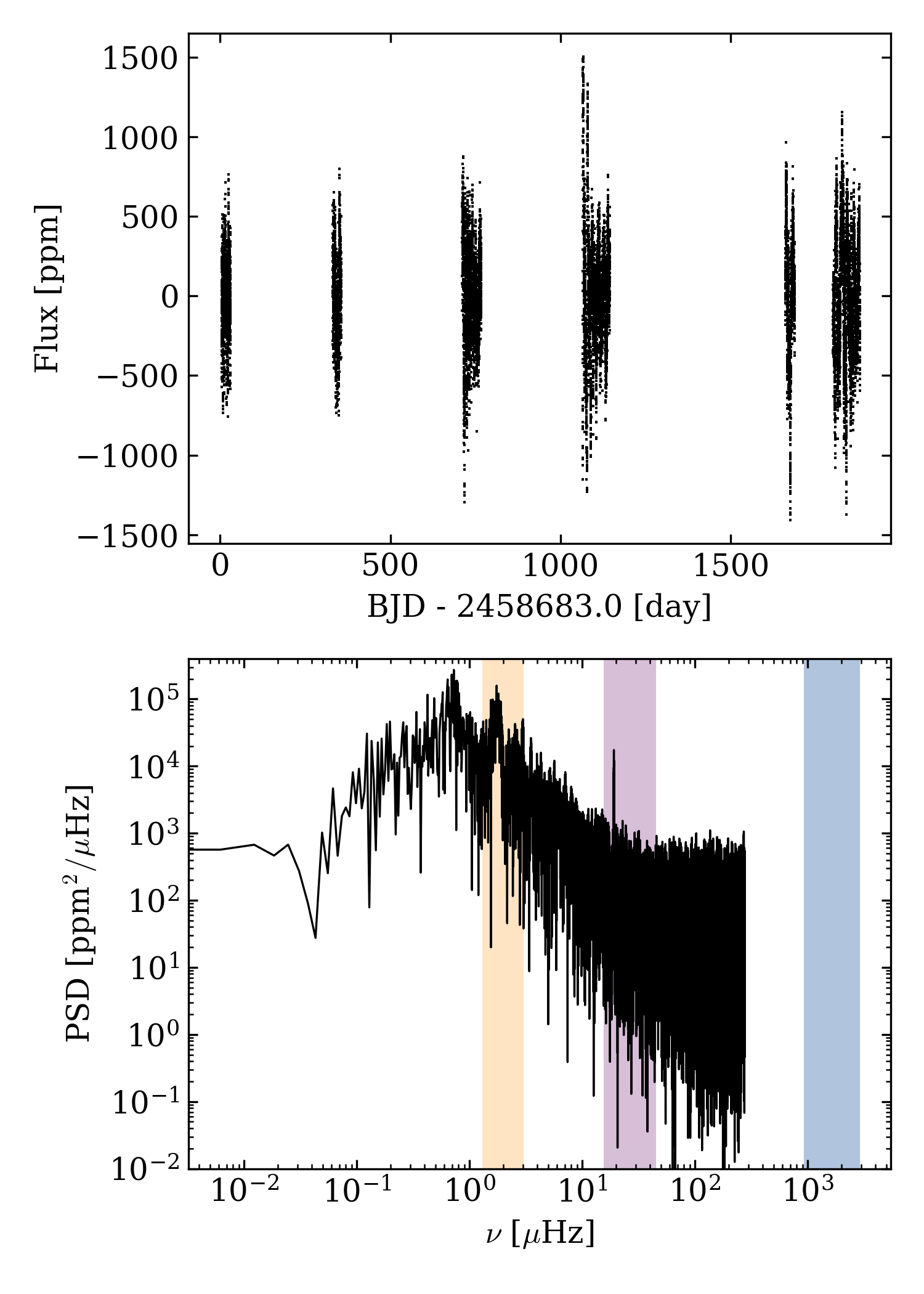}
    \caption{{TESS photometric signal of KIC~9139163 along side the power spectral density.} \textit{Top:} TESS light curves obtained for KIC~9139163. \textit{Bottom:} Corresponding PSD. As in Fig.~\ref{fig:kic9139163_lc_psd}, the frequency region of the surface rotational modulation is highlighted in light orange, the 0.6-day signal and its first harmonic in purple and the p-mode hump in blue.}
    \label{fig:tess_lc_psd}
\end{figure}

{It is also expected that the TESS photometry for KIC~9139163 is contaminated by the presence of its wide binary companion in the field. Indeed,} the binary star, KIC 9139151, is separated from KIC 9139163 by 29.3 arcsec\footnote{See \url{https://aladin.cds.unistra.fr/AladinLite/} for example.}. The TESS camera has a pixel field of view of 21 arcsec$^2$ \citep{vanderspek_tess_instrument_handbook_2018}, with an ensquared energy of the PSF up to 50 \% in one pixel. To capture at least 90 \% of the ensquared energy, there is a need to take the neighbouring pixels into account when extracting the PSF \citep{vanderspek_tess_instrument_handbook_2018}. With a separation of 29.3 arcsec between the two stars, the binary is therefore contaminating the TESS time-series of KIC 9139163.  On the contrary, the \textit{Kepler} pixel field of view is 3.98 arcsec$^2$ and the PSF FWHM generally between 3.1 and 7.5 arcsec \citep{cleve_kepler_instrument_handbook_16042217_2016}. We can therefore make sure that there is no contamination from the binary star in the \textit{Kepler} time series of KIC 9139163. 

\section{Asteroseismic analysis \label{appendix:asteroseismic_analysis}}

The \texttt{apollinaire} module \citep{Breton2022apollinaire} uses the \texttt{emcee} Ensemble Sampler \citep{Foreman-Mackey2013} to sample posterior probability distributions of stellar background and p-mode model parameters in the PSD of seismic solar-type targets. The Ensemble Samplers from \texttt{emcee} correspond to an improvement of the traditional Metropolis scheme \citep{Metropolis1953} designed to sample posterior probability with MCMC.  

\begin{figure}[ht!]
    \centering
    \includegraphics[width=.48\textwidth]{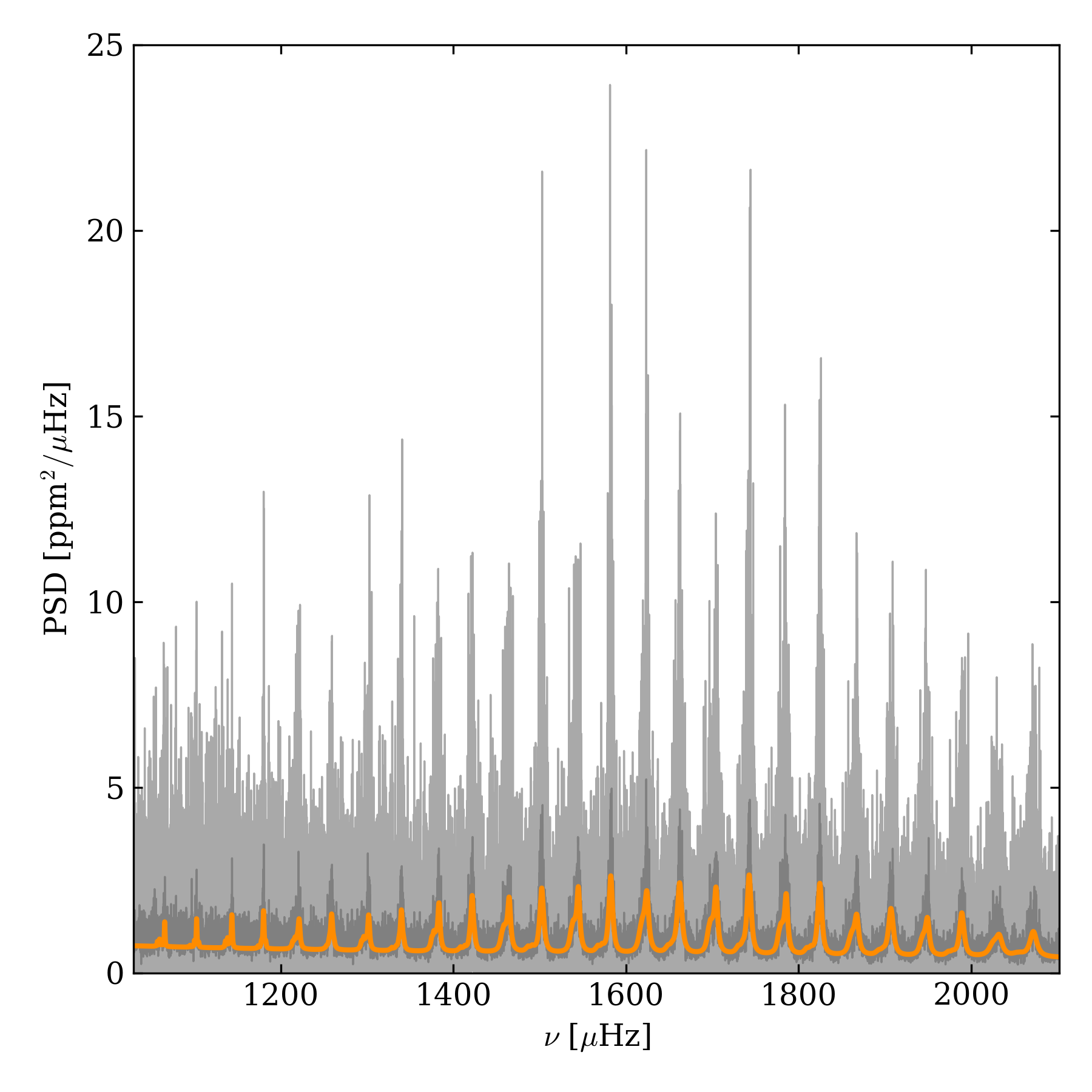}
    \caption{Short cadence PSD (grey) of KIC~9139163 around the p-mode hump, and corresponding model (orange) obtained with \texttt{apollinaire}. The smoothed PSD is shown in dark grey to emphasise the agreement with the fitted model.}
    \label{fig:p_mode_kic9139163}
\end{figure}

\begin{figure}[ht!]
    \centering
    \includegraphics[width=.48\textwidth]{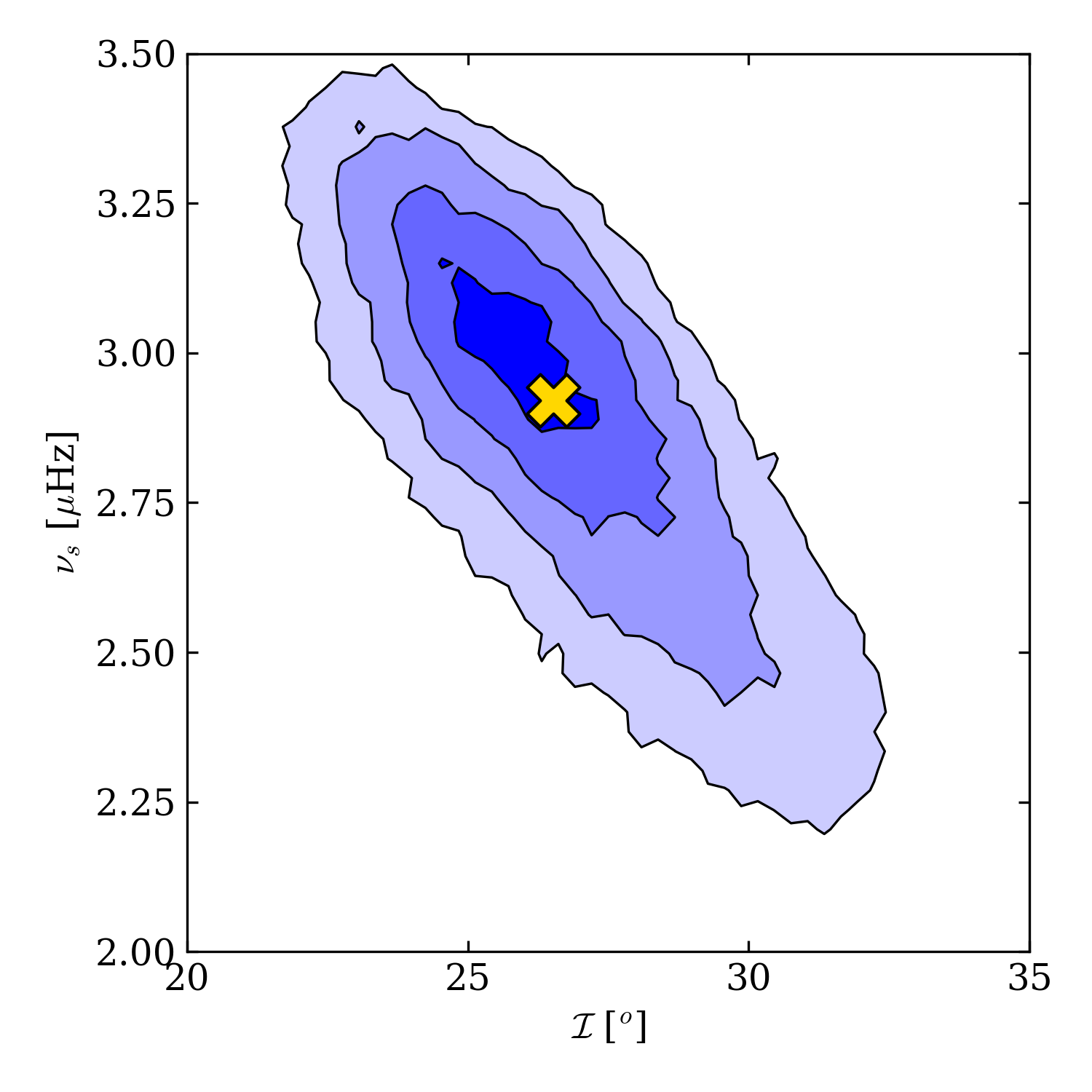}
    \caption{Posterior distribution obtained with the MCMC analysis performed with \texttt{apollinaire} and marginalised over $\mathcal{I}$ and $\nu_s$. The yellow cross indicates the position of the median of both marginalised distribution.}
    \label{fig:kic9139163_angle_splittings}
\end{figure}

In this work, the background is modelled as the sum of two Harvey's profiles \citep{Harvey1985}, the $p$-mode Gaussian envelope and a high-frequency constant taking the photon noise into account \citep{Mathur2010,Kallinger2014}. 
{In particular, from this fit, we obtain the following value for the frequency of maximum power, $\nu_{\rm max} = 1710 \pm 10$~$\mu$Hz, which corresponds to a period of approximately 10 minutes.}
The $p$-mode pattern is then fitted with a global parametrisation in order to obtain good initial values for mode frequencies, amplitudes and width \citep[see][for details]{Breton2022apollinaire}. 
We consider modes of degree $\ell = 1$, 2, and 3, and radial orders $n$ in the interval $11 < n < 24$. 
The posterior probability for the parameters of the 52 considered modes is then sampled at once. The inclination angle of the stellar rotation axis with respect to the observer, $\mathcal{I}$, and the mean rotational splitting, $\nu_s$, are included in our model as common parameters. 
The prior distributions $p$ are the same as those described in Section~3.1 from \citet{Breton2022apollinaire}. In particular, we ensure isotropy for the prior on the inclination angle by setting $p(\mathcal{I}) = \sin \mathcal{I}$. The bounds of the uniform distribution for $\nu_s$ are 0 and 4~$\mu$Hz.
To ensure convergence for the $\mathcal{I}$ and $\nu_s$ distribution (convergence is significantly faster for the other parameters), we use 500 walkers run over 15000 sampling steps, the 3000 first steps being discarded as burnt-in.
We extract the median of each marginalised distribution and the difference with the 16$^{\rm th}$ and 84$^{\rm th}$ percentiles of the distribution as summary statistics for our sampling. 

\begin{figure}[ht!]
    \centering
    \includegraphics[width=.48\textwidth]{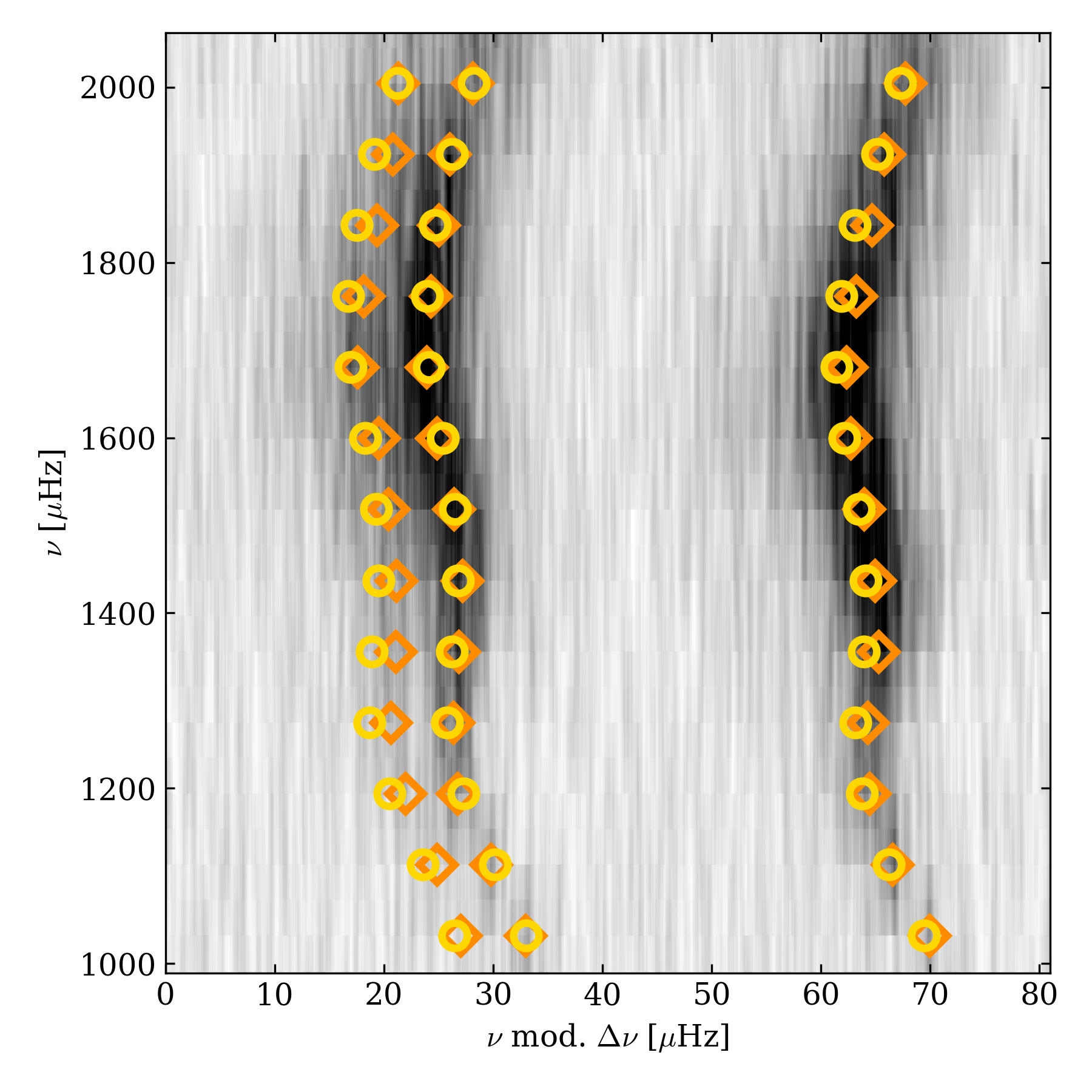}
    \caption{
    Echelle diagram obtained from the asteroseismic analysis of KIC~9139163. Only frequencies for modes $\ell=0$, 1, and 2 are shown, ordered, from left to right in a given row, as $\ell=2$, 0, and 1. The fitted frequencies are shown with the orange diamonds and the model frequencies (including surface effect corrections) with the yellow circles. The fitted-frequency error bars are smaller than the marker width and are not represented here.
    }
    \label{fig:echelle_diagram}
\end{figure}

The fitted model is represented in Fig.~\ref{fig:p_mode_kic9139163} where we compare it to the PSD of KIC~9139163 in the p-mode frequency range. Comparing the smoothed PSD with the model allows us to visually check the good agreement between the model and the data. The fitted mode frequencies are summarised in Table~\ref{tab:mode_frequency}.
Followingly, we show in Fig.~\ref{fig:kic9139163_angle_splittings} the posterior probability distribution, thinned by a factor ten and marginalised over $\mathcal{I}$ and $\nu_s$. 
The correlation between $\mathcal{I}$ and $\nu_s$ is clearly visible \citep{Ballot2006,Ballot2008}. From this analysis, we are however able to obtain a good constraint on the value of $\mathcal{I}$.

Only $\ell = 0$, 1, and 2 modes are used for the modelling step, which is performed using a model grid computed with MESA. The echelle diagram \citep[see e.g.][and references therein]{Garcia2019} comparing the fitted mode frequencies and the model frequencies, accounting for surface effect corrections, is shown in Fig.~\ref{fig:echelle_diagram}. We adopt $\Delta \nu = 81$~$\mu$Hz to compute the echelle diagram, where $\Delta \nu$ is the large-frequency separation of the modes. 

\begin{table}
\label{tab:mode_frequency}
\caption{Fitted mode order $n$, degree $\ell$, and frequency $\nu$ for KIC~9139163.}
\centering
\begin{tabular}{rrl}
\hline
\hline
$n$ & $\ell$ & $\nu$ ($\mu$Hz) \\
\hline
11 & 2 & $1059.41_{-1.18}^{+1.78}$ \\
12 & 0 & $1065.36_{-0.31}^{+0.21}$ \\
11 & 3 & $1097.20_{-2.07}^{+1.42}$ \\
12 & 1 & $1102.34_{-0.13}^{+0.11}$ \\
12 & 2 & $1138.22_{-0.81}^{+1.17}$ \\
13 & 0 & $1143.17_{-0.17}^{+0.16}$ \\
12 & 3 & $1176.24_{-1.74}^{+1.56}$ \\
13 & 1 & $1179.93_{-0.13}^{+0.13}$ \\
13 & 2 & $1216.35_{-0.58}^{+0.87}$ \\
14 & 0 & $1221.12_{-0.25}^{+0.28}$ \\
13 & 3 & $1255.57_{-2.62}^{+2.28}$ \\
14 & 1 & $1258.84_{-0.23}^{+0.23}$ \\
14 & 2 & $1296.01_{-0.95}^{+0.97}$ \\
15 & 0 & $1301.72_{-0.23}^{+0.23}$ \\
14 & 3 & $1331.12_{-1.77}^{+2.00}$ \\
15 & 1 & $1339.66_{-0.22}^{+0.22}$ \\
15 & 2 & $1377.41_{-0.58}^{+0.71}$ \\
16 & 0 & $1383.20_{-0.20}^{+0.19}$ \\
15 & 3 & $1410.73_{-1.49}^{+1.74}$ \\
16 & 1 & $1421.66_{-0.18}^{+0.17}$ \\
16 & 2 & $1458.49_{-0.59}^{+0.65}$ \\
17 & 0 & $1464.53_{-0.22}^{+0.22}$ \\
16 & 3 & $1488.89_{-1.38}^{+1.49}$ \\
17 & 1 & $1502.29_{-0.18}^{+0.18}$ \\
17 & 2 & $1538.77_{-0.49}^{+0.53}$ \\
18 & 0 & $1544.73_{-0.18}^{+0.19}$ \\
17 & 3 & $1569.83_{-1.32}^{+1.87}$ \\
18 & 1 & $1582.28_{-0.17}^{+0.16}$ \\
18 & 2 & $1618.86_{-0.53}^{+0.55}$ \\
19 & 0 & $1624.16_{-0.20}^{+0.21}$ \\
18 & 3 & $1651.05_{-1.41}^{+1.22}$ \\
19 & 1 & $1662.06_{-0.18}^{+0.19}$ \\
19 & 2 & $1697.92_{-0.44}^{+0.48}$ \\
20 & 0 & $1704.25_{-0.19}^{+0.19}$ \\
19 & 3 & $1732.50_{-1.40}^{+1.21}$ \\
20 & 1 & $1742.64_{-0.18}^{+0.18}$ \\
20 & 2 & $1779.43_{-0.51}^{+0.54}$ \\
21 & 0 & $1785.61_{-0.20}^{+0.20}$ \\
20 & 3 & $1811.89_{-1.80}^{+1.90}$ \\
21 & 1 & $1824.55_{-0.17}^{+0.17}$ \\
21 & 2 & $1861.63_{-0.65}^{+0.76}$ \\
22 & 0 & $1867.33_{-0.24}^{+0.24}$ \\
21 & 3 & $1895.08_{-1.70}^{+1.66}$ \\
22 & 1 & $1906.95_{-0.23}^{+0.23}$ \\
22 & 2 & $1944.12_{-0.53}^{+0.59}$ \\
23 & 0 & $1949.34_{-0.29}^{+0.27}$ \\
22 & 3 & $1975.97_{-1.88}^{+1.85}$ \\
23 & 1 & $1989.08_{-0.22}^{+0.22}$ \\
23 & 2 & $2025.58_{-0.74}^{+0.96}$ \\
24 & 0 & $2032.41_{-0.46}^{+0.43}$ \\
23 & 3 & $2054.29_{-1.84}^{+1.34}$ \\
24 & 1 & $2071.99_{-0.30}^{+0.32}$ \\
\hline
\end{tabular}
\end{table}

\section{Comparison with previous stellar modelling of KIC9139163 \label{appendix:modelling_comparison}}

In this appendix, we compare the stellar modelling results obtained with the IACgrid with those previously published by \citet{SilvaAguirre2017}\footnote{The catalogue is available on Vizier: \url{https://vizier.cds.unistra.fr/viz-bin/VizieR?-source=J/ApJ/835/173}.} using different modelling pipelines. In Fig.~\ref{fig:stellar_modelling_comparison}, we show the mass, radius, and age values from the IACgrid against the ones obtained from the Aarhus Stellar Evolution Code fitting method \citep[ASTFIT,][]{Christensen-Dalsgaard2008a,Christensen-Dalsgaard2008b}, the Bayesian Stellar Algorithm \citep[BASTA,][]{SilvaAguirre2015}, the \texttt{Cesam2k} Stellar Model Optimization pipeline \citep[C2kSMO][]{Lebreton2014}, the Goettingen pipeline \citep[GOE,][]{Appourchaux2015,Reese2016}, the V\&A pipeline \citep{Verma2014}, and the Yale Monte Carlo Method \citep[YMCM,][]{SilvaAguirre2015}. As  visible, the parameters inferred with the IACgrid lay between the extremal values from \citet{SilvaAguirre2017}.

\begin{figure}
    \centering
    \includegraphics[width=0.99\linewidth]{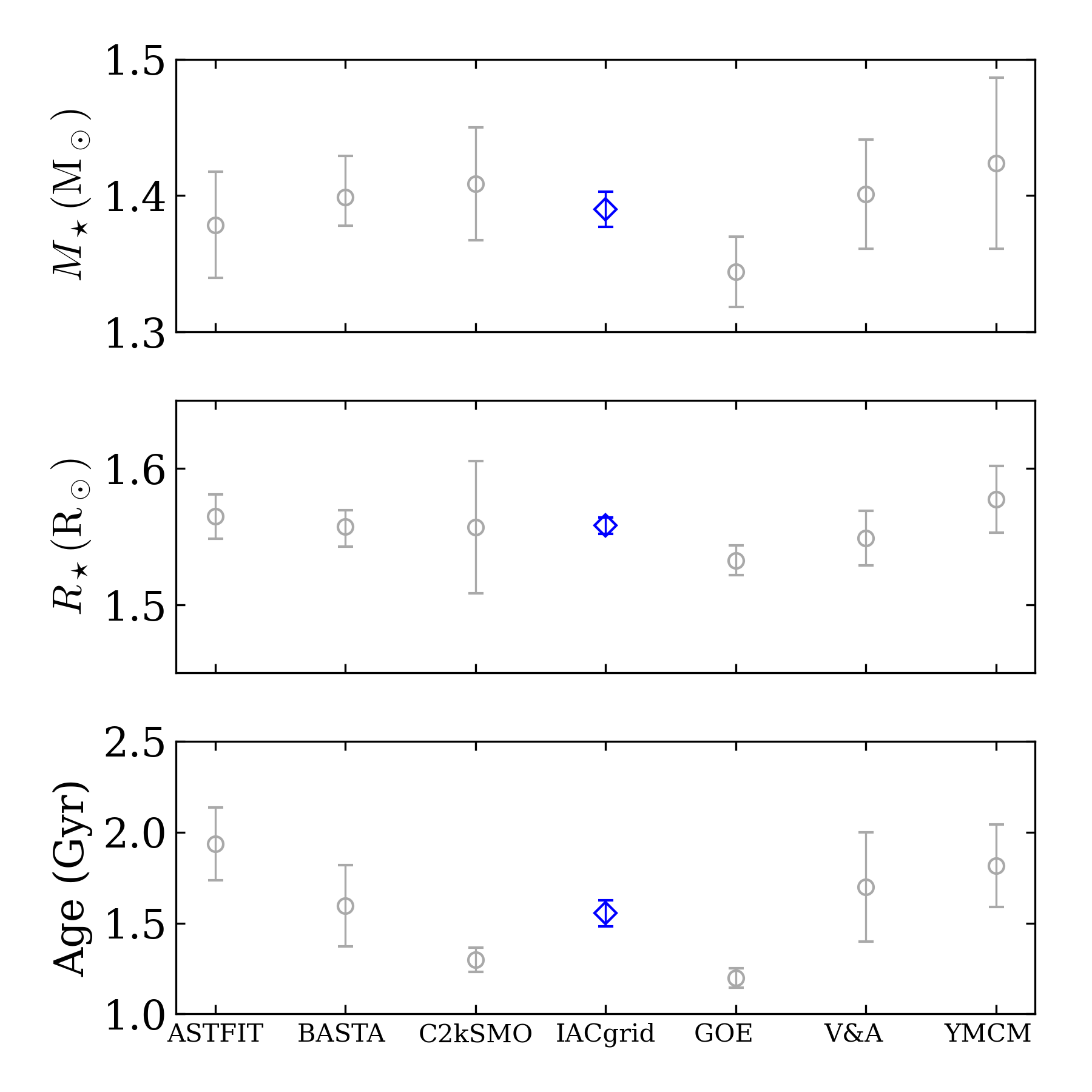}
    \caption{Comparison between the stellar mass $M_\star$, radius $R_\star$, and age obtained in this work (IACgrid, blue) and the values from the ASTFIT, BASTA, C2kSMO, GOE, V\&A, and YMCM pipelines published in \citet{SilvaAguirre2017}, in grey.}
    \label{fig:stellar_modelling_comparison}
\end{figure}

\section{Radial velocity}

\subsection{RV data and additional sampling diagnostic \label{appendix:rv_analysis}}

The SN and bp method for RV extraction used in this work was presented in Sect.~\ref{sec:rv_extraction_analysis}. 
{The $\rm RV_{SN}$ and $\rm RV_{bp}$ time series obtained with this approach are listed in Table~\ref{tab:RVsnbp}. Additionally, the periodogram of the $\rm RV_{bp}$ time series is shown in Fig.~\ref{fig:rv_periodogram}, with the peak structure corresponding to the 0.60474 period highlighted. The periodogram of the residuals after removing the best fitting model is also shown.

Finally, the chain autocorrelation function for the parameters of the sampled distribution is shown in Fig.~\ref{fig:rv_autocorrelation} for the parameters $K_2$, $P$, and $t_{\rm inf}$. In practice, the parameter $K_2 = K \sqrt{1 - e^2} P^{1/3}$ is sampled instead of $K$ in order to reduce correlation between parameters \citep[see][]{bonfanti2020}. As visible, the autocorrelation time becomes and remains very low after just a few steps, supporting the statement of an efficient exploration of the target distributions. 
}

\begin{figure}[ht!]
    \centering
    \includegraphics[width=0.9\linewidth]{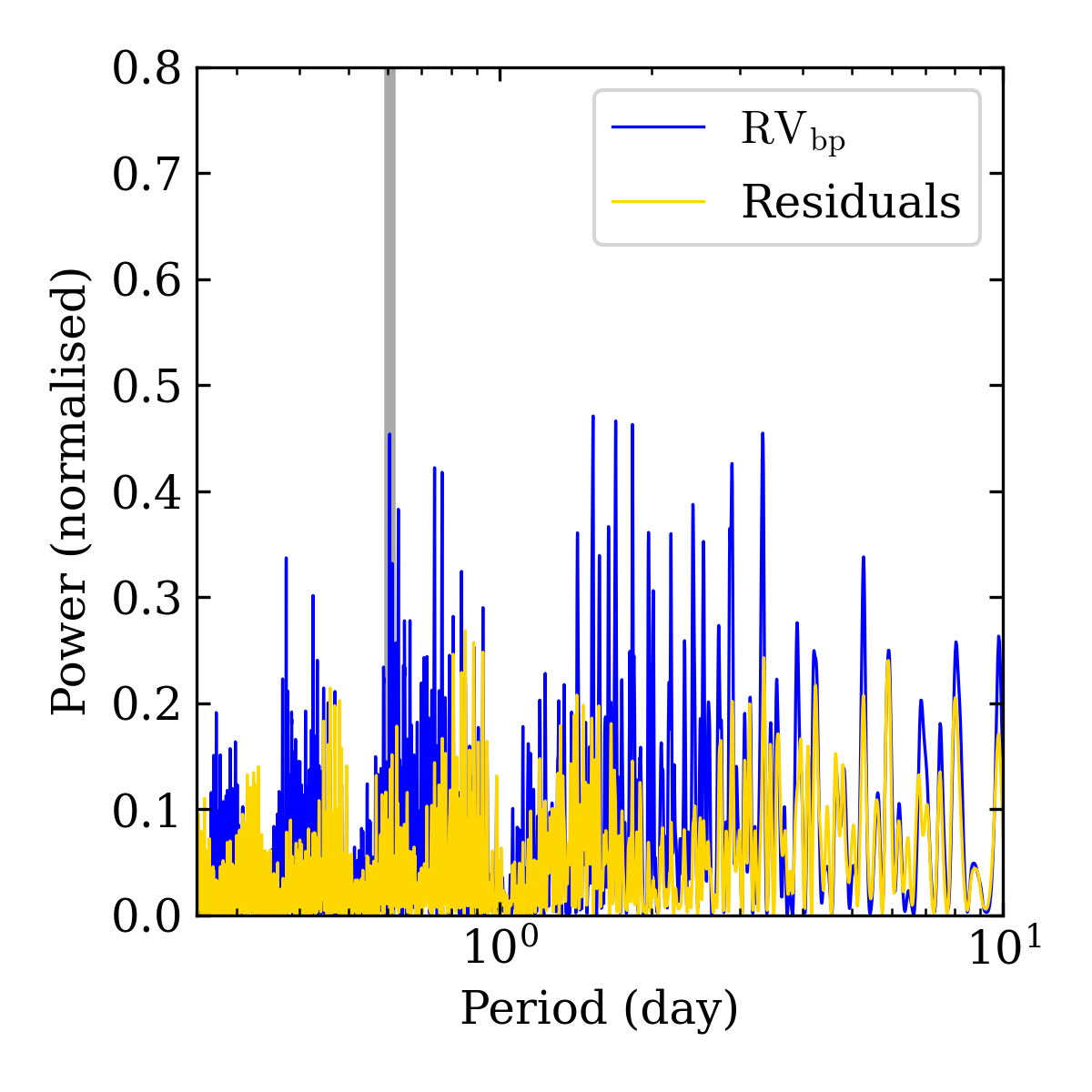}
    \caption{Periodogram of the $\rm RV_{bp}$ (blue) and of the residuals that are obtained after subtracting the best-fit Keplerian signal from the detrended RV time series (yellow). The vertical grey line corresponds to the 0.60474 orbital period of the candidate planet.} 
    \label{fig:rv_periodogram}
\end{figure}

\begin{figure}[ht!]
    \centering
    \includegraphics[width=0.9\linewidth]{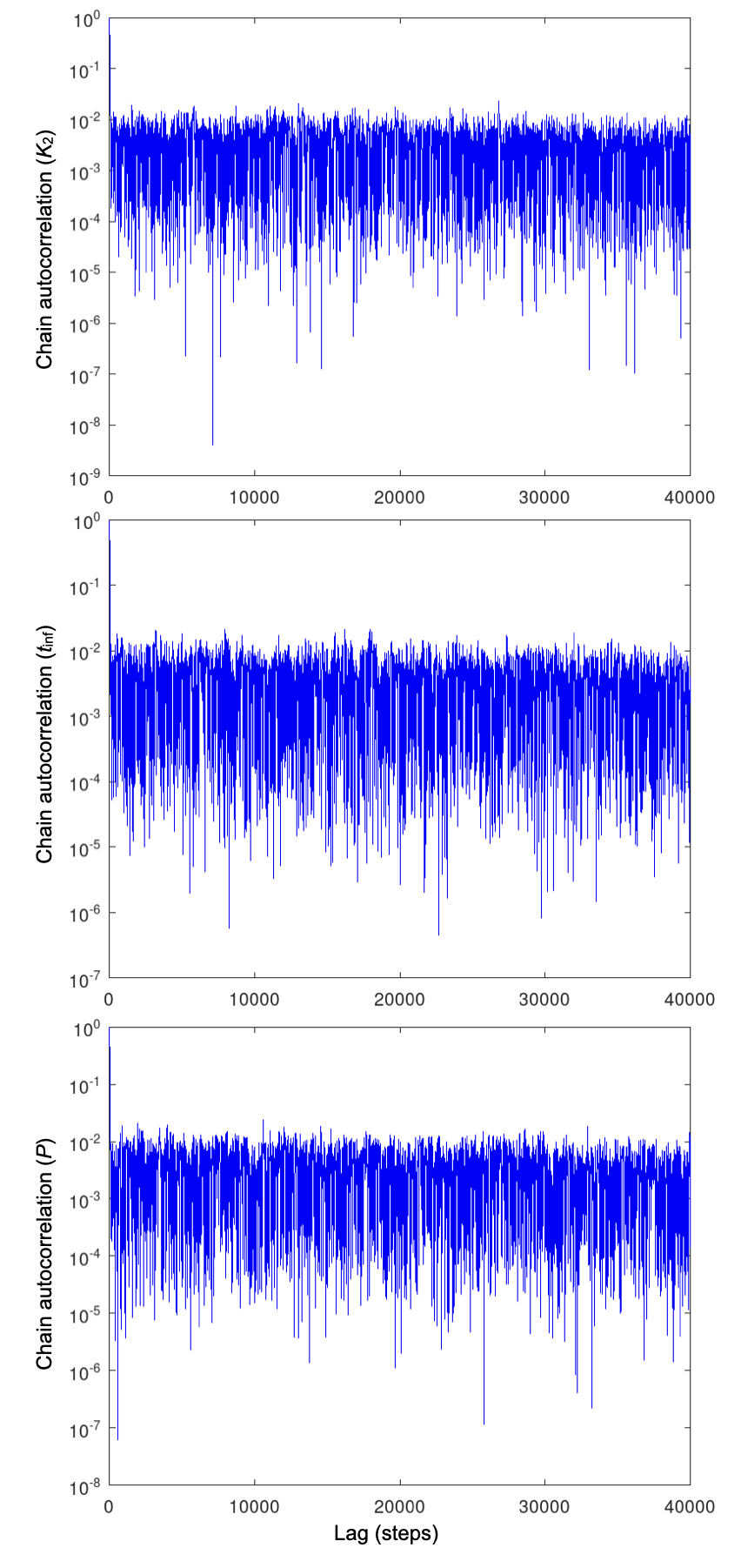}
    \caption{Chain autocorrelation functions for the three orbital parameters sampled in the RV analysis presented in Sect.~\ref{sec:rv_extraction_analysis}, $K_2$, $P$, and $t_{\rm inf}$.} 
    \label{fig:rv_autocorrelation}
\end{figure}

\onecolumn
\begin{table}[h!]
\centering
\caption{Radial velocities ${\rm RV_{SN}}$ extracted from the centred HARPS-N CCFs with their errors $\sigma_{\mathrm{RV}}$, followed by the hyperparameters inferred from the SN fit onto the CCFs (i.e. $\mathrm{FWHM_{SN}}$, $A_{\mathrm{SN}}$, and $\gamma$) and by the \emph{bp}-detrended RV values ($\rm RV_{\mathrm{bp}}$) with their errors which also account for the jitter ($\sigma_{\mathrm{RV(bp+jitter)}}$).}
\label{tab:RVsnbp}
\begin{tabular}{rrrrrrrr}
\hline\hline
\multicolumn{1}{c}{$\mathrm{BJD_{TDB}}$} & \multicolumn{1}{c}{$\rm RV_{SN}$} & \multicolumn{1}{c}{$\sigma_{\mathrm{RV}}$} & \multicolumn{1}{c}{$\mathrm{FWHM_{SN}}$} & \multicolumn{1}{c}{$A_{\mathrm{SN}}$} & \multicolumn{1}{c}{$\gamma$} & \multicolumn{1}{c}{$\rm RV_{\mathrm{bp}}$} & \multicolumn{1}{c}{$\sigma_{\mathrm{RV(bp+jitter)}}$} \\
\multicolumn{1}{c}{[$\mathrm{JD-2\,450\,000}$]} & \multicolumn{1}{c}{[m\,s$^{-1}$]} & \multicolumn{1}{c}{[m\,s$^{-1}$]} & \multicolumn{1}{c}{[km\,s$^{-1}$]} & \multicolumn{1}{c}{[\%]} & \multicolumn{1}{c}{} & \multicolumn{1}{c}{[m\,s$^{-1}$]} & \multicolumn{1}{c}{[m\,s$^{-1}$]} \\
\hline
  9664.691479 & $ 13.065$ & 2.233 & 8.800 & 30.658 & $ 0.0709$ & $ 8.063$ & 3.676 \\
  9676.642688 & $  7.464$ & 1.690 & 8.774 & 30.772 & $ 0.0764$ & $ 4.367$ & 3.374 \\
  9676.695248 & $  7.064$ & 1.385 & 8.765 & 30.857 & $ 0.0755$ & $ 4.917$ & 3.232 \\
  9676.742645 & $ 10.905$ & 1.629 & 8.761 & 30.739 & $ 0.0724$ & $ 7.117$ & 3.343 \\
  9677.646973 & $  3.339$ & 1.752 & 8.795 & 30.693 & $ 0.0789$ & $ 0.212$ & 3.405 \\
  9677.739893 & $  3.973$ & 1.598 & 8.795 & 30.572 & $ 0.0741$ & $ 0.190$ & 3.329 \\
  9702.599246 & $  7.424$ & 1.763 & 8.812 & 30.388 & $ 0.0639$ & $ 8.568$ & 3.411 \\
  9702.651204 & $  1.868$ & 1.583 & 8.810 & 30.632 & $ 0.0685$ & $ 0.392$ & 3.322 \\
  9702.731508 & $  5.366$ & 1.281 & 8.815 & 30.565 & $ 0.0683$ & $ 4.042$ & 3.189 \\
  9703.591076 & $ -4.441$ & 1.676 & 8.814 & 30.524 & $ 0.0703$ & $-5.542$ & 3.367 \\
  9703.641344 & $ -9.540$ & 1.650 & 8.811 & 30.495 & $ 0.0691$ & $-10.388$ & 3.354 \\
  9703.730722 & $ -5.117$ & 1.333 & 8.815 & 30.519 & $ 0.0705$ & $-6.164$ & 3.210 \\
  9720.569605 & $ -8.334$ & 1.382 & 8.788 & 30.646 & $ 0.0799$ & $-7.682$ & 3.231 \\
  9720.625834 & $ -7.579$ & 1.279 & 8.789 & 30.759 & $ 0.0755$ & $-6.781$ & 3.188 \\
  9720.720027 & $ -2.563$ & 1.294 & 8.775 & 30.681 & $ 0.0734$ & $-2.615$ & 3.194 \\
  9722.573244 & $  1.616$ & 1.447 & 8.771 & 30.639 & $ 0.0710$ & $ 1.502$ & 3.259 \\
  9722.629160 & $  4.434$ & 2.105 & 8.785 & 30.453 & $ 0.0748$ & $ 5.629$ & 3.600 \\
  9722.711906 & $  5.981$ & 1.372 & 8.770 & 30.685 & $ 0.0731$ & $ 6.064$ & 3.226 \\
  9765.516186 & $  3.554$ & 1.325 & 8.802 & 30.704 & $ 0.0734$ & $ 7.910$ & 3.207 \\
  9765.544161 & $  2.161$ & 1.249 & 8.792 & 30.702 & $ 0.0693$ & $ 6.401$ & 3.176 \\
  9765.592333 & $  1.035$ & 1.541 & 8.814 & 30.379 & $ 0.0661$ & $ 7.773$ & 3.301 \\
  9765.657680 & $ -2.059$ & 1.633 & 8.783 & 30.552 & $ 0.0655$ & $ 2.274$ & 3.346 \\
  9765.721443 & $ -4.184$ & 1.424 & 8.806 & 30.400 & $ 0.0657$ & $ 2.110$ & 3.249 \\
  9766.510929 & $ -6.840$ & 1.359 & 8.790 & 30.879 & $ 0.0738$ & $-0.434$ & 3.221 \\
  9766.538845 & $ -8.030$ & 1.171 & 8.794 & 30.804 & $ 0.0758$ & $-2.533$ & 3.146 \\
  9766.586600 & $ -9.944$ & 1.617 & 8.791 & 30.594 & $ 0.0738$ & $-5.844$ & 3.338 \\
  9766.651740 & $-13.508$ & 2.012 & 8.798 & 30.427 & $ 0.0754$ & $-7.794$ & 3.546 \\
  9785.558834 & $ -7.263$ & 2.657 & 8.801 & 30.574 & $ 0.0702$ & $-1.379$ & 3.948 \\
  9785.624621 & $  2.520$ & 1.784 & 8.795 & 30.575 & $ 0.0719$ & $ 8.346$ & 3.422 \\
  9786.538263 & $-12.888$ & 4.031 & 8.809 & 30.622 & $ 0.0537$ & $-3.818$ & 4.978 \\
  9786.572777 & $ -5.022$ & 2.084 & 8.802 & 30.561 & $ 0.0658$ & $ 1.294$ & 3.587 \\
  9786.600022 & $-27.477$ & 8.650 & 8.704 & 31.191 & $ 0.0390$ & $-2.724$ & 9.130 \\
  9786.636538 & $ -5.606$ & 2.152 & 8.805 & 30.643 & $ 0.0684$ & $ 0.452$ & 3.627 \\
  9810.365074 & $-14.889$ & 1.478 & 8.775 & 30.861 & $ 0.0745$ & $-4.936$ & 3.273 \\
  9810.480304 & $-16.356$ & 2.020 & 8.791 & 30.472 & $ 0.0699$ & $-7.563$ & 3.551 \\
  9810.509840 & $-10.152$ & 2.113 & 8.786 & 30.392 & $ 0.0736$ & $-0.101$ & 3.604 \\
  9810.559388 & $-13.444$ & 1.490 & 8.783 & 30.593 & $ 0.0711$ & $-5.541$ & 3.278 \\
  9810.602709 & $-13.392$ & 1.947 & 8.790 & 30.608 & $ 0.0784$ & $-4.890$ & 3.509 \\
  9811.362880 & $ -6.099$ & 1.620 & 8.772 & 30.488 & $ 0.0719$ & $ 2.402$ & 3.339 \\
  9811.477380 & $ -8.721$ & 1.910 & 8.773 & 30.449 & $ 0.0668$ & $ 0.468$ & 3.489 \\
  9811.506037 & $ -7.880$ & 2.217 & 8.758 & 30.364 & $ 0.0730$ & $ 2.556$ & 3.666 \\
  9811.555272 & $ -1.870$ & 2.052 & 8.775 & 30.463 & $ 0.0716$ & $ 6.965$ & 3.569 \\
  9811.597702 & $ -3.956$ & 2.878 & 8.790 & 30.502 & $ 0.0728$ & $ 4.624$ & 4.100 \\
  9812.452361 & $ -4.149$ & 1.781 & 8.778 & 30.656 & $ 0.0703$ & $ 3.937$ & 3.420 \\
  9812.602312 & $  0.000$ & 1.804 & 8.794 & 30.498 & $ 0.0775$ & $ 9.136$ & 3.432 \\
  9813.363917 & $-28.495$ & 8.898 & 8.807 & 31.866 & $ 0.0791$ & $ 5.314$ & 9.365 \\
  9813.453800 & $  6.766$ & 2.354 & 8.772 & 30.580 & $ 0.0721$ & $ 2.209$ & 3.750 \\
  9813.497109 & $  1.145$ & 2.643 & 8.773 & 30.534 & $ 0.0710$ & $ 0.764$ & 3.939 \\
  9813.542178 & $ 12.444$ & 3.899 & 8.782 & 30.737 & $ 0.0858$ & $-1.022$ & 4.872 \\
  9814.364186 & $  8.516$ & 1.515 & 8.777 & 30.778 & $ 0.0750$ & $-2.743$ & 3.289 \\
  9814.482031 & $ 10.836$ & 1.886 & 8.781 & 30.633 & $ 0.0732$ & $ 3.462$ & 3.476 \\
  9814.505804 & $  5.360$ & 1.933 & 8.787 & 30.563 & $ 0.0695$ & $ 4.116$ & 3.502 \\
  9814.554136 & $ 15.013$ & 2.396 & 8.796 & 30.650 & $ 0.0855$ & $ 4.662$ & 3.777 \\
  9814.599343 & $  8.951$ & 2.149 & 8.793 & 30.599 & $ 0.0750$ & $ 3.678$ & 3.625 \\
  9882.310783 & $ 19.017$ & 1.948 & 8.747 & 31.136 & $ 0.0801$ & $ 4.514$ & 3.510 \\
  9882.360237 & $ 14.921$ & 1.908 & 8.764 & 31.098 & $ 0.0750$ & $ 4.182$ & 3.488 \\
  9883.308535 & $  0.148$ & 1.180 & 8.808 & 30.727 & $ 0.0631$ & $-2.904$ & 3.149 \\
  9883.352237 & $  1.813$ & 1.369 & 8.808 & 30.621 & $ 0.0675$ & $ 0.032$ & 3.225 \\
  9883.381148 & $  4.454$ & 1.804 & 8.811 & 30.646 & $ 0.0707$ & $ 0.528$ & 3.432 \\
\hline
\end{tabular}
\end{table}

\twocolumn



\subsection{Additional RV analysis}

We performed a second analysis of the RV$_{\rm SN}$ time series by employing the Floating-Chunk Offset (FCO) method \citep{2010A&A...520A..93H,2014A&A...568A..84H}.This method was developed specifically for the determination of RV amplitudes of short-period planets of $P \lesssim $ 1 day. It requires that two or more RV observations be acquired per observing night, and the nightly blocks of RVs are then treated as independent chunks of data with unknown (free) RV offsets. The RV offsets for all chunks are then independently fitted against an RV model for a circular orbit of given period, amplitude, and phase. Hence, only RV variations within a given night contribute to detected signals;  the FCO method therefore acts like a high-pass filter that ignores any RV variations between different observing nights.

We adapted the procedure described in \citet{Deeg2023} by implementing the FCO method into a MCMC framework. 
The model combined a circular Keplerian orbit and an ensemble of offset parameters $c_i$, each corresponding to one night. The framework was implemented using the \texttt{pymc} module \citep{exoplanet:pymc3} and the Keplerian orbit solver from the \texttt{exoplanet} module \citep{exoplanet:joss}. A standard normal likelihood was chosen and before running the Hamiltonian Monte Carlo exploration \citep[HMC,][]{Duane1987,Neal2011}, a Maximum A-Posteriori (MAP) search of a likelihood maximum was performed.

The fit failed to converge and therefore to constrain the time of inferior conjunction $t_{\rm inf}$ and the RV semi-amplitude $K$. KIC~9139163 being an F-type star, it is impossible to neglect the contribution of granulation and oscillation variability over one night \citep[e.g.][]{Sulis2024}. Orbital modulation is therefore not dominating the RVs and is blurred, thus explaining the poor fit result. In this case, the offset will filter out the totality of the modulation, favouring a RV model without any orbital modulation. 

In our case, the root-mean-square (RMS) value of the $RV_{\rm SN}$ time series is 9.6 m s$^{-1}$. For comparison, considering the global stellar parameters derived in Sect.~\ref{sec:stellar_modelling} and the scaling laws provided in Eq.~(7) from \citet{Kjeldsen1995}, we expect to have a contribution to the signal from the $p$ modes of $\sim 0.6$~m/s. Indeed, the HARPS-N exposure time is comprised between 5 and 8 minutes, which is shorter than the $p$-mode characteristic period of ten minutes (see Sect.~\ref{RV_main} and Appendix~\ref{appendix:asteroseismic_analysis}). In addition, from Eq.~(6) and (7) from \citet{Sulis2022}, we expect the RMS contribution from the granulation to the RV signal to lay between 1.3 and 5.2 m/s, which is significant in regard of the observational RMS of our time series. Combined to the fact that we do not have a pre-determined knowledge on $t_{\rm inf}$ due to the absence of transit, this supports the statement that it is difficult to draw conclusions from the application of the FCO on our dataset. 

\section{Additional grid-fitting results}
\label{appendix:grid-fitting}

\subsection{Robustness of the grid-based phase-curve fit}


To illustrate the posterior distributions of the fitted parameters, we show the corner plot that displays the marginal distributions of the three joint parameters: the planetary radius $R_p$, the phase offset $\delta_{\rm phase}$ and the inclination of the orbit $i$, as well as the parameters fitted for each dataset (\textit{Kepler} and TESS) independently: the geometric albedo $A_g$, the heat redistribution efficiency $\varepsilon$, the cloud offset $\eta_c$, and the photometric offset $\Delta A$. Fig.~\ref{fig:corner_joint_fit_photometry} shows the corner plot, highlighting the parameter correlations and 1$\sigma$ uncertainties.

\begin{figure*}[htbp!]
    \centering
    \includegraphics[width=\linewidth]{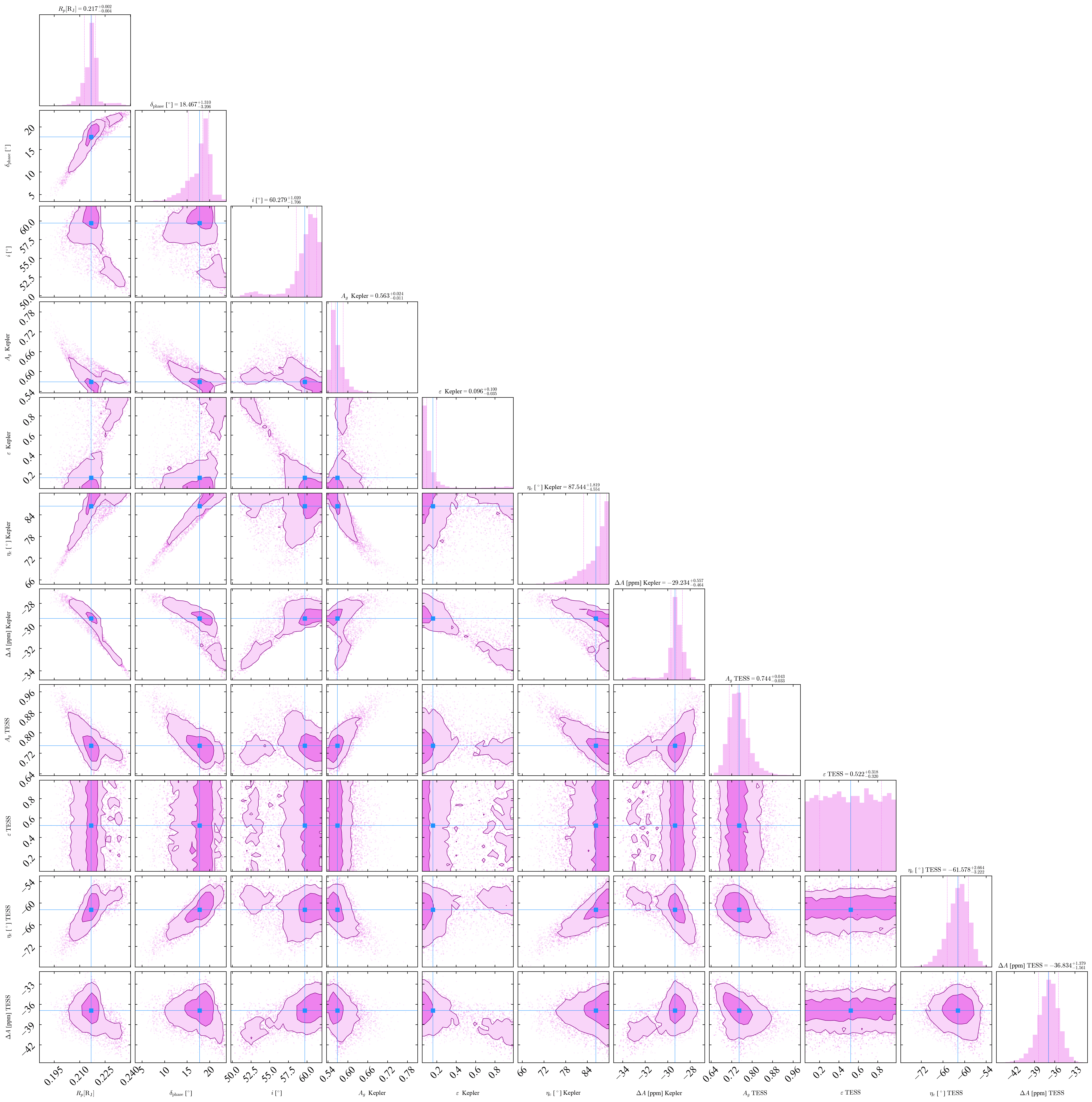}
    \caption{Corner plot of the \textit{Kepler} and TESS joint fit, showing the posterior distributions of $R_p$, $\delta_{\rm phase}$, $i$ as joint parameters, and $A_g$, $\varepsilon$, $\eta_c$ and $\Delta A$ as independent parameters for each dataset. The 1D histograms on the diagonal show the marginal distributions, while the 2D histograms display the covariances between parameters.}
    \label{fig:corner_joint_fit_photometry}
\end{figure*}

\subsection{Reflection and emission maps of the planetary candidate}

{By injecting the median values of the retrieved parameters displayed in Table~\ref{tab:grid_fit_results} into the model described in Sect.~\ref{sec:photometric_planet_model}, we computed the longitude and latitude reflection and emission maps of the candidate planet, displayed in Fig.~\ref{fig:emission_reflection_maps_Kepler} and Fig.~\ref{fig:emission_reflection_maps_TESS}. The period was set to $P=0.60474 $ day. The planetary emission is calculated based on the atmospheric temperature $T_{\rm atm}$ that depends on the heat redistribution efficiency parameter $\varepsilon$ (see Eq.~\ref{eq:ODE_P}). To compare the intrinsic reflective properties of the atmosphere independently of stellar irradiation, the reflected radiance is normalised by the incident stellar flux scaling, $(1 \; {\rm AU}/a)^2$, where $a = 0.0158 \; {\rm AU}$ (see Table~\ref{tab:grid_fit_results}). We estimated the reference Earth reflected radiances assuming Lambertian scattering $\mathcal{I} \approx (A F_{\rm band}) /\pi$, where $A$ represents an effective broadband reflectivity. Using an Earth mean albedo of $A \approx 0.3$ and the solar flux integrated over the \textit{Kepler} and TESS bandpasses, yielding approximate Earth-equivalent reflected radiances of $\sim 90~{\rm W \; m^{-2} \; sr^{-1}}$ in the \textit{Kepler} bandpass and $\sim 60~{\rm W \; m^{-2} \; sr^{-1}}$ in the TESS bandpass. This normalisation, $\mathcal{I_{\oplus} \sim \mathcal{I}\times \rm{(1} \; {\rm AU}/a)^2}$, is equivalent to comparing the planet to an Earth-like atmosphere placed at the same orbital separation. As a result, the reflection map displays the highly reflective westward offset as a possible explanation for the shift of the maximum of the photometric phase-folded \textit{Kepler} data, with a maximum reflection offset of $\eta_{c \rm , Kepler} \sim 86.4^{\circ}$ and a geometric albedo of $A_{g \rm , Kepler} ~ 0.52$. By contrast, the reflection map displays a reflective eastward offset as a possible explanation for the shift of the maximum of the photometric phase-folded TESS data, with a maximum reflection offset of $\eta_{c \rm , TESS} \sim -62^{\circ}$ and a geometric albedo of $A_{g \rm , TESS} ~ 0.75$.  As our fits favour a candidate planet with low efficient day-to-night heat transport, the temperature maxima are globally centred around the substellar point, with a slight eastward offset towards the evening terminator induced by the longitudinal asymmetry of the \cite{cowan_model_2011} formalism. This asymmetry is less pronounced, and the temperature is less uniform, in the fit to the \textit{Kepler} data because the retrieved redistribution efficiency is lower.} 

\begin{figure}[htbp!]
    \centering
    \includegraphics[width=0.48\textwidth]{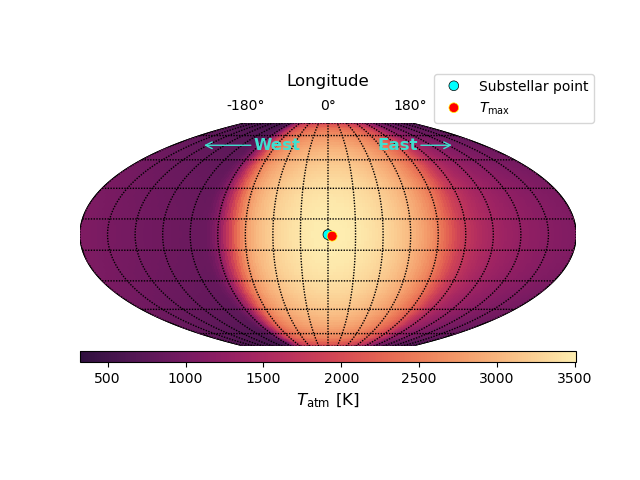}
    \includegraphics[width=0.48\textwidth]{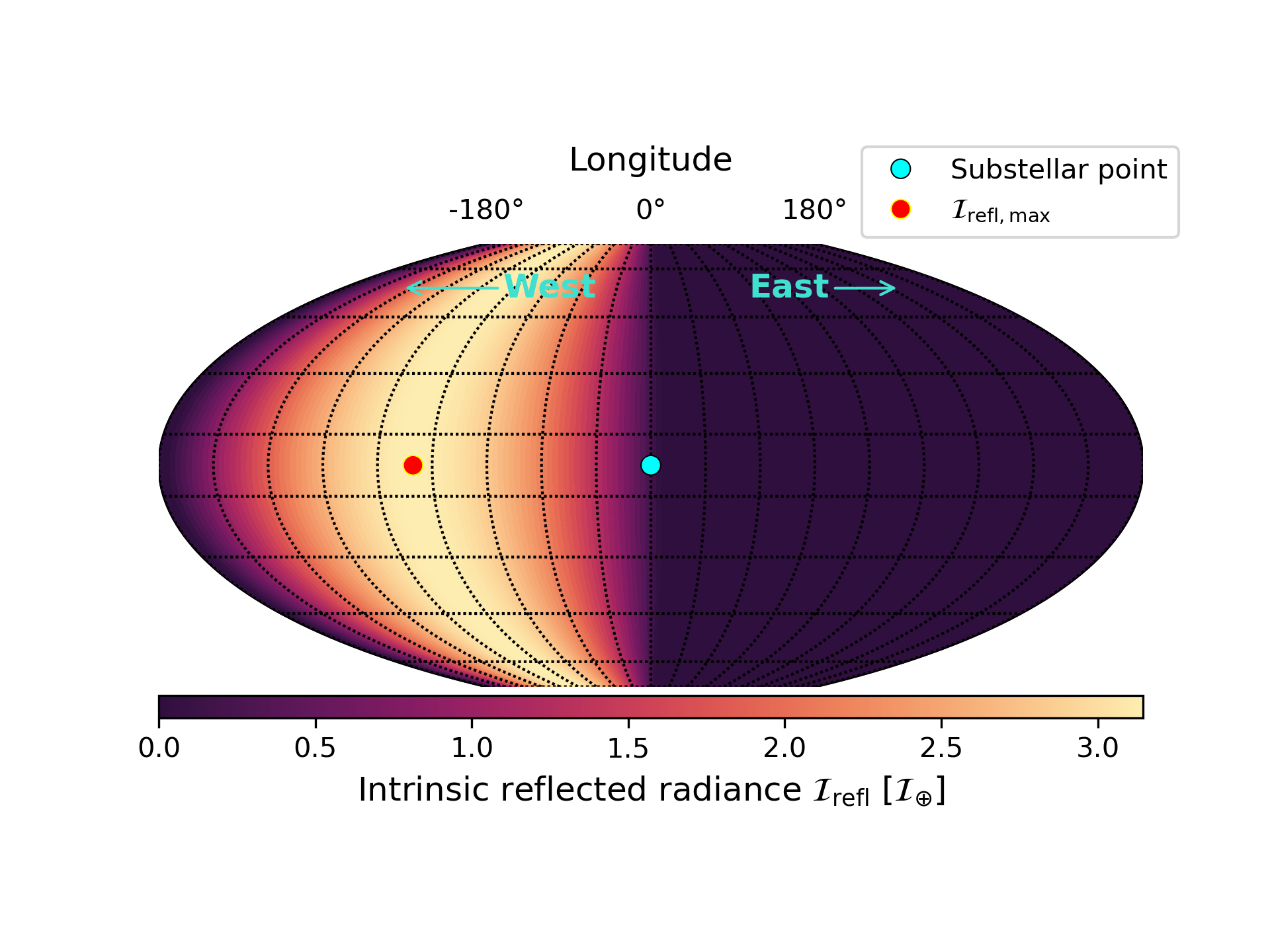}
    \caption{\textit{Kepler} longitude-latitude maps of the candidate planet assuming the best-fit parameters $\varepsilon$, $A_g$, $\eta_c$, $R_p$ inferred from the phase-curve fitting (Table~\ref{tab:grid_fit_results}). \textit{Top:} Temperature map with low heat redistribution, $\varepsilon=0.16$ and an intrinsic temperature of $T_{\rm int} = 100$ K. 
    \textit{Bottom:} Reflection map normalised to the intrinsic Earth radiance with a geometric albedo of $A_g = 0.52$ and a cloud offset of $\eta_c = 86.4^{\circ}$.}
    \label{fig:emission_reflection_maps_Kepler}
\end{figure}

\begin{figure}[htbp!]
    \centering
    \includegraphics[width=0.48\textwidth]{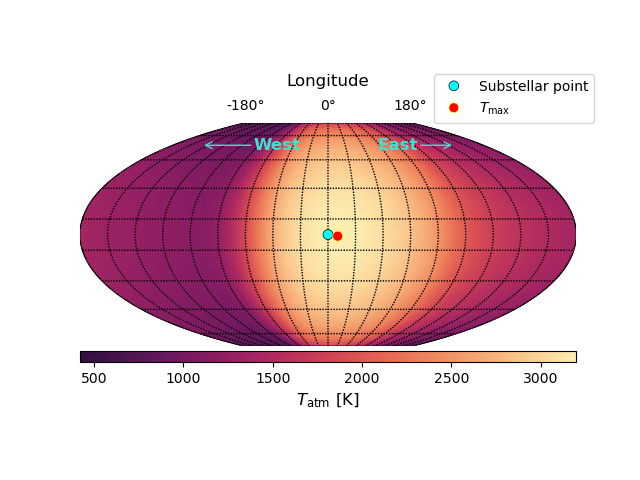}
    \includegraphics[width=0.48\textwidth]{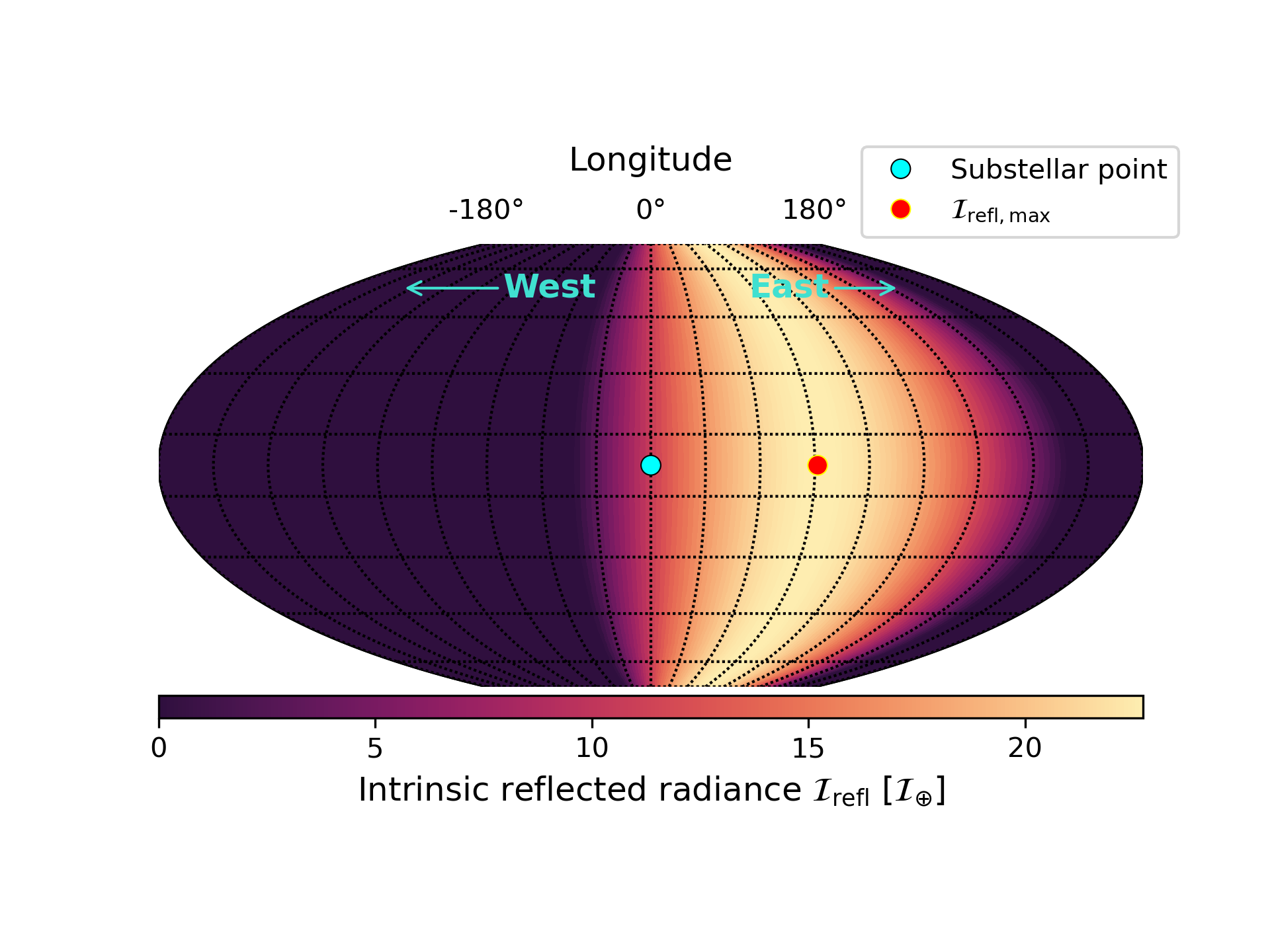}
    \caption{TESS longitude-latitude maps of the candidate planet assuming the best-fit parameters $\varepsilon$, $A_g$, $\eta_c$, $R_p$, inferred from the phase-curve fitting (Table~\ref{tab:grid_fit_results}). \textit{Top:} Temperature map with moderate heat redistribution, $\varepsilon=0.52$ and an internal temperature of $T_{\rm int} = 100$ K. 
    \textit{Bottom:} Reflection map normalised to the intrinsic Earth radiance with a geometric albedo of $A_g = 0.75$ and a cloud offset of $\eta_c = -62^{\circ}$.}
    \label{fig:emission_reflection_maps_TESS}
\end{figure}

\end{document}